\documentclass[paper]{JHEP3}
\usepackage{epsfig}
\usepackage{axodraw}
        \newdimen\eqskip
        \newdimen\txtskip
        \baselineskip=\txtskip
        \newdimen\mysep                
        \newdimen\hmysep
  \newcommand{\ccaption}[2]{
    \begin{center}
    \parbox{0.85\textwidth}{
      \caption[#1]{\small{{#2}}}
      }
    \end{center}
    }

\def    \be             {\begin{equation}}
\def    \ee             {\end{equation}}
\def    \ba             {\begin{eqnarray}}
\def    \ea             {\end{eqnarray}}

\def    \=              {\;=\;}
\def    \frac           #1#2{{#1 \over #2}}

\def    \bra#1          {\mbox{$\langle #1 |$}}
\def    \ket#1          {\mbox{$| #1 \rangle$}}

\def    \gev            {\mbox{$\mathrm{GeV}$}}

\def \sss {\scriptscriptstyle}
  
\def    \mZ             {\mbox{$m_Z$} }

\def    \mtsq             {\mbox{$m_{\sss{top}}^2$}}

\def    \pt             {\mbox{$p_{\sss T}$}}
\def    \ptjet             {\mbox{$p^{\sss{jet}}_{\sss T}$}}
\def    \ptpart             {\mbox{$p^{\sss{part}}_{\sss T}$}}
\def    \ptjet#1         {\mbox{$p_{\sss T, #1}$}}

\def    \etapart             {\mbox{$\eta_{\sss{part}}$}}
\def    \etamax         {\mbox{$\eta_{\sss{max}}$}}
\def    \pttop             {\mbox{$p^{\sss{top}}_{\sss T}$}}
\def    \pttopsq             {\mbox{$p^2_{\sss t, \sss T}$}}

\def    \pttbarsq             {\mbox{$p^2_{\overline{\sss t},\sss T}$}}
\def    \ptpair             {\mbox{$p^{\sss t\bar{\sss t}}_{\sss T}$}}
\def    \ptmin             {\mbox{$p_{\sss T}^{\sss{min}}$}}
\def    \Rmin             {\mbox{$R_{\sss{min}}$}}

\def    \dR             {\mbox{$\Delta R$}}
\def    \dphitt        {\mbox{$\Delta \phi^{\sss t \overline{\sss t}}$}}
\def    \dphil        {\mbox{$\Delta \phi^{\sss{lept}}$}}

\def    \et             {\mbox{$E_T$}}

\def    \etclus         {\mbox{$E^{\sss{clus}}_{\sss T}$}}
\def    \etaclmax       {\mbox{$\eta^{\sss{clus}}_{\sss{max}}$}}

\def    \Rmat           {\mbox{$R_{\sss{match}}$}}

\def    \rcone          {\mbox{$R_{\sss{cone}}$}}
\def    \rclus          {\mbox{$R_{\sss{clus}}$}}

\def    \njet           {\mbox{$N_{\sss{jet}}$}} 
\def    \nclus           {\mbox{$N_{\sss{clus}}$}} 

\def    \exc            {\mbox{$_{\sss{exc}}$}}
\def    \inc            {\mbox{$_{\sss{inc}}$}}

\newcommand {\ptja}{\mbox {$p_{\sss T,1}$}}

\newcommand {\yjet}{\mbox {$y_{{\sss 1}}$}}

\newcommand {\ytop}{\mbox{$ y_{\sss {top}}$}}
\newcommand {\phitt}{\dphitt} 
        
\newcommand     \MSB            {\ifmmode {\overline{\rm MS}} \else 
                                 $\overline{\rm MS}$  \fi}

\def    \mufsq          {\mbox{$\mu^2_{\rm F}$}}

\def    \mursq          {\mbox{$\mu^2_{\rm R}$}}

\def    \as             {\ifmmode \alpha_s \else $\alpha_s$ \fi}

\def \oacube {\mbox{$ {\cal O}(\alpha_s^3)$}}

\def \oatwo {\mbox{$ {\cal O} (\alpha_s^2)$}}

\def \oas   {\mbox{$ {\cal O}(\alpha_s)$}}

\def\rt1{\raisebox{-1ex}{\rlap{$\; \rho \to 1 \;\;$}}
\raisebox{.4ex}{$\;\; \;\;\simeq \;\;\;\;$}}

\def\herwig{{\small HERWIG}}
\def\herwigs{{\small HERWIG} \ }

\def\pythia{{\small PYTHIA}}

\def\ALPGEN{{\small ALPGEN}}
\def\ALPGENs{{\small ALPGEN} \ }
\def\alpgen{{\small ALPGEN}}
\def\alpgens{{\small ALPGEN} \ }
\def\mcnlo{{\small MC@NLO}}
\def\mcnlos{{\small MC@NLO} \ }
\def\ppbar{\mbox{$p \bar{p}$}}
\def\ttbar{\mbox{$t \bar{t}$}}

\def\met{$\rlap{\kern.2em/}E_T$}

\newcommand{\ben}{\begin{enumerate}}
\newcommand{\een}{\end{enumerate}}
\newcommand{\bit}{\begin{itemize}}
\newcommand{\eit}{\end{itemize}}

\title{Matching matrix elements and shower evolution for
      top-quark production in hadronic collisions}
\vfill                                                       
\author{
  Michelangelo L. MANGANO\\
CERN, TH-PH, CH~1211 Geneva 23, Switzerland
\\ E-mail: \email{michelangelo.mangano@cern.ch}}
\author{Mauro MORETTI\\
Dipartimento di Fisica, Universit\`{a} di 
Ferrara, and INFN, Ferrara, Italy \\
E-mail: \email{mauro.moretti@fe.infn.it}}
\author{Fulvio PICCININI \\ 
INFN, Sezione di Pavia, v. A. Bassi 6, I~27100, Pavia, Italy \\
E-mail: \email{fulvio.piccinini@pv.infn.it}}
\author{Michele TRECCANI\\
Dipartimento di Fisica, Universit\`{a} di 
Ferrara, and INFN, Ferrara, Italy \\
E-mail: \email{michele.treccani@fe.infn.it}}
\abstract{We study the matching of multijet matrix elements and shower
  evolution in the case of top production in hadronic collisions at
  the Tevatron and at the LHC. We present the
  results of the matching algorithm implemented in the \alpgen\
  Monte Carlo generator, and compare them with results obtained at the
  parton level, and with the predictions of the MC@NLO approach. We
  highlight the consistency of the matching algorithm when applied to
  these final states, and the excellent agreement obtained with MC@NLO
  for most inclusive quantities. We nevertheless identify also a
  remarkable difference in the rapidity spectrum of the leading jet
  accompanying the top quark pair, and comment on the likely origin
  of this discrepancy.}

\preprint{CERN-PH-TH/2006-232\\FNT/T-2006/09\\hep-ph/0611129 \\ November 2006}

\begin{document}
\section{Introduction}
\label{sec:intro}
Following the excellent performance and large statistics accumulated
by the Tevatron collider, and in anticipation of the forthcoming
beginning of the LHC operations, the recent years have witnessed an
impressive progress in the development of improved tools for the
simulation of the complex final states produced in hard hadronic
collisions~\cite{Mangano:2005dj,Dobbs:2004qw}.  The leading themes of
these advances have been the inclusion of next-to-leading-order (NLO)
matrix elements in the shower Monte Carlo (MC)
codes~\cite{Frixione:2002ik,Nason:2004rx}, and the consistent merging
of shower MCs with the leading-order (LO) calculations for final
states with many hard
partons~\cite{Kuhn:2000cr}-\cite{Hoche:2006ph}. These two directions
provide alternative approaches to the common goal of improving the
accuracy of the description of multijet final states. In the first
case, known as \mcnlo, the emphasis is on achieving the NLO
accuracy in the description of the inclusive rates for a given final
state $F$, accompanied by the exact LO description of the emission of
one extra jet ($F+$jet). Concrete implementations so far include the
cases where $F$ is a pair of gauge
bosons~\cite{Frixione:2002ik,Nason:2006hf}, a heavy-quark
pair~\cite{Frixione:2003ei,Nason:2006XX}, a single vector boson or a
Higgs~\cite{Frixione:2003vm}, and single
top~\cite{Frixione:2005vw}. In the second approach, the goal is to
maintain a consistent leading-logarithmic (LL) accuracy in the
prediction of a final state $F$ accompanied by a varying number of
extra jets. This means that the cross sections
for each jet multiplicity $N$ are calculated using the LO matrix
elements for $N$ hard partons, followed by the full shower evolution
obtained with a shower MC. The removal of double counting of jet
configurations that would appear both from hard emissions during
the shower evolution and from the inclusion of the higher-order matrix
elements is achieved in the \mcnlo\ technique through the explicit
subtraction from the NLO matrix elements of the \oas\ emission
probabilities generated by the shower. It is instead achieved in the
second approach via the inclusion in the LO matrix elements of the
appropriate Sudakov form factors, and by vetoing shower evolutions
leading to multiparton final states already described by the matrix
element computation.  This procedure is known as a {\em matching
  algorithm} for matrix elements and parton shower. While the LL
matching algorithm approach cannot be expected to accurately reproduce
the inclusive NLO cross section and its stability w.r.t. scale variations,
the exact LO description of higher multiplicity partonic final states
will give it a better accuracy for the distributions of two or more
jets produced in addition to $F$. In this respect the two approaches
are complementary, both in goals and in expected
performance. Furthermore, we expect that they should give comparable
results for $F$ and $F$+1 jet inclusive distributions, up to a possible
{\em overall} NLO $K$ factor to be applied to the LL results.

The aim of this note is to compare the results of the two approaches
in the case of top quark production. We use the code developed by
Frixione et al~\cite{Frixione:2003ei} to generate  the \mcnlo\
results, and the \alpgen\ code~\cite{Mangano:2002ea} 
with the so-called {\em \small MLM}
  matching~\cite{mlmfnal,Mangano-ta,Hoche:2006ph} to generate the LL
distributions. Section~2 will briefly review this algorithm. Section~3
will show in detail the results of the \alpgen\ calculations for
\ttbar\ production, presenting a series of consistency checks of the
apporach and  discussing its systematic uncertainties. Section~4
covers the comparison between \alpgen\ and \mcnlo, and Section~5 will
present our conclusions.

\section{Review of the matching algorithm}
\label{sec:matching}
The main requirements and
features of a good matching algorithm are:
\begin{itemize}
\item The removal of double counting of equivalent phase-space configurations.
\item The ability to merge together samples with different hard-parton
  multiplicity, generating fully inclusive event samples leading to
  predictions for physical
  observables that are independent of the partonic generation cuts. 
 In particular, this means that the physical definition of
  a jet will not be a required input for the generation, and that any
  definition of jet can be used at the analysis level.
\end{itemize}

The algorithm used in this work 
 is defined by the following rules (for more details, see~\cite{Mangano-ta}):
\begin{enumerate}
\item Generate parton-level configurations for all final-state parton
  multiplicities up to $N$, with partons constrained by 
\be \label{eq:cuts}
 \ptpart>\ptmin\; , \quad \vert \etapart \vert < \etamax \; , \quad 
 \dR>\Rmin \; ,
\ee
where \ptpart\ and \etapart\ are the transverse momentum and
pseudorapidity of the light final-state partons, 
and \dR\ is their minimal separation in the
$(\eta,\phi)$ plane. 
Each of the samples will be called the $n$-parton sample
  ($n=0,\dots,N$), and the one with  $n=0$ will also be referred to as
   the {\em lowest-order} sample. 
\item Perform the shower evolution on each $n$-parton sample, 
using e.g. the standard \herwig~\cite{Marchesini:1988cf}
 or \pythia~\cite{Sjostrand:1994yb}
  shower MC codes.
\item For each event, apply a cone jet algorithm to all partons
  resulting from the shower evolution, before hadronization. We call
  {\em clusters} the resulting jets; they are defined by a minimum
  \et, \etclus, and by a jet cone size \rclus, parameters which are
  related but not necessarily identical to the partonic generation
  parameters \ptmin\ and \Rmin.
\item Associate each parton from the PL event to one and only one of
  the reconstructed clusters:
  \begin{itemize}
  \item Starting from the highest-\pt\ parton, select the cluster with
    minimum distance \dR\ from it; if $\dR < \Rmat$, where \Rmat\
    is a fixed parameter called the matching radius, then we say that
    the parton is {\em matched}.
  \item Remove the cluster from the list of clusters, go to the next
    parton and iterate until all hard partons have been processed. 
    Since clusters are removed from the list after they are matched,
 a given cluster can only be matched to a single parton.
  \end{itemize}
\item If each  parton is matched to a cluster, the event ``matches'', and
  is kept for further scrutiny, else it is rejected.
\item In the case of $n<N$, matched events with a number of clusters
  $\nclus>n$ are rejected. This leaves a exclusive sample with
  $\nclus=n$.
\item If $n=N$, the largest parton multiplicity for which we generated
  PL events, accept matched events where $\nclus>N$, provided the non-matched
  clusters (namely those remaining in the cluster list after all clusters
  matching partons have been removed) are softer than each of the
  matched clusters.
\item After matching, combine the exclusive event samples from each partonic
  multiplicity $n=0,\dots,N-1$ and the inclusive event sample with
  $n=N$ into a single event sample, which will define the fully
  inclusive sample.
\end{enumerate}
We shall use
 the implementation of this matching algorithm given in the \alpgen\
code, but the algorithm can be implemented in
any other matrix-element based programme. 
For the shower evolution we use \herwig~\cite{Marchesini:1988cf}, version
6.510~\cite{Corcella:2002jc}. For a more direct comparison with
the PL results, we stopped the evolution after the perturbative phase,
and our results do not therefore include the effects of cluster
splitting and hadronization.

\section{Consistency studies of the matching algorithm}
\label{sec:sanity}
In this section we study the overall consistency of
the matching algorithm applied to \ttbar\ final states. 
We need to verify the following:
\begin{itemize}
\item Inclusive distributions obtained after the matching should
  reproduce inclusive quantities as calculated at the PL.
\item Physical observables should be stable w.r.t. variations of the
  phase-space cuts applied in the generation of the PL samples (the
  {\em generation parameters}), and  w.r.t. variations of the
  parameters chosen for the parton-jet matching and extra-jet vetoes  (the
  {\em matching parameters}).
\end{itemize}
To start with, we introduce our sets of generation and matching
parameters.  We shall consider \ttbar\ production at the Tevatron
(\ppbar\ collisions at $\sqrt{S}=1.96$~TeV) and at the LHC ($pp$
collisions at $\sqrt{S}=14$~TeV). The generation parameters for the
light partons are defined by the kinematical cuts given in
eq.(\ref{eq:cuts}), while no cuts are applied to the top quarks. The
numerical values chosen for the generation of the default event
samples at the Tevatron (LHC) are given by: \ptmin=20 (30) \gev,
\etamax=4 (5) and \Rmin=0.7 (0.7).
The top quarks are assumed to be stable (with the exception of the
study of spin correlations in Section~\ref{sec:topdec}), and
therefore all jets coming from the decay of top quarks are neglected.

For all generations we chose the parton distribution function set
{\small MRST2001J}\cite{Martin:2001es}, with renormalization and
factorization scales squared set equal to the sum of the
squared transverse
masses of the final state partons:
$\mursq=\mufsq=\sum_{i=t,\bar{t},\mathrm{jets}} \; [m_i^2+(p_{\sss T}^i)^2]$.

The default matching parameters are defined by the following set of relations:
\be \label{eq:matchparams}
\etclus=\rm{max}(\ptmin+5~\gev,1.2\times \ptmin)\; , 
\quad  \etaclmax = \etamax \; , \quad 
\Rmat = 1.5 \times \Rmin \; ,
\ee
where \etclus\ is the minimum transverse energy of the jet clusters used
for the jet-parton matching, \etaclmax\ is their maximum $\vert \eta
\vert$ and \Rmat\ is the maximum separation between parton and jet
cluster required for the parton-jet pair to match. Jet clusters are
defined by the cone algorithm provided by the
{\small GETJET} package \cite{getjet}, which represents a simplified
jet cone algorithm a la UA1. Variations of these
default choices will be defined when exploring the parameter
dependence of the results.

The event samples emerging after the showering, matching and veto are
 defined by the multiplicity of the light partons present in the PL
 sample, $n$, and by the presence (or absence) of the extra-jet veto:
 $n\exc$ ($n\inc$). For example, the sample $1\exc$ refers to
 the event sample obtained after the showering, matching and extra-jet
 veto of a set of $\ttbar+1-$parton PL events.
The event sample obtained by combining
 $0\exc+1\exc+\dots+(n-1)\exc+n\inc$ will be referred to as the $S_n$ sample.

The sample $S_n$ constructed according to the above prescriptions can
then be used for arbitrary analyses of the final states.
The analysis phase is independent of the way the sample was generated;
in particular one is allowed to
choose an arbitrary jet-finding algorithm ($k_\perp$, cone, mid-point cone,
 $\dots$), possibly different than the algorithm used to carry out the
matching. 
It's for a mere matter of convenience that we adopt here the
same clustering algorithm that we have used at the matching stage,
namely the {\small GETJET} jet definition. Jet observables are built
out of the  partons emerging form the shower in the rapidity range
 $\vert\eta\vert \le 6$. The jet cone size is set to
$\rcone=0.7$ and the minimum transverse momentum to define a jet is
15~\gev\ at the Tevatron, and 20~\gev\ at the LHC.

\subsection{Comparison with parton-level results}
\label{sec:PL}
In this section we compare inclusive distributions obtained after the
shower evolution and matching with the distributions derived at the
pure PL.  Here we confine ourselves to the case of 0 and 1 light
partons, analyzing the results of the $S_1=0\exc +1\inc$ combined
sample.

The cross-section results obtained for the Tevatron and LHC are shown
in table~\ref{tab:sigma}. For the results after shower and matching,
we quote the individual rates obtained for the $0\exc$ and $1\inc$
samples, together with their sum, which is the only physical
quantity. For the PL results, we quote separately the \oatwo, Born
level cross section, and the full NLO, \oacube, cross
section~\cite{Mangano:1991jk}. We notice that the matching 
algorithm reproduces very well the inclusive LO cross section, where the
rate reduction of the \ttbar\ PL process due to the exclusive veto
that removes jet events after the shower is properly compensated by
the rate of the \ttbar+1 parton process (the almost complete agreement
in the LHC case should be taken as accidental). 

\begin{table}
\begin{center}
\begin{tabular}{|l|ll|lll|}
\hline
Collider & LO         & NLO  & $0\exc$  & $1\inc$ & $0\exc+1\inc$  \\
\hline
Tevatron & {\bf 4.37} & 6.36 &  3.42    &  0.78   & {\bf 4.20} \\
LHC      & {\bf 471}  & 769  &  217     &  252    & {\bf 469} \\
\hline
\end{tabular}
\ccaption{}{\label{tab:sigma} Cross sections (in pb) at the LO
(\oatwo) and NLO (\oacube) PL, compared with the rates of the $S_1$
samples.}
\end{center}
\end{table}

\begin{figure}
\begin{center}
\includegraphics[height=0.22\textheight,clip]{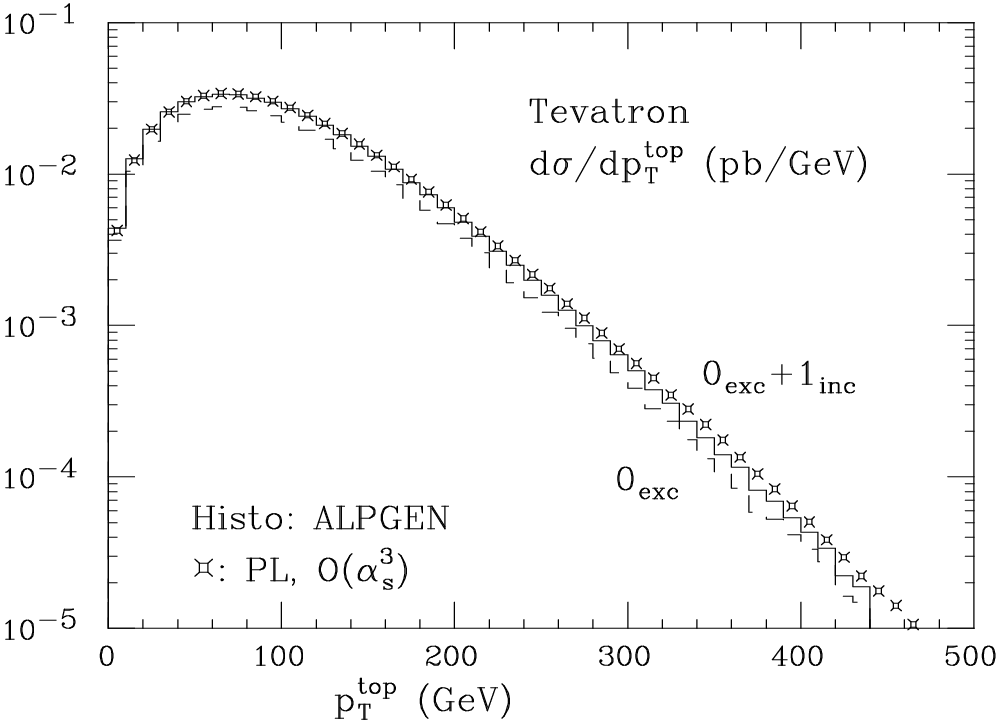}
\hspace*{0.5cm}
\includegraphics[height=0.22\textheight,clip]{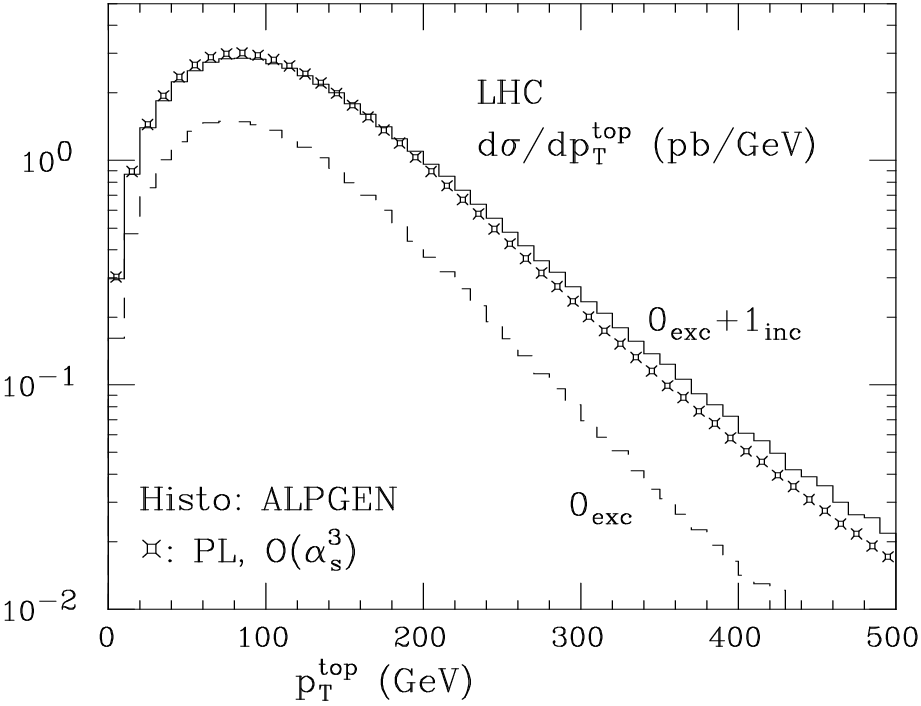}
\\
\includegraphics[height=0.22\textheight,clip]{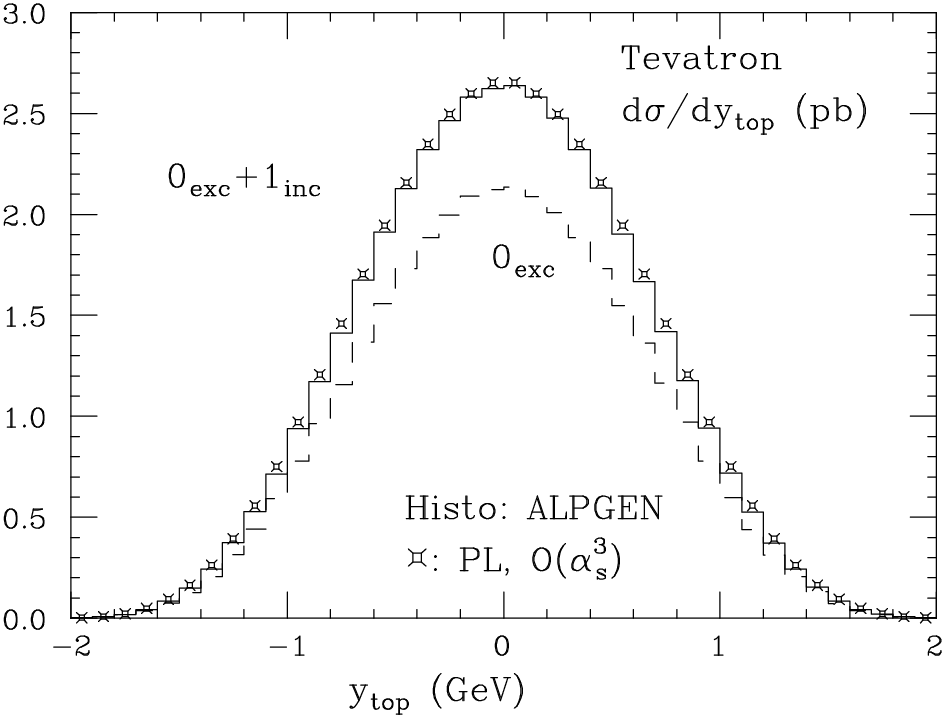}
\hspace*{0.5cm}
\includegraphics[height=0.22\textheight,clip]{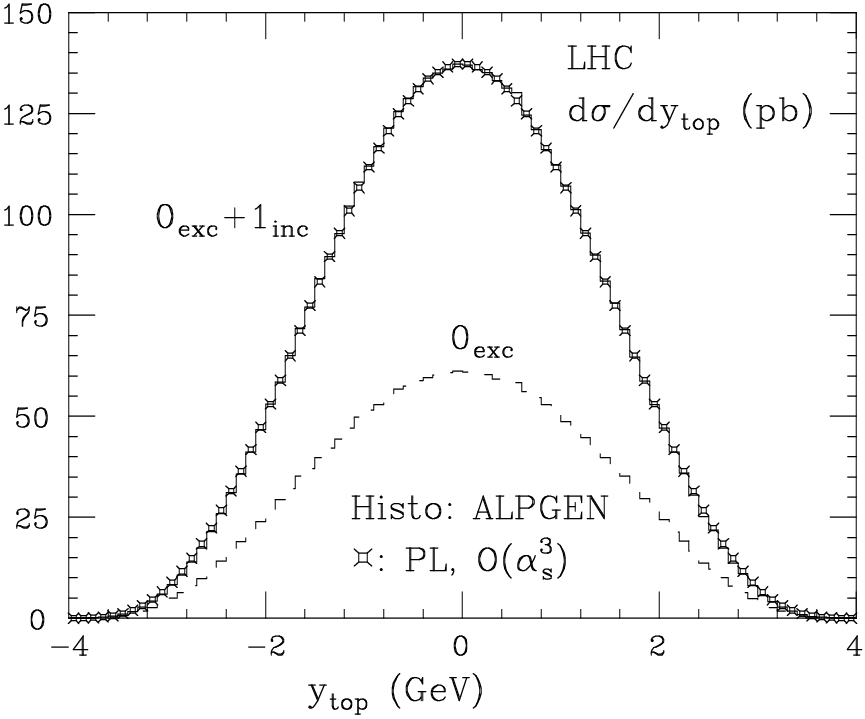}
\\
\includegraphics[height=0.22\textheight,clip]{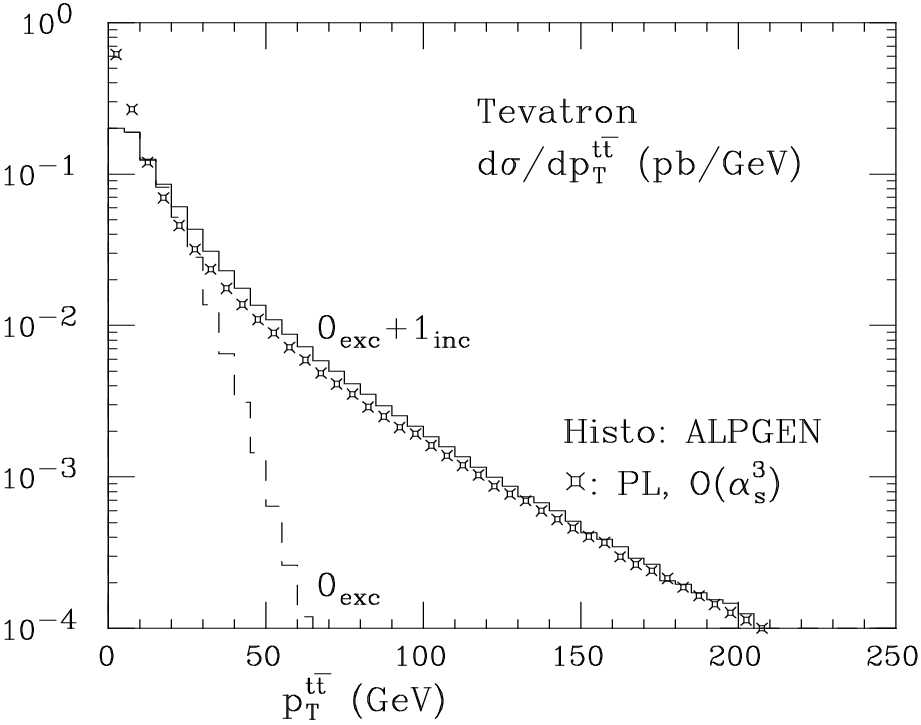}
\hspace*{0.5cm}
\includegraphics[height=0.22\textheight,clip]{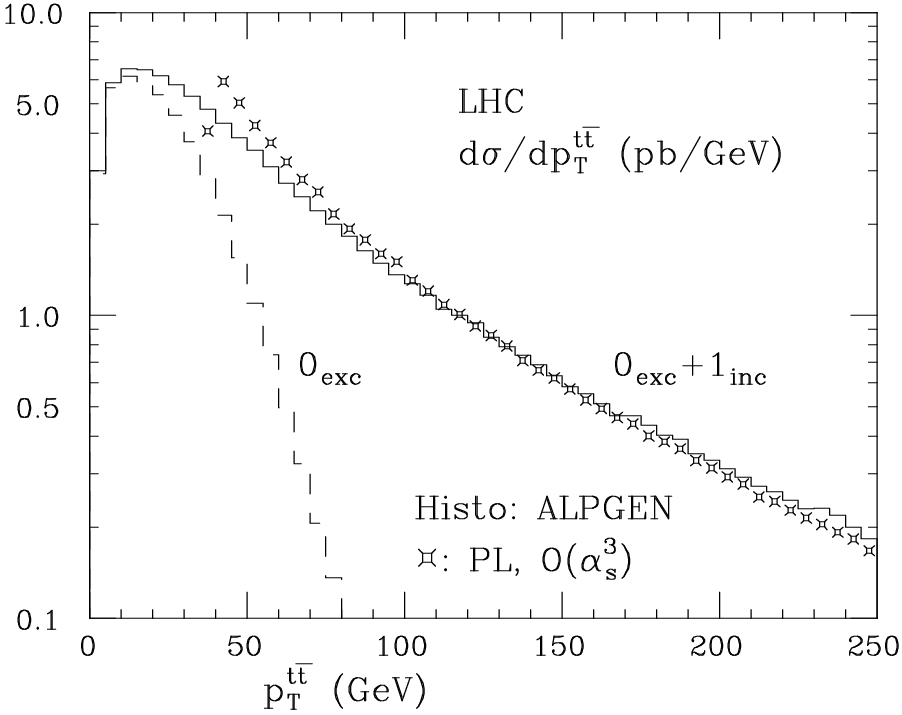}
\\
\includegraphics[height=0.22\textheight,clip]{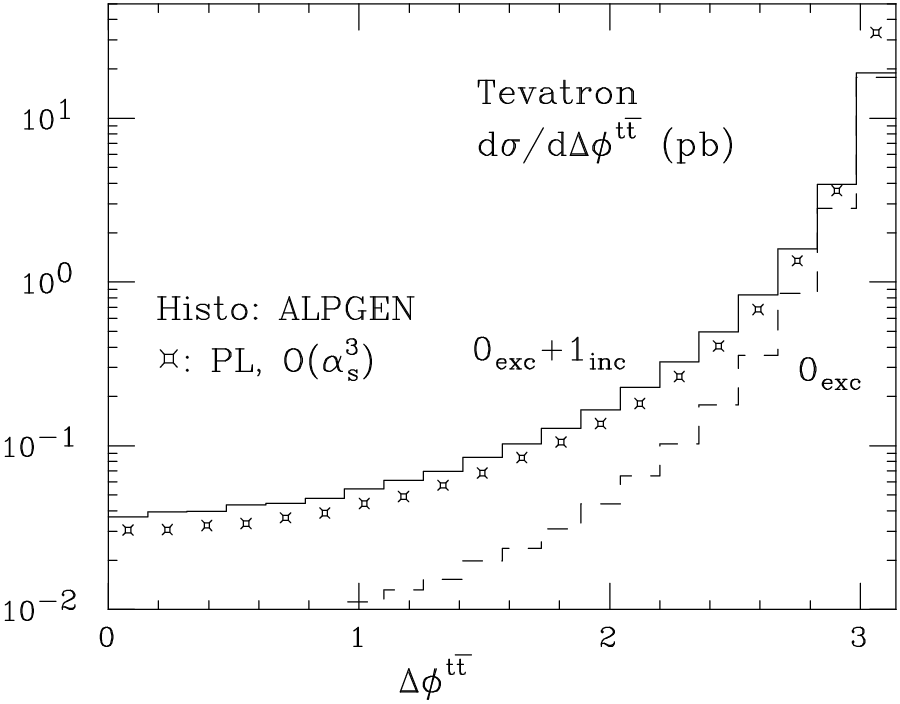}
\hspace*{0.5cm}
\includegraphics[height=0.22\textheight,clip]{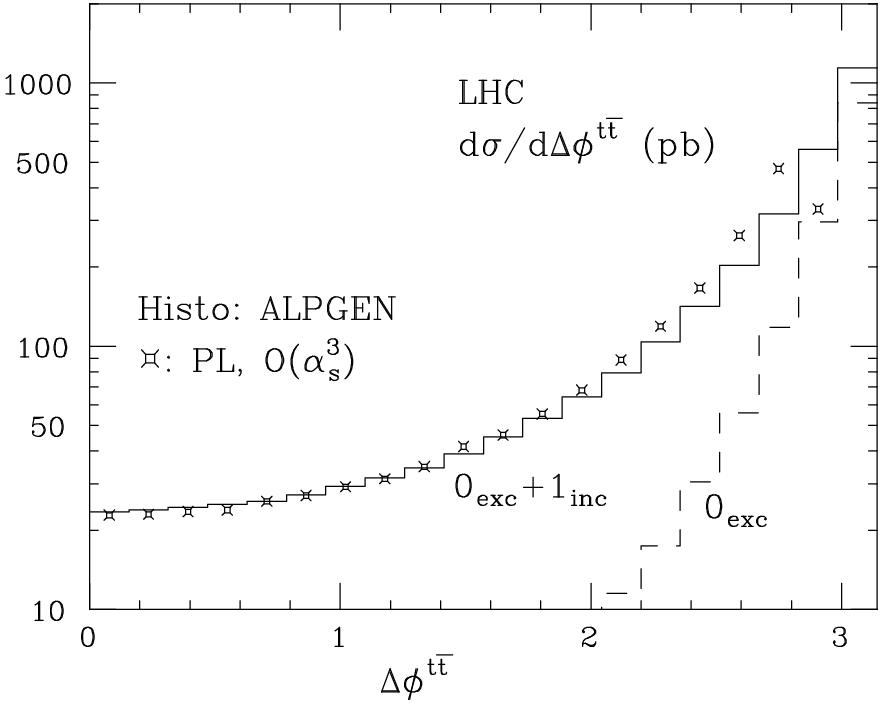}
\ccaption{}{\label{fig:pt_PL} Comparison of the \alpgen\ $S_1$ results
and the LO PL spectra for the inclusive transverse momentum and
rapidity of top
quarks, for the transverse momentum of the \ttbar\ pair, and for their
azimuthal correlations. All distributions are absolutely
normalized. The contribution of the 0\exc\ sample is shown by the
dashed line. The plots on the left are for the Tevatron, those on the right for
the LHC.}
\end{center}
\end{figure}

Figure~\ref{fig:pt_PL} shows a comparison in absolute rates, for the
Tevatron and the LHC, of four inclusive observables evaluated at the
PL and after shower evolution.  We plot the transverse momentum
(\pttop) and rapidity spectra (\ytop) of the top quark, , the spectrum
of the transverse momentum of the \ttbar\ pair, \ptpair, and the
azimuthal correlation \dphitt\ between the $t$ and $\bar{t}$
quarks. For \pttop\ and \ytop\ we compare the result of the $S_1$
sample with the PL Born spectrum. The sub-contribution coming from the
0\exc\ subset is also shown, as a dashed histogram. The \ptpair\ and
\dphitt\ distributions are
 non-trivial only starting at \oacube\, and therefore we
compare the $S_1$ results with the NLO calculation of
ref.~\cite{Mangano:1991jk}, in which the divergent terms at \ptpair=0
and at $\dphitt=\pi$ are cancelled between the real and virtual
contributions.

The agreement for \pttop\ is excellent. Likewise, there is excellent
agreement for \ptpair\ and \dphitt\ as soon as we move away from the
regions dominated by Sudakov effects (\ptpair=0 and $\dphitt=\pi$),
effects which are incorporated in the $S_1$ sample but which are not
present in the NLO calculation. 

Notice that at the LHC the Sudakov effects are much
stronger, as shown in the plot of the \ptpair\ variable. The first few
empty
bins in the NLO result are due to the
complete cancellation between the negative virtual rate at \ptpair=0 and the
positive \oacube\ rate integrated up to approximately 35 GeV.
Above this threshold, the \alpgen\ $S_1$ result and the NLO one agree
very well. These large Sudakov effects indicate that a fixed-order,
\oacube, calculation with parton \pt\ below 40-50 GeV is not
reliable. Nevertheless, in spite of the fact that we generated the
1-parton sample with a threshold of \ptmin=30~GeV, the combination of
matching and jet veto leads to a smooth interpolation between the soft
and hard \ptpair\ regions, as will be confirmed in a later section
with the comparison with the full NLO+shower treatment of \mcnlo.

These results give us good confidence
that the matching algorithm allows to merge the 0-parton and
1-parton samples with the proper removal of double counting, and the
accurate description of the hard-jet emission probability.
A more complete comparison with a
NLO calculation including the Sudakov effects is given below, where we
analyze \ttbar\ production using \mcnlo.

\clearpage

\begin{table}
\begin{center}
\begin{tabular}{|l|ll|}
\hline
      & Tevatron & LHC \\
\hline
0\exc & 3.42 & 216.6 \\ 
1\exc & 0.66 & 149.9 \\
2\exc & 0.09 & \phantom{0}65.8 \\
3\inc & 0.010 & \phantom{0}29.9 \\
\hline
Total & 4.18 & 462.2 \\
\hline
\end{tabular}
\ccaption{}{\label{tab:xsect} Cross sections (in pb),
for $t\bar t$ production at the Tevatron and LHC. The contribution of the
different parton samples, for the default generation and matching
options.
The relative numerical integration precision is at the permille level.}
\end{center}
\end{table}

\subsection{Impact of higher-order parton processes}
\label{sec:0123}
In this section we introduce higher-multiplicity final states in the
matrix element (ME) calculation. In particular, we generate PL samples
with up to 3 final-state partons in addition to the \ttbar\ pair.
After showering and matching, these events are combined into the fully
inclusive $S_3$ sample, contributing with the cross-sections given in
table~\ref{tab:xsect}. Notice that the overall rates are well
consistent with those obtained with the $S_1$ sample, in
table~\ref{tab:sigma}. This indicates that the matching algorithm
correctly ensures that the 1-parton inclusive rate, $\sigma(1\inc)$,
is reproduced by the sum of the partial contributions,
$\sigma(1\exc)+\sigma(2\exc)+\sigma(3\inc)$.

\begin{figure}
\begin{center}
\includegraphics[height=0.22\textheight,clip]{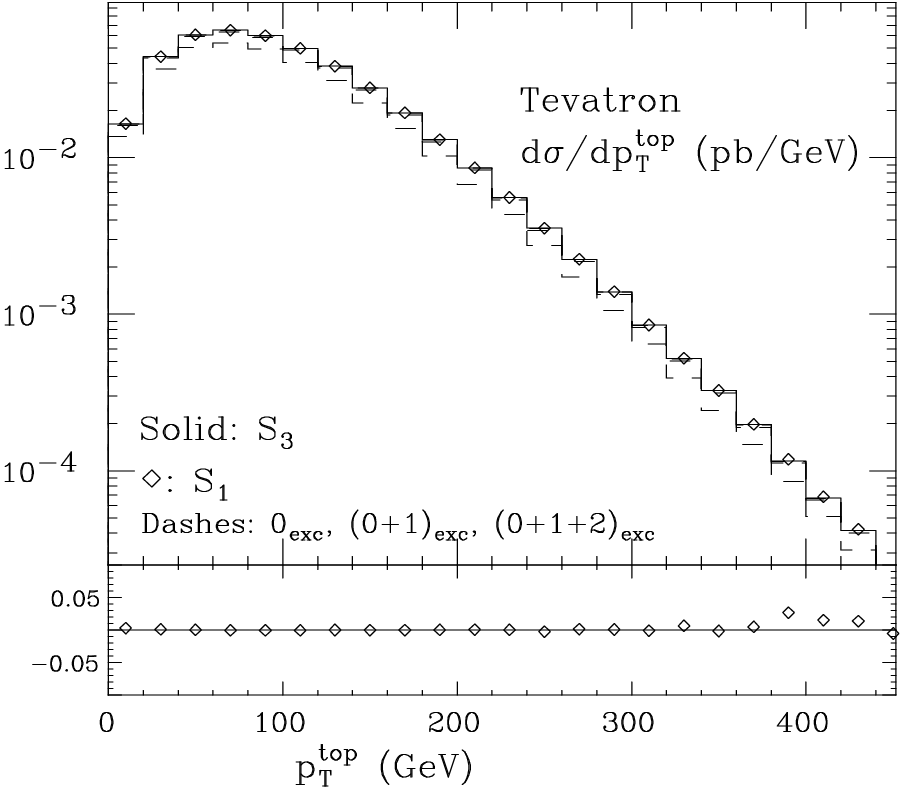}
\hspace*{0.5cm}
\includegraphics[height=0.22\textheight,clip]{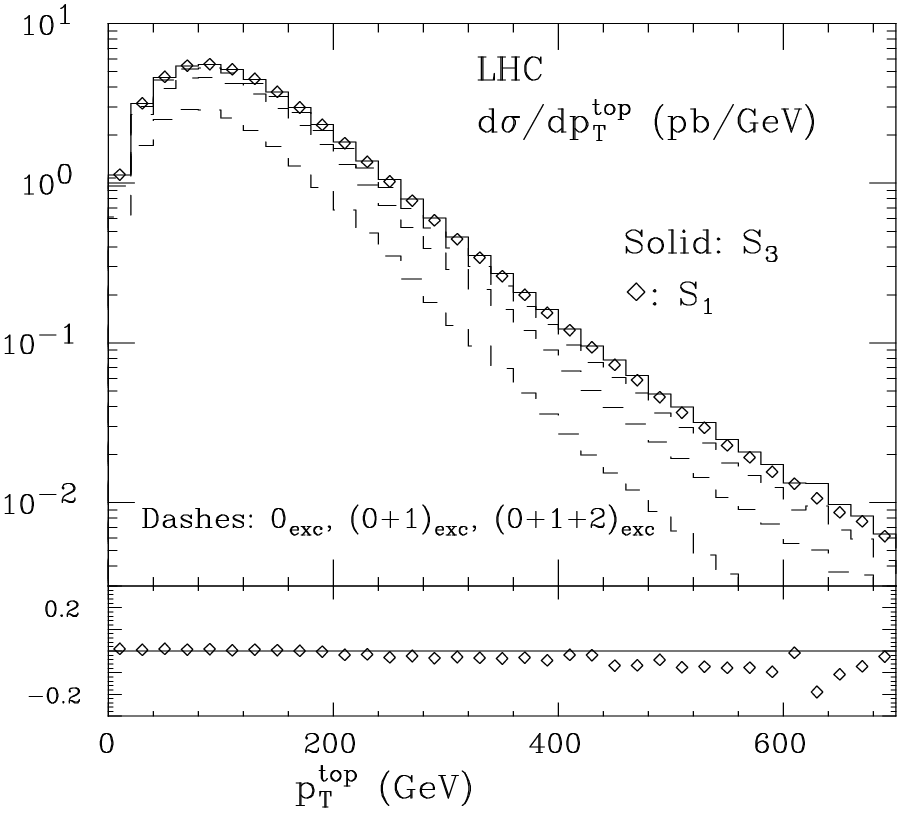}
\\
\includegraphics[height=0.22\textheight,clip]{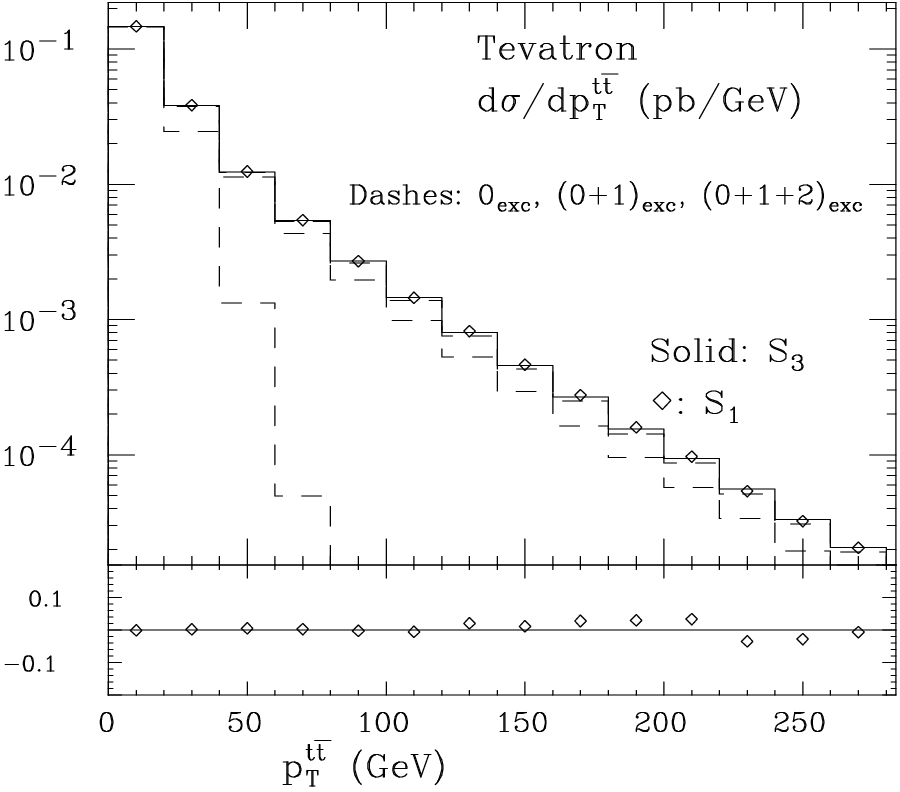}
\hspace*{0.5cm}
\includegraphics[height=0.22\textheight,clip]{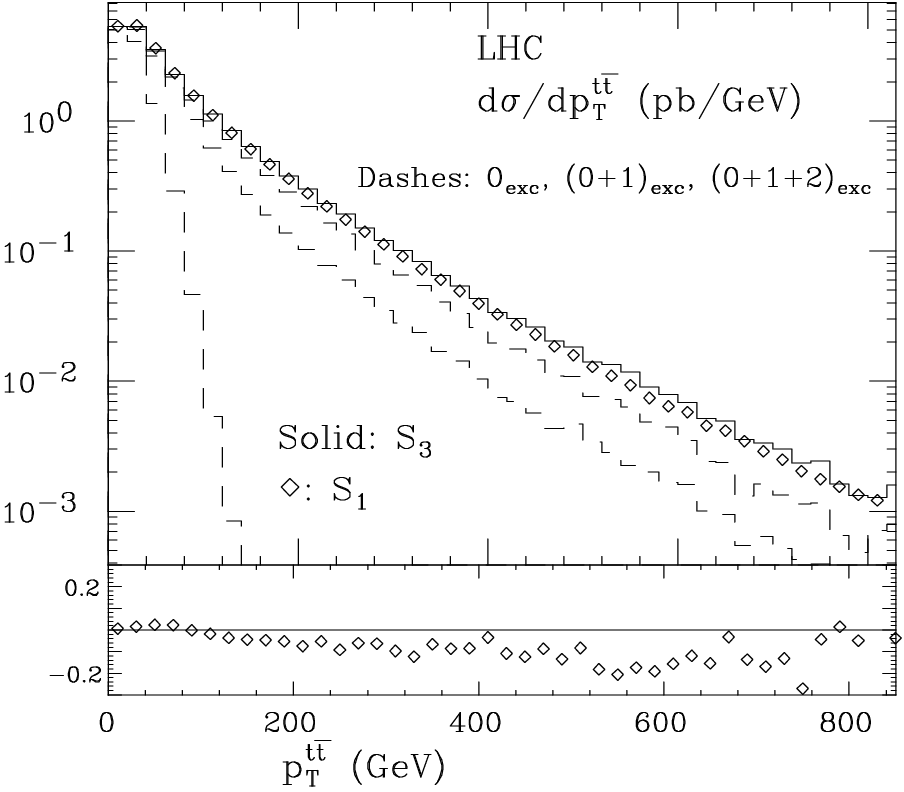}
\\
\includegraphics[height=0.22\textheight,clip]{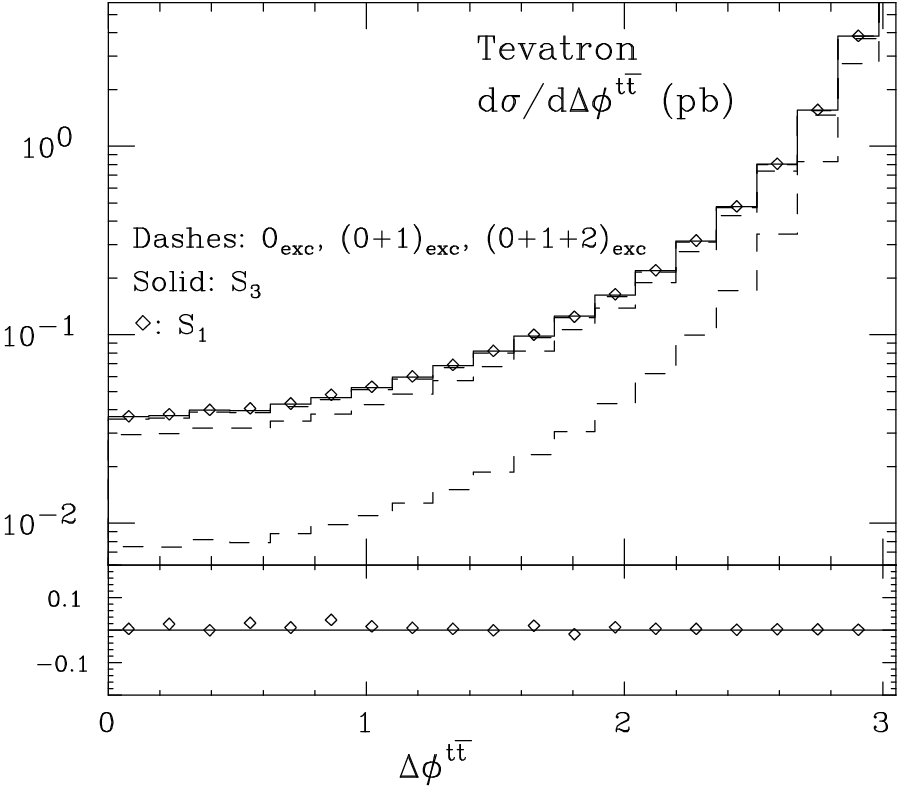}
\hspace*{0.5cm}
\includegraphics[height=0.22\textheight,clip]{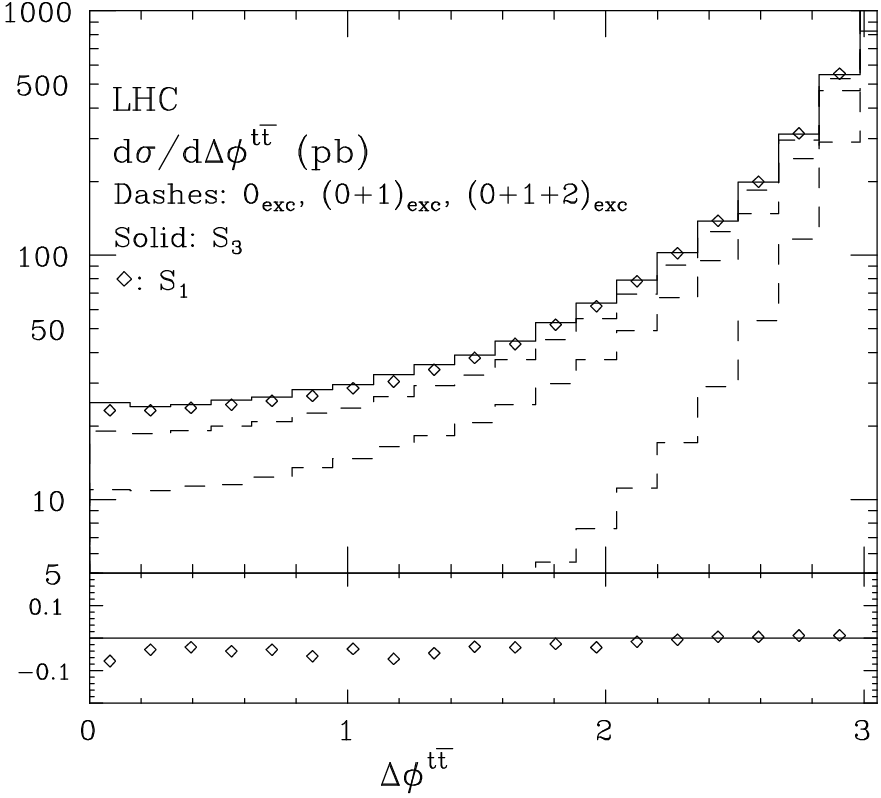}
\\
\includegraphics[height=0.22\textheight,clip]{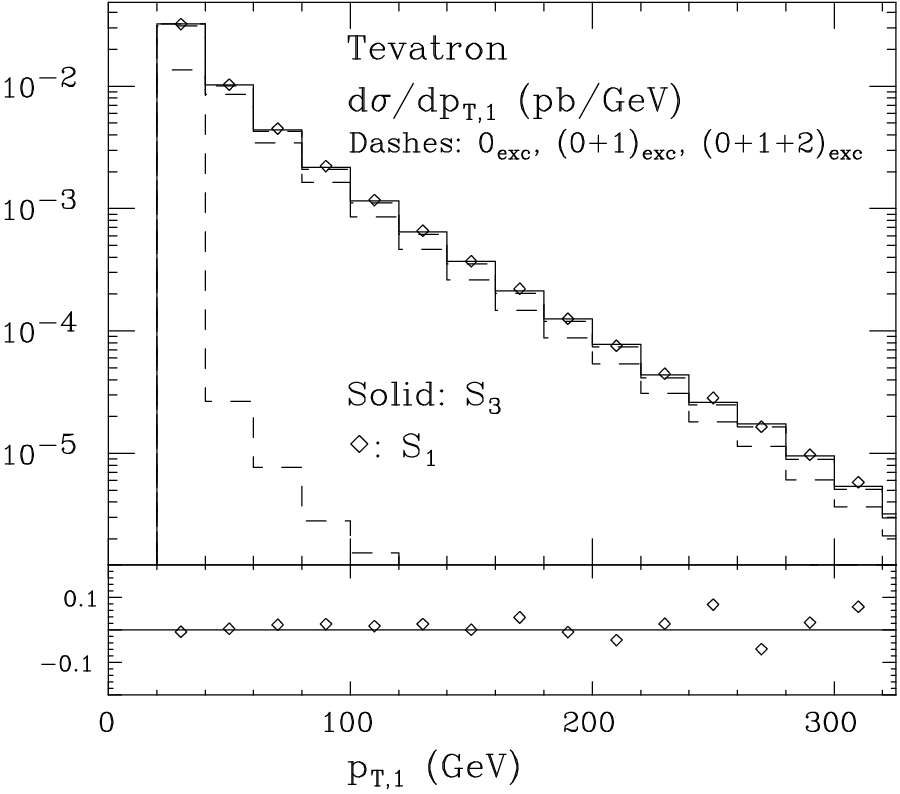}
\hspace*{0.5cm}
\includegraphics[height=0.22\textheight,clip]{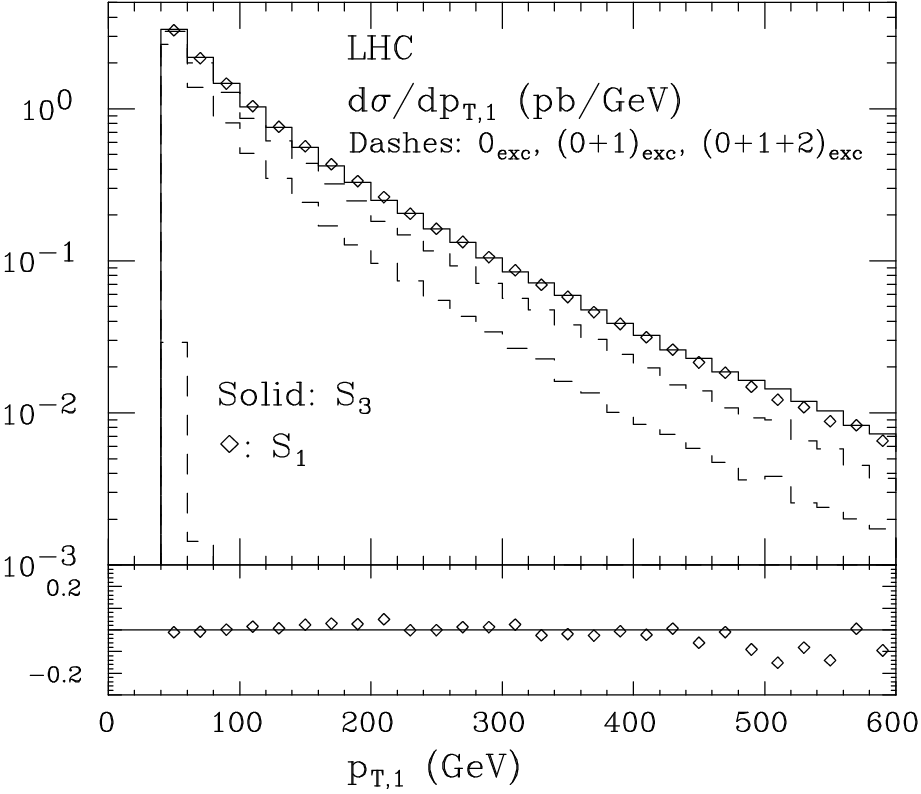}
\\
\ccaption{}{\label{fig:pt_01} Comparison between the distributions
obtained from the $S_1$ event samples (0\exc+1\inc) and from the $S_3$
event samples (0\exc+1\exc+2\exc+3\inc), for various ($\le 1$)-parton
observables at the Tevatron (left-hand side) and LHC (right-hand
side). Cumulative contributions from the 0\exc, 1\exc\ and 2\exc\
subsamples are shown by the dashed histograms.}
\end{center}
\end{figure}

This consistency is maintained, at the level of spectra, for the
inclusive distributions that receive their LO contributon from the 0-
and 1-parton final states. This is shown in the first three plots of
fig.~\ref{fig:pt_01}, where we compare the predictions for \pttop,
\ptpair, \dphitt\ and \ptjet{1} (the leading-jet \pt) obtained, for
the Tevatron, with the $S_1$ and $S_3$ event samples. 
In the figures we also
display the incremental contribution given by the various subsets,
0\exc, 1\exc\ and 2\exc.
The lower insets represent the relative
difference between the two results,
$[d\sigma(S_1)-d\sigma(S_3)]/d\sigma(S_3)$. As one can see these
differences are at the few-\% level at most, except in the
high-momentum tail of
the \pt\ distributions at the LHC, where the $S_3$ spectrum is harder than
$S_1$.

More interesting is the case of observables that receive their LO
contribution from final states with more than 1 extra hard parton.
The comparison between the predictions of the $S_1$ and $S_3$ samples 
 can tell us more about possible limitations of the shower
MC in describing hard emissions leading to extra jets.
\begin{figure}
\begin{center}
\includegraphics[height=0.22\textheight,clip]{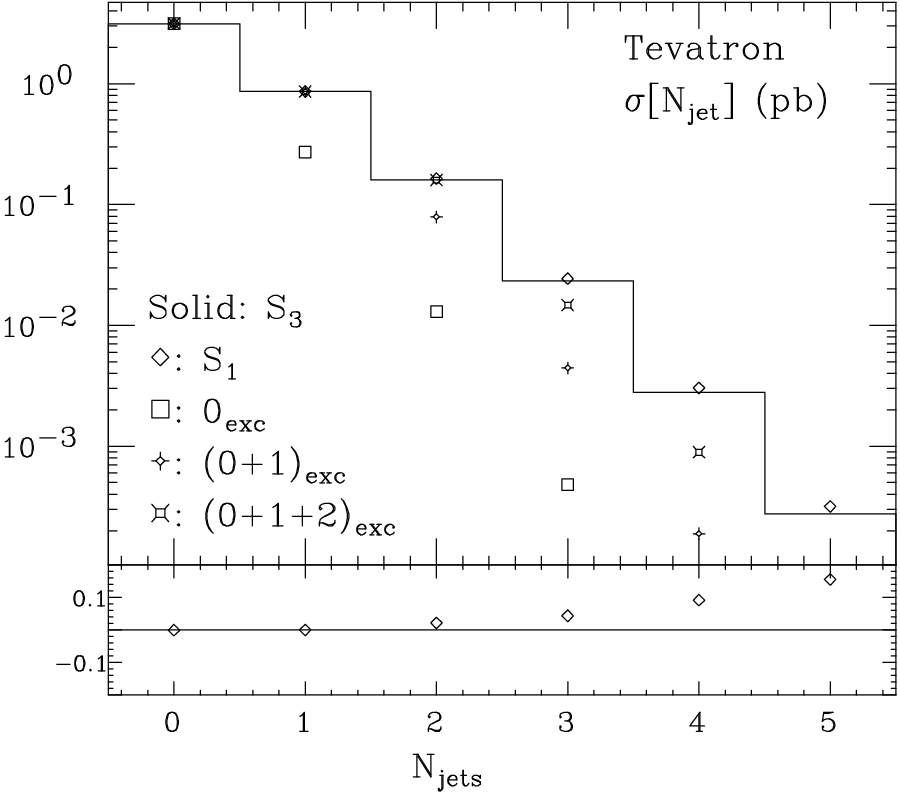}
\hspace*{0.5cm}
\includegraphics[height=0.22\textheight,clip]{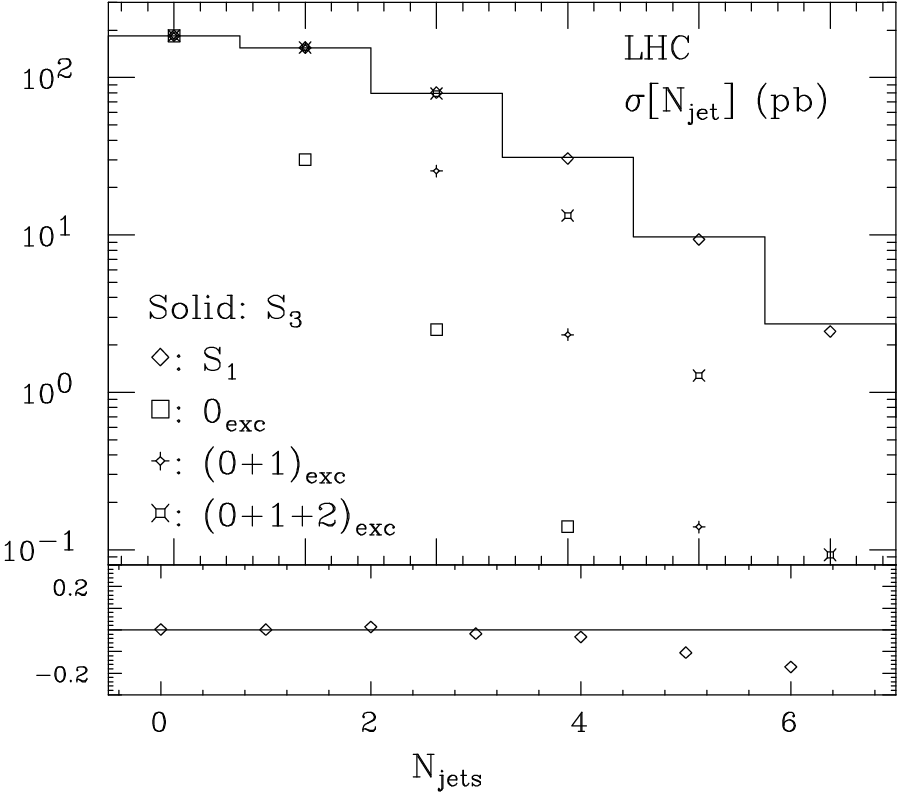}
\\
\includegraphics[height=0.22\textheight,clip]{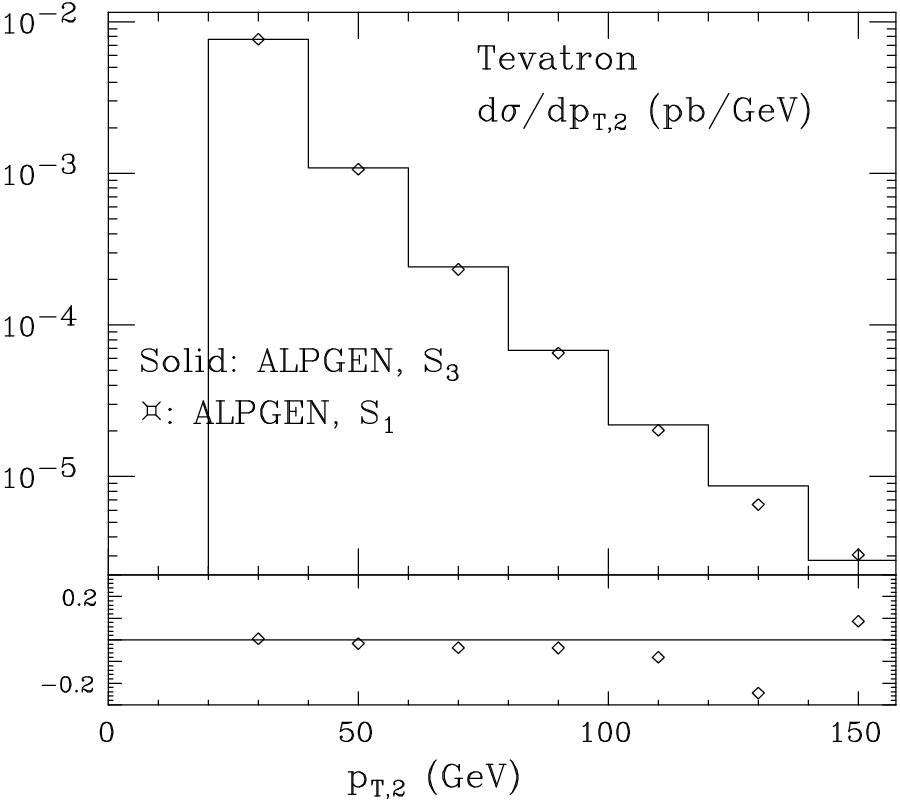}
\hspace*{0.5cm}
\includegraphics[height=0.22\textheight,clip]{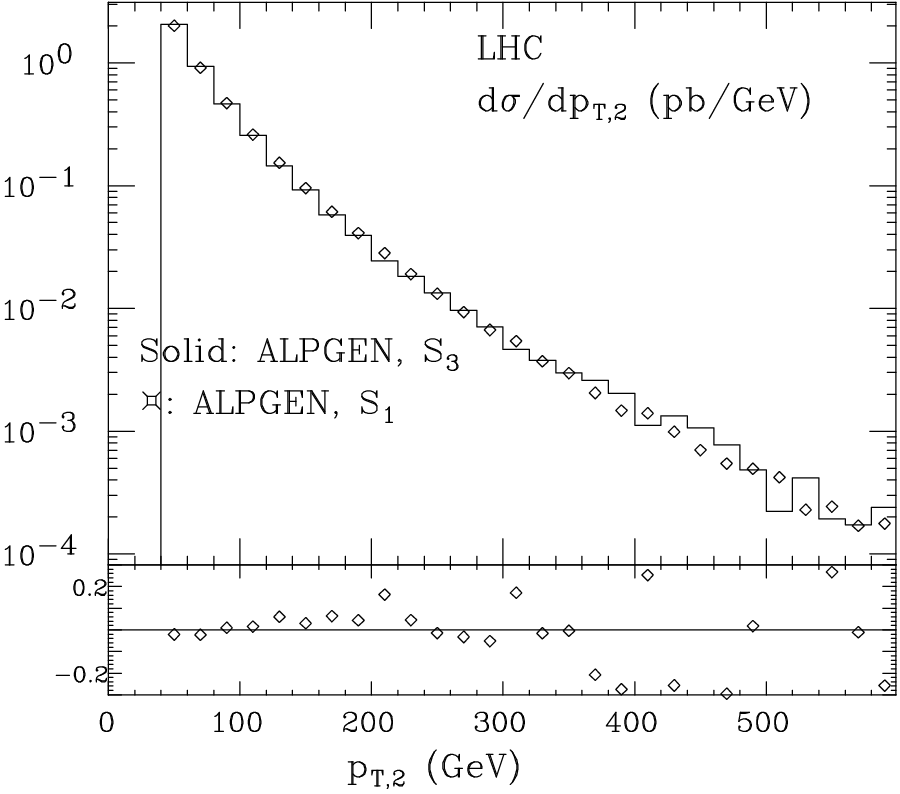} 
\\
\includegraphics[height=0.22\textheight,clip]{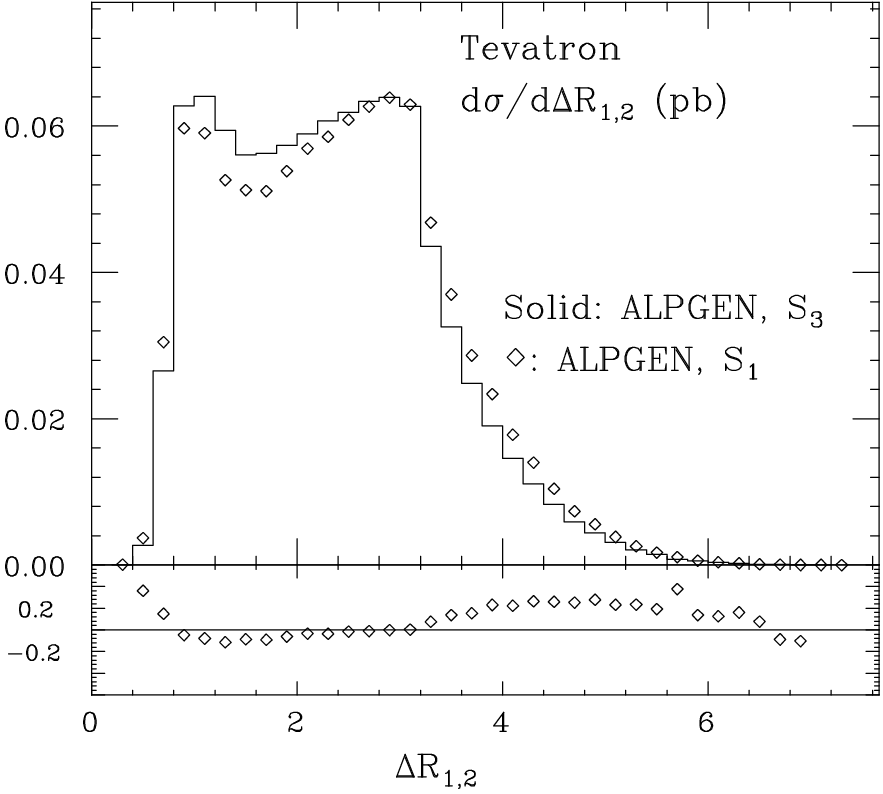}
\hspace*{0.5cm}
\includegraphics[height=0.22\textheight,clip]{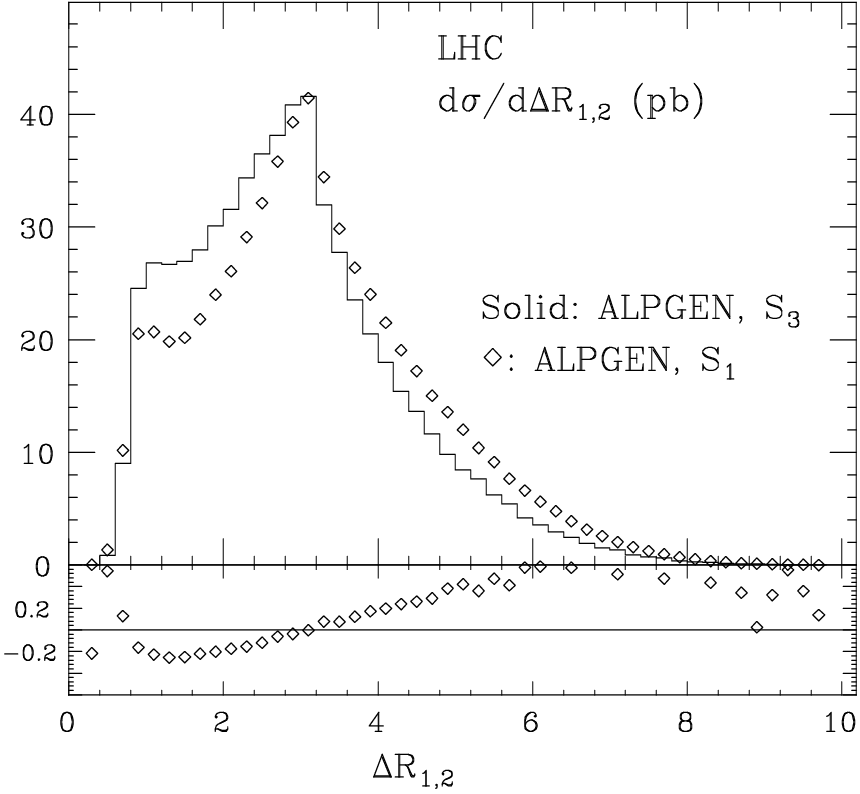}
\\
\includegraphics[height=0.22\textheight,clip]{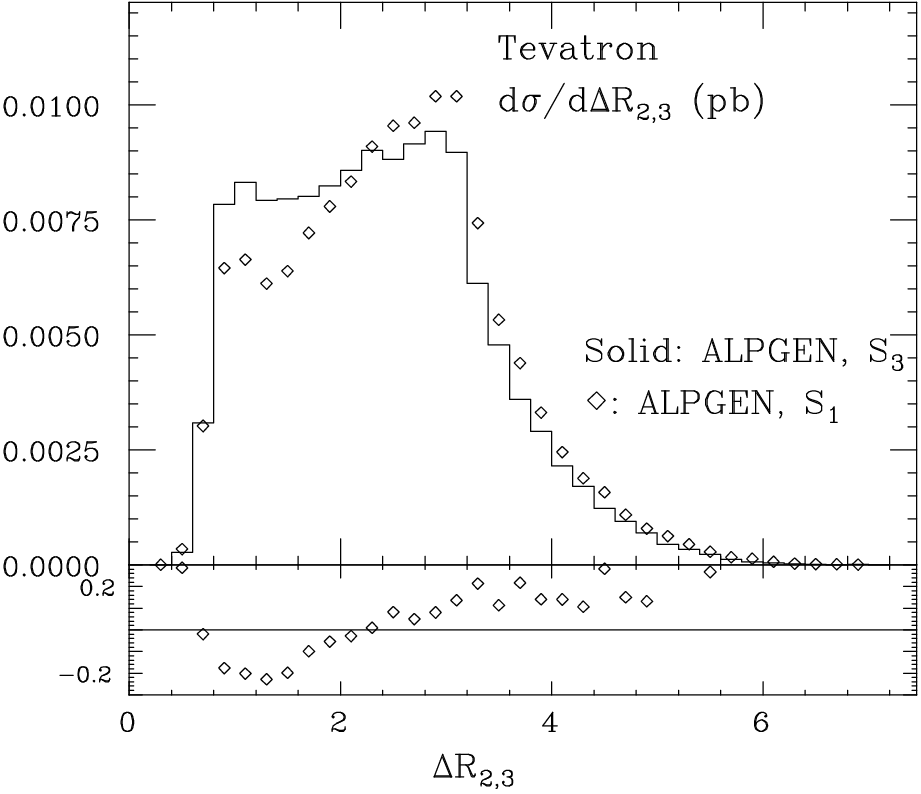}
\hspace*{0.5cm}
\includegraphics[height=0.22\textheight,clip]{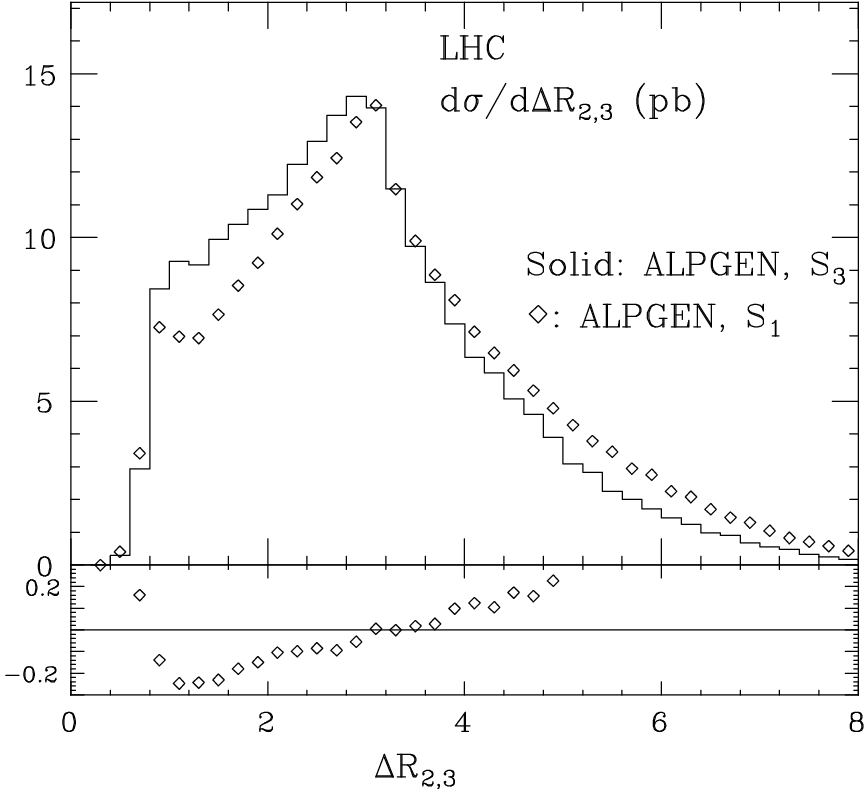}
\ccaption{}{\label{fig:Njet_01_TEV} Comparison between the distributions
obtained from the $S_1$ event samples (0\exc+1\inc) and from the $S_3$
event samples (0\exc+1\exc+2\exc+3\inc), for various higher-order parton
observables at the Tevatron (left-hand side) and at the LHC
(right-hand side).}
\end{center}
\end{figure}
We start by plotting
 the jet multiplicity distribution, \njet, in the
upper panels of fig.~\ref{fig:Njet_01_TEV}. The agreement between
the two calculations is remarkable, at the level of better than 20\%
 even for the   large jet
multiplicities.
This can be justified by the fact that the 20 GeV jets
we are considering here are rather soft objects when compared with the
total amount of energy involved in a \ttbar\ event. Therefore the
soft-approximation used in the shower evolution correctly describes
the emission rate of these multijet events. This is confirmed by the
plots in the second row, showing the 
 \pt\ spectrum of the 2nd leading jet in the event. Once again the
agreement between the two calculations is excellent, up to very large
values of the jet \et. Where the shower approximation appears to be
less reliable is in the description of the kinematical correlations
between the jets. The lower plots of the figure show the $(\eta,\phi)$
correlations between the 1st and 2nd and between the 2nd and 3rd jets.
Clear differences in the shapes are evident.

\clearpage

\subsection{Study of the generation and matching systematics}
\label{sec:syst}
In this section we explore the systematic uncertainties due to the
variation of generation and matching parameters. These uncertainties
reflect the underlying fact that this approach relies on the
LO evaluation of the hard ME and on the
LL accuracy in the removal of double counting and in
the description of the shower evolution. As mentioned in the
beginning, the ultimate goal of this approach is to enable the
generation of fully inclusive event samples that offer LL accuracy
throughout phase-space, including configurations with many
jets. In this section we shall show that the size of the resulting
uncertainties 
is  consistent with what can be expected in
such a LL approach in the case of \ttbar\ production. A more thorough
discussion of matching systematics can be found in ref.~\cite{Mangano-ta}.

It should  be remarked that the presence in this approach of extra
parameters -- such as the matching parameters -- compared to the usual
PL or shower-only approaches is not necessarily a curse. The ultimate
use of LL event generators is not to incorporate and enable
high-precision predictions of QCD, but rather to provide the most
faithful representation of the data, so that the experimental searches
for and studies of new phenomena can be built on a solid
foundation. The uncertainties introduced by the possibility to change
the matching and generation parameters should therefore be seen as an
opportunity to optimize, via their fitting, the agreement between the
generator and the data.

In our examples here we consider two independent variations of the
generation and of two of the matching cuts, keeping fixed our
definition of the physical objects (the jets) and of the
observables. For the generation variations we maintain the relation
between generation and matching cuts given in
eq.~(\ref{eq:matchparams}), and consider a lowering and an increase of
the \ptmin\ thresholds. For the matching variations we keep fixed the
generation parameters, and consider a change in the \et\ threshold for
the clusters, \etclus, and a change in the minimal separation
\Rmat\ required
for a parton and a cluster to match. The numerical values
are detailed in table~\ref{tab:cuts_syst}.

We start by discussing the cross sections, which are given in
tables~\ref{tab:xsect-tev} and~\ref{tab:xsect-lhc}.
While the contributions of the individual partonic samples can vary by
a large amount, the total cross sections are very stable, with the
maximum excursion between minimum and maximum being of the order of
5\%. For comparison, the rate of the 0\exc\ samples, which are the
dominant ones, can vary by up to 35\%.  

\begin{figure}
\begin{center}
\includegraphics[height=0.22\textheight,clip]{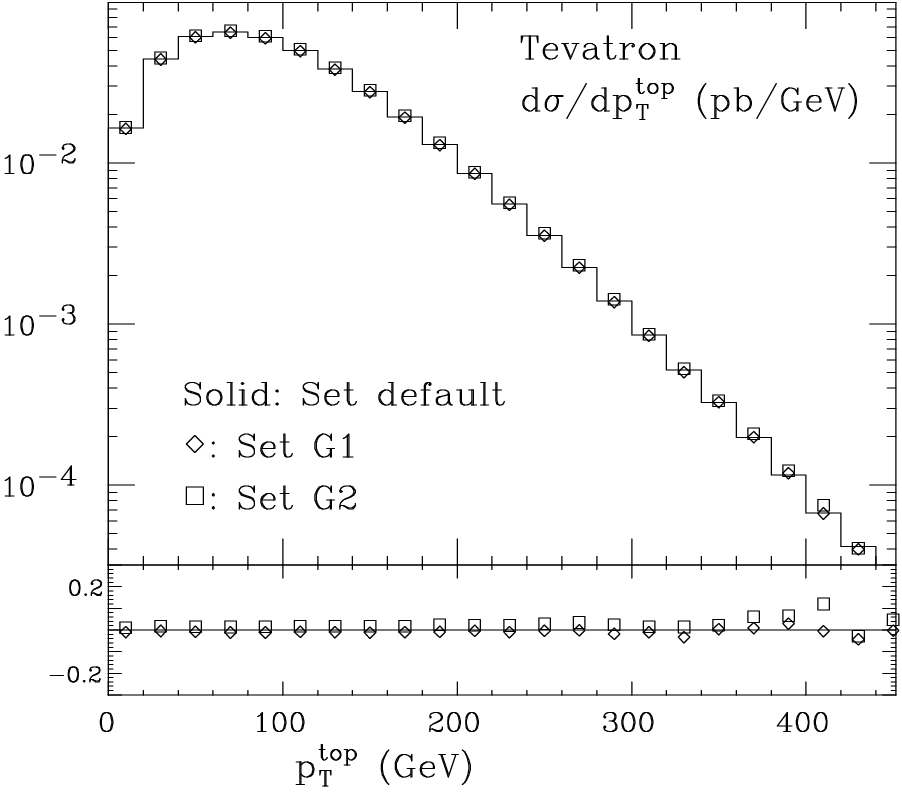}
\hspace*{0.5cm}
\includegraphics[height=0.22\textheight,clip]{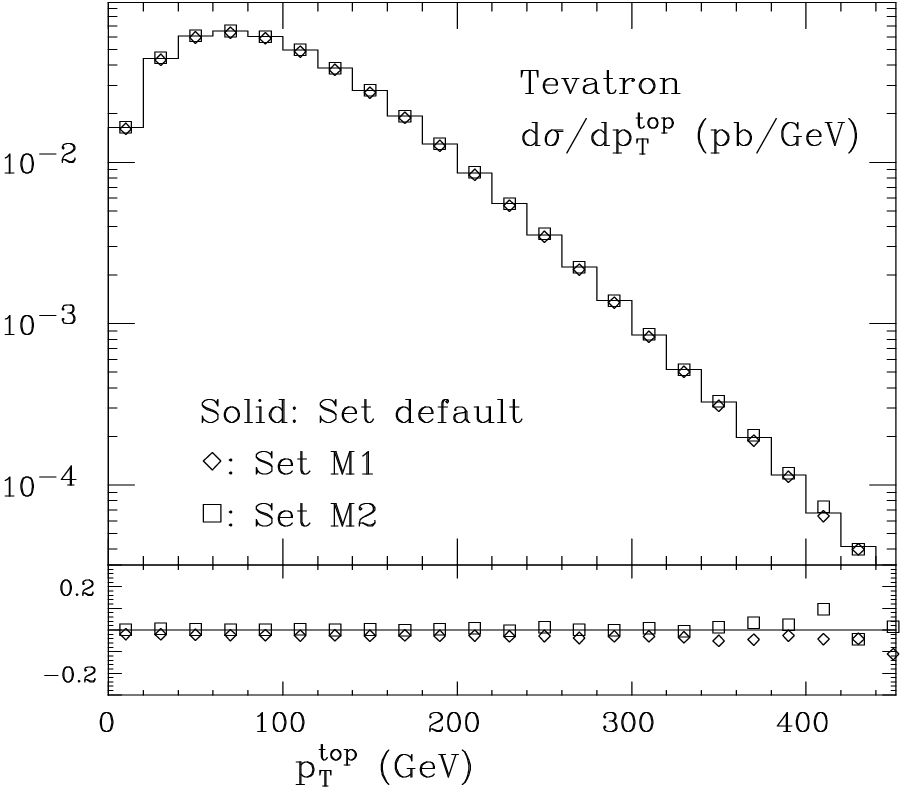}
\\
\includegraphics[height=0.22\textheight,clip]{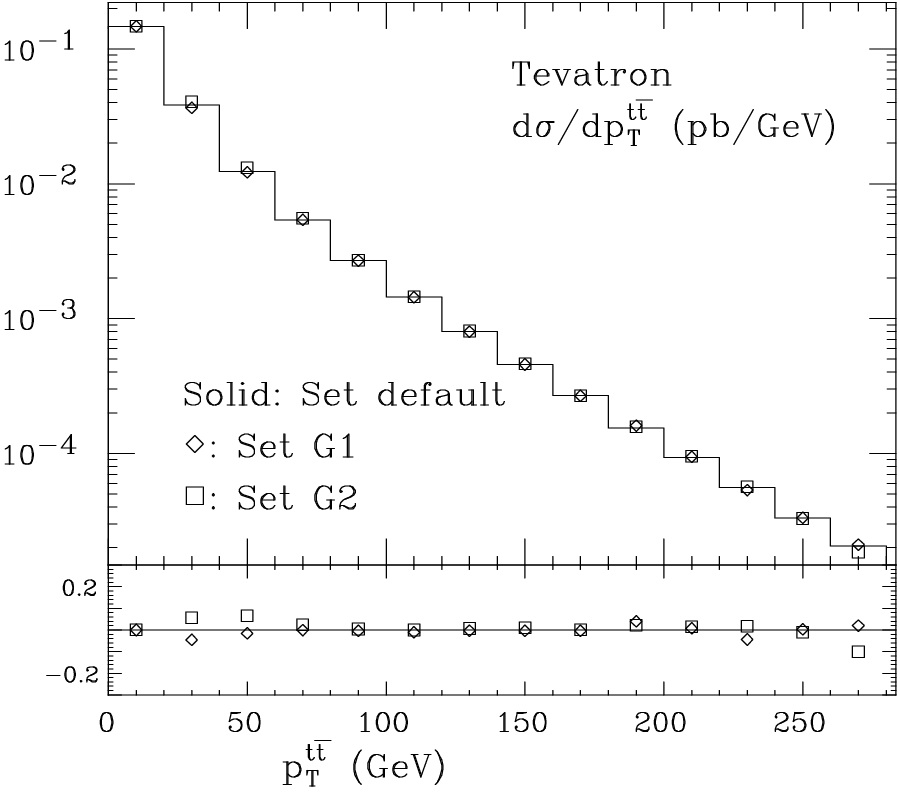}
\hspace*{0.5cm}
\includegraphics[height=0.22\textheight,clip]{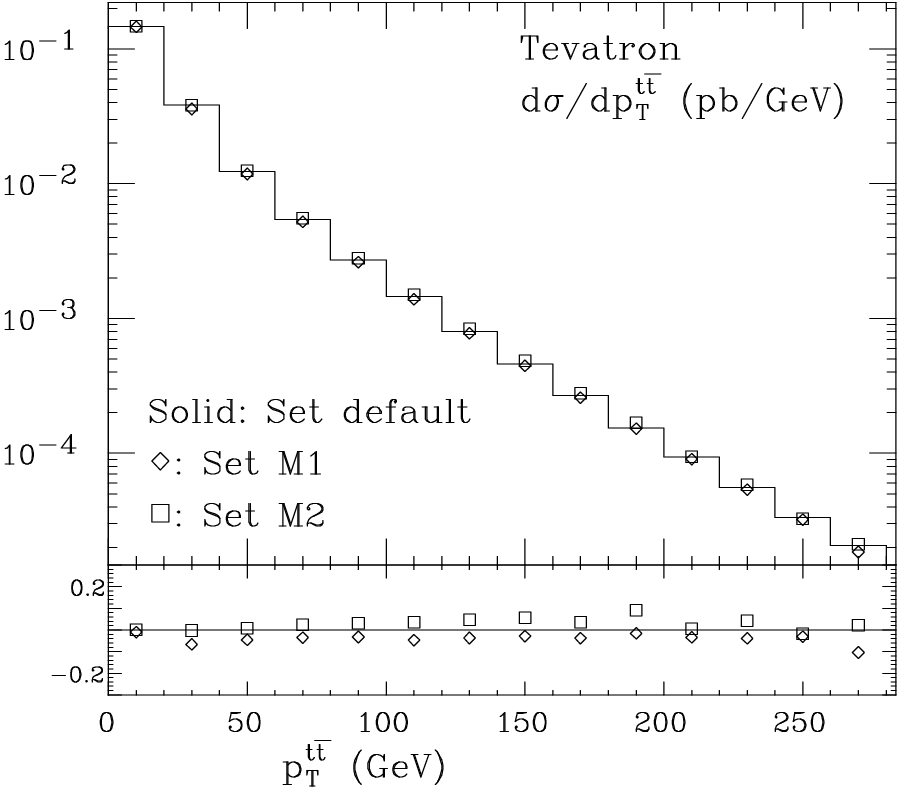}
\\
\includegraphics[height=0.22\textheight,clip]{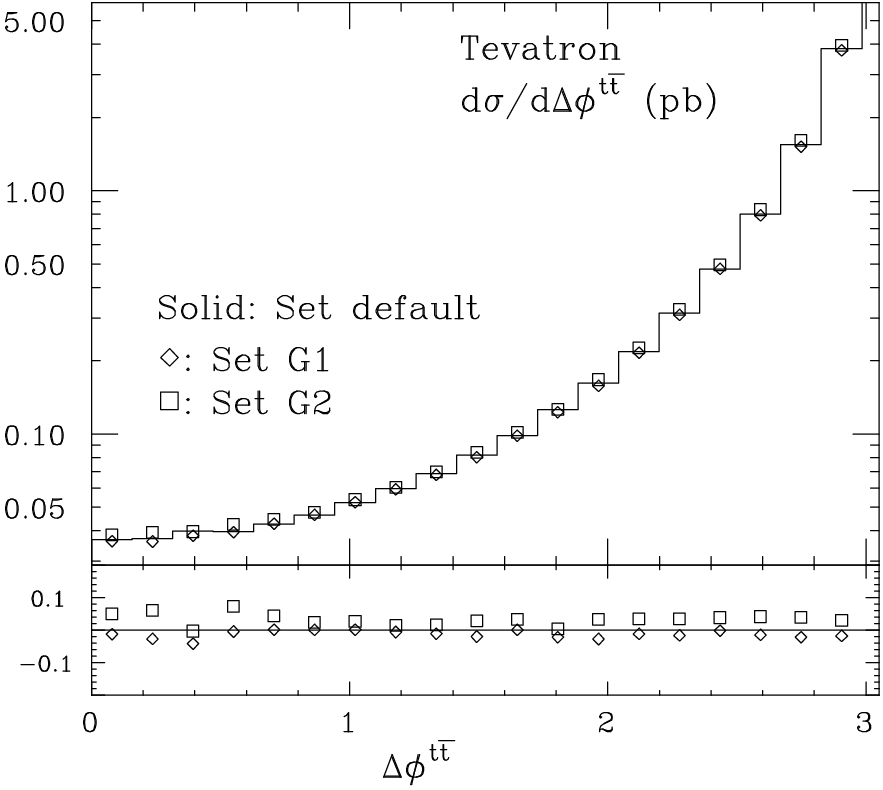}
\hspace*{0.5cm}
\includegraphics[height=0.22\textheight,clip]{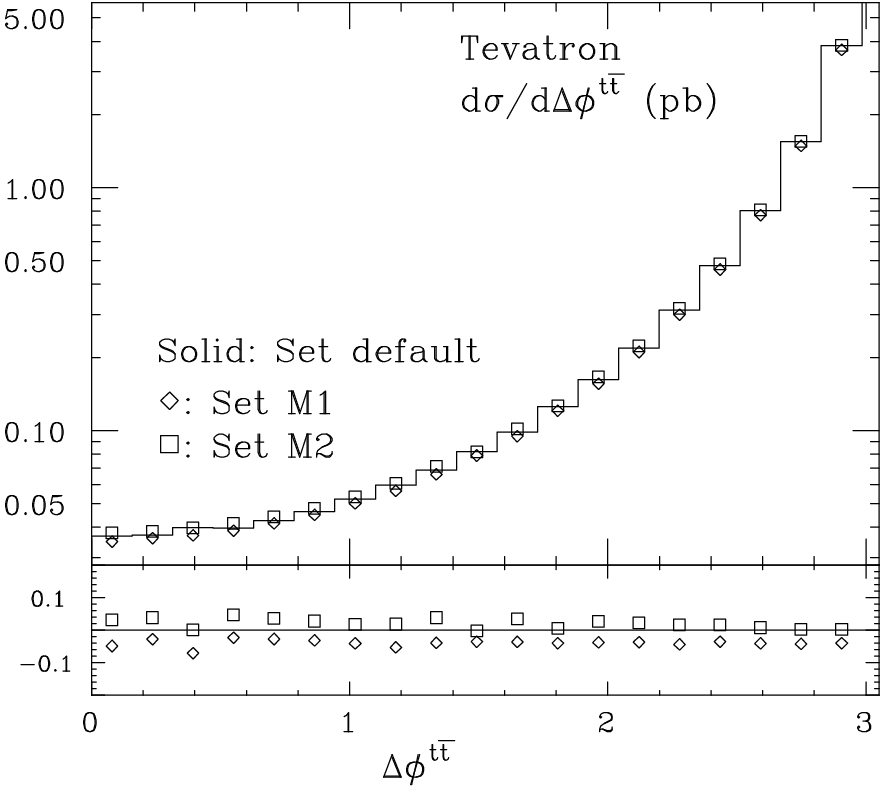}
\\
\includegraphics[height=0.22\textheight,clip]{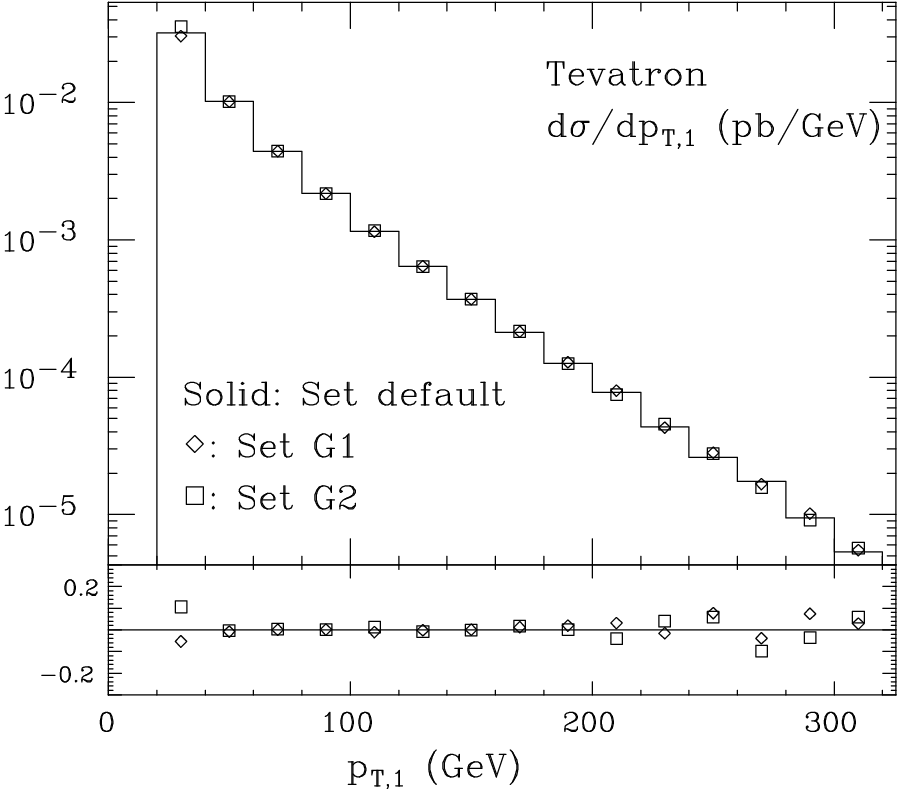}
\hspace*{0.5cm}
\includegraphics[height=0.22\textheight,clip]{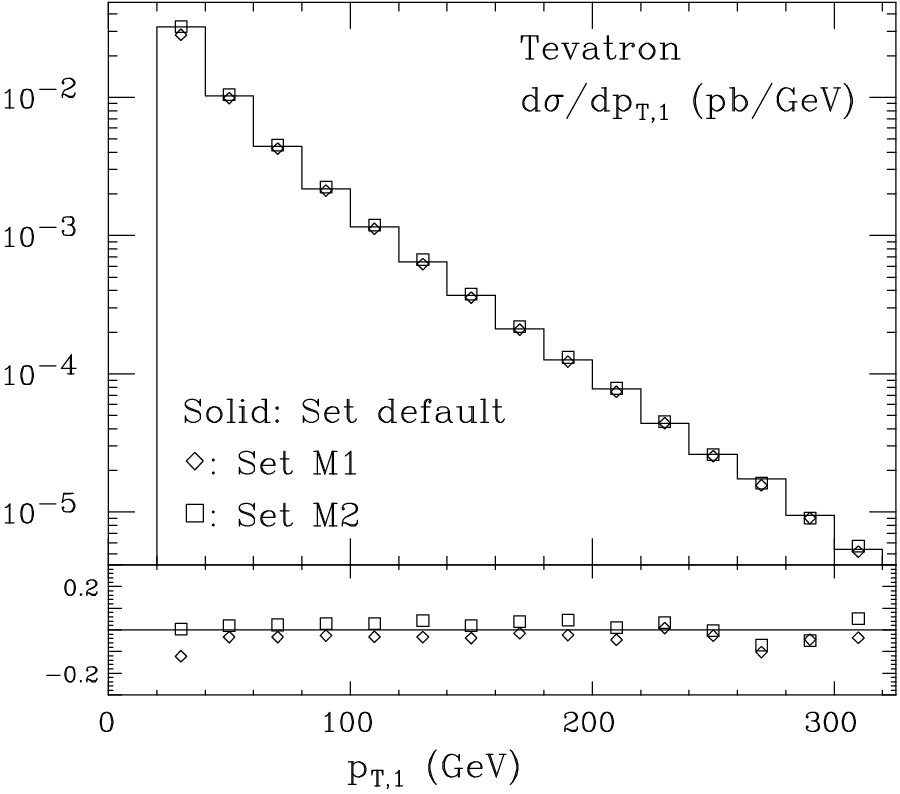} \\
\ccaption{}{\label{fig:pt_GM} Comparison between the three
alternative sets of generation (left) and matching (right) parameters
given in table~\ref{tab:cuts_syst}, at the Tevatron.}
\end{center}
\end{figure}

\begin{figure}
\begin{center}
\includegraphics[height=0.22\textheight,clip]{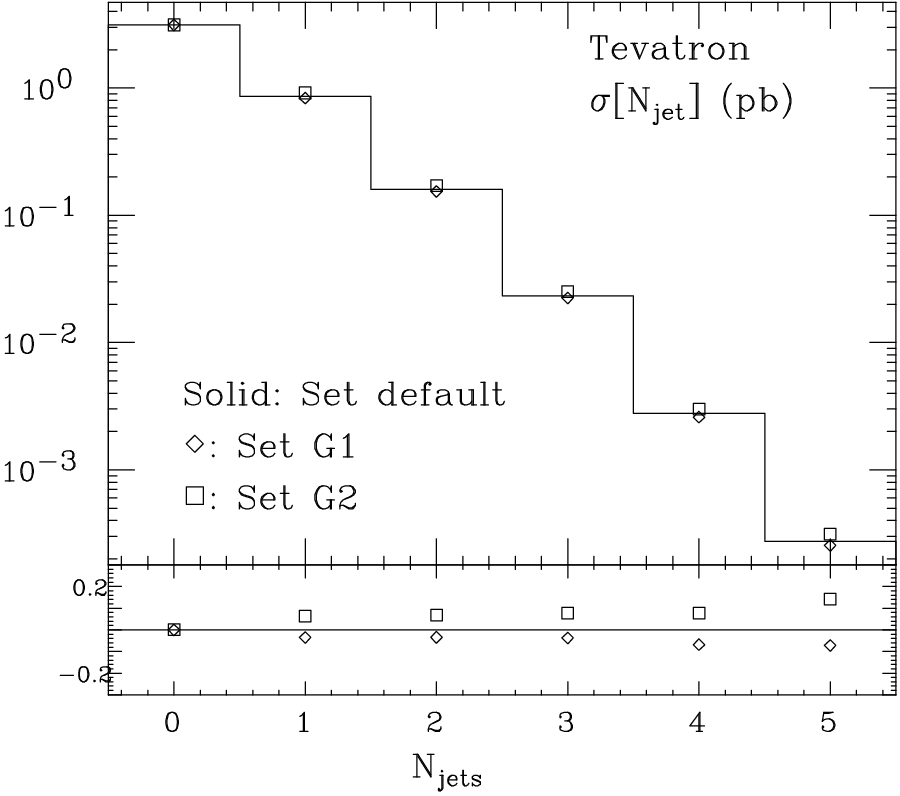}
\hspace*{0.5cm}
\includegraphics[height=0.22\textheight,clip]{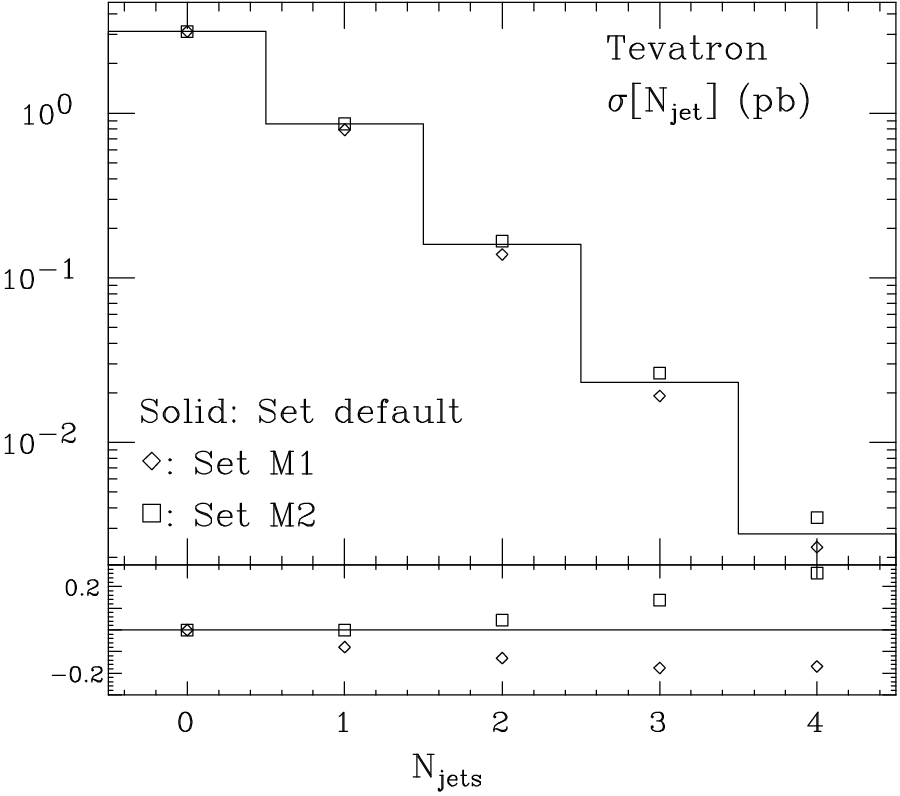}
\\
\includegraphics[height=0.22\textheight,clip]{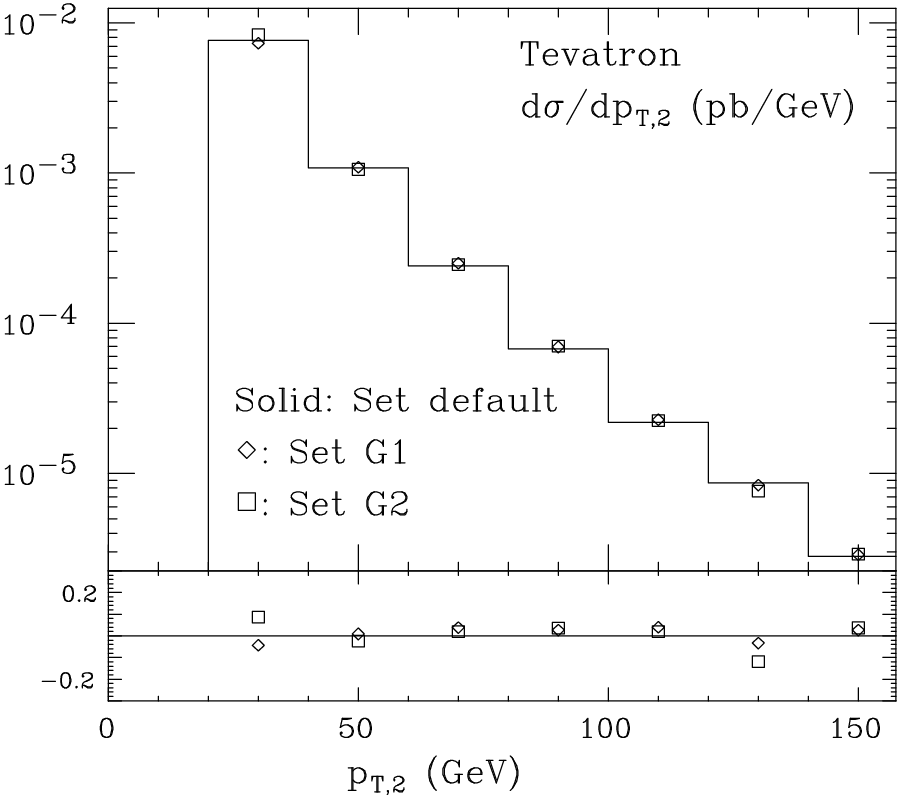}
\hspace*{0.5cm}
\includegraphics[height=0.22\textheight,clip]{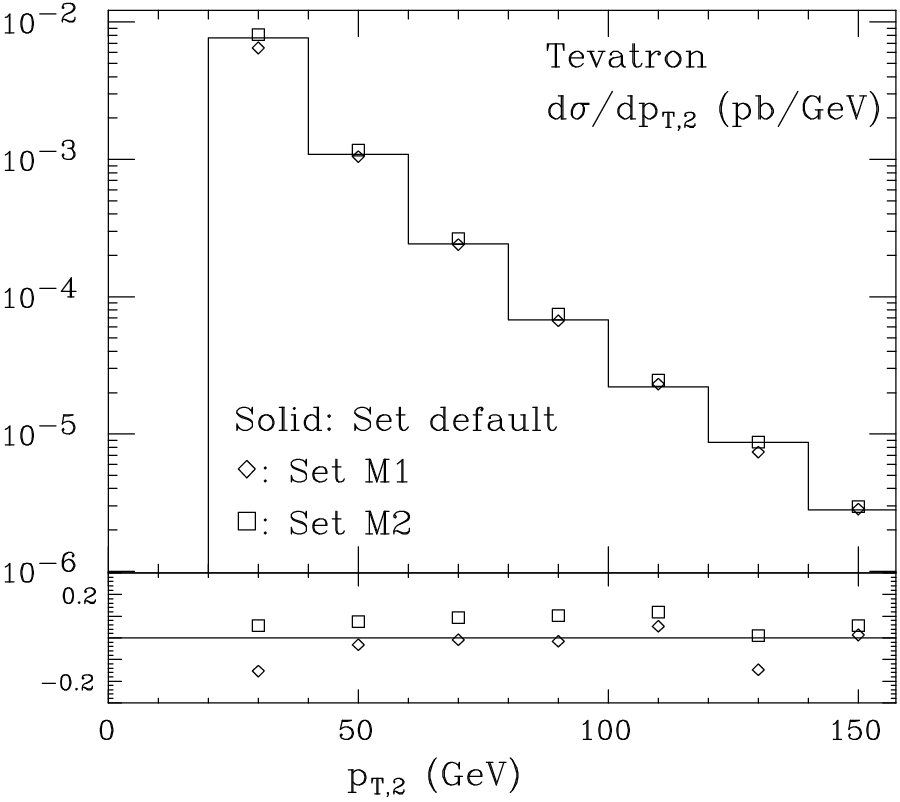}
\\
\includegraphics[height=0.22\textheight,clip]{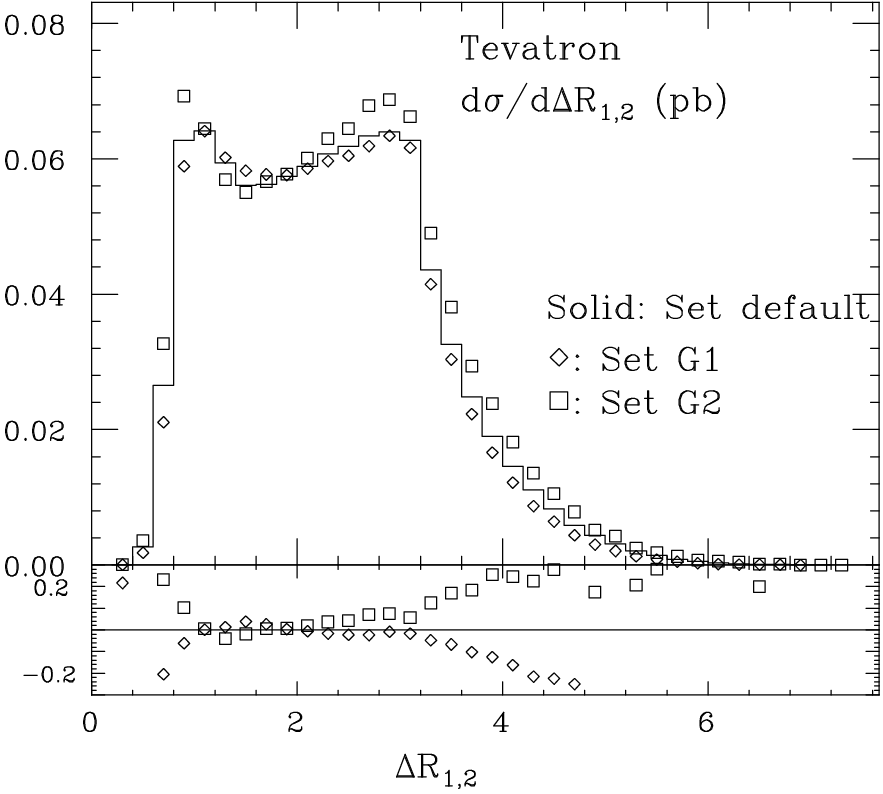}
\hspace*{0.5cm}
\includegraphics[height=0.22\textheight,clip]{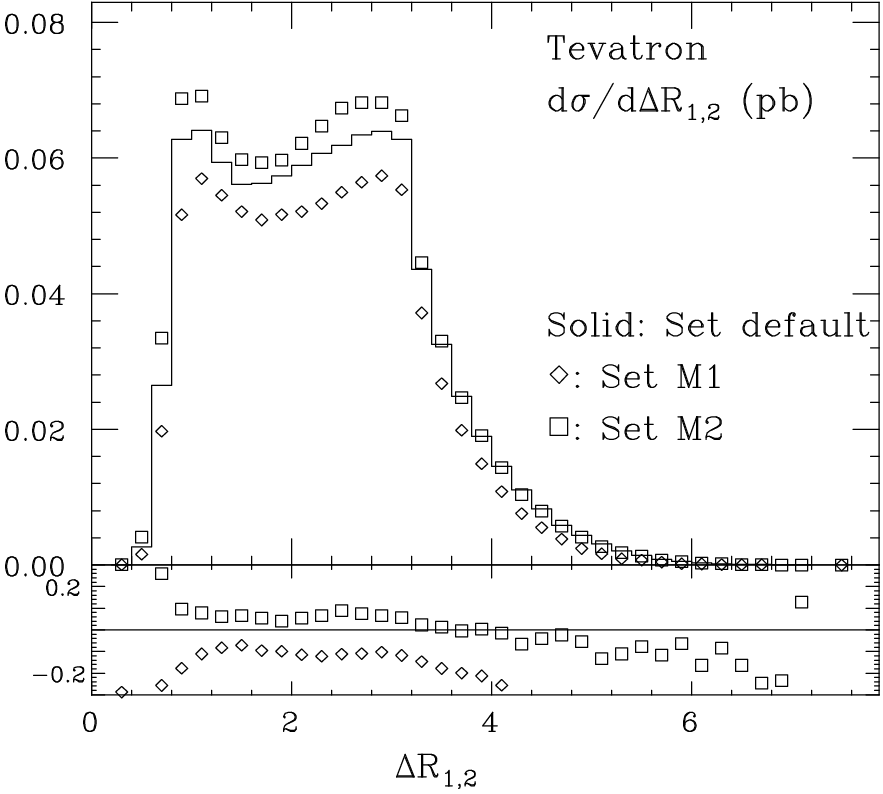}
\\
\includegraphics[height=0.22\textheight,clip]{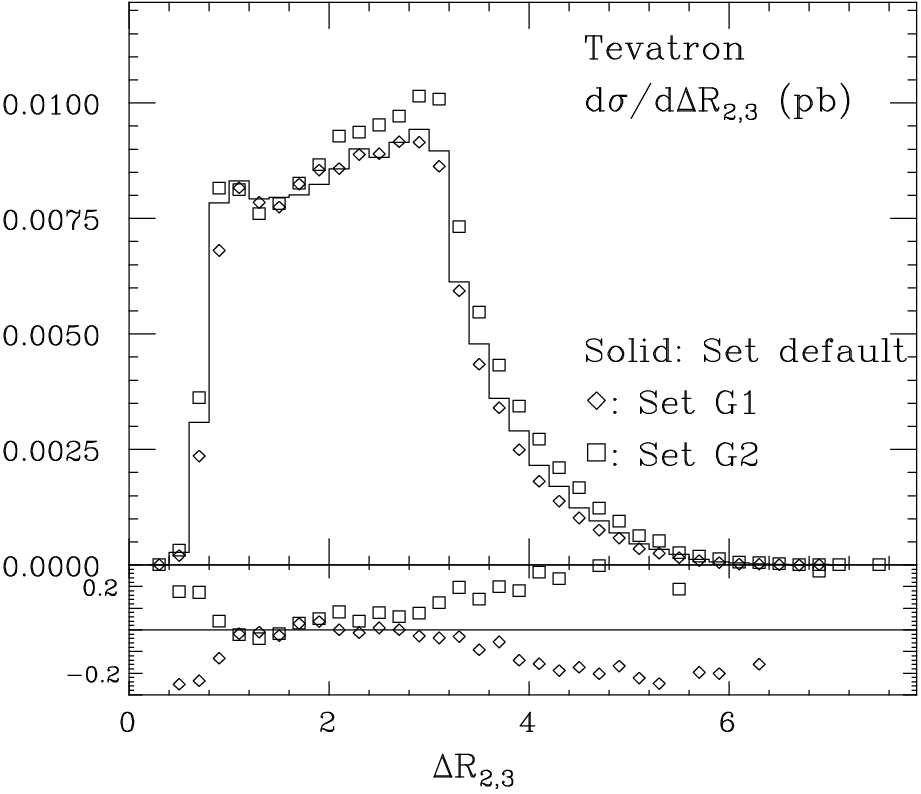}
\hspace*{0.5cm}
\includegraphics[height=0.22\textheight,clip]{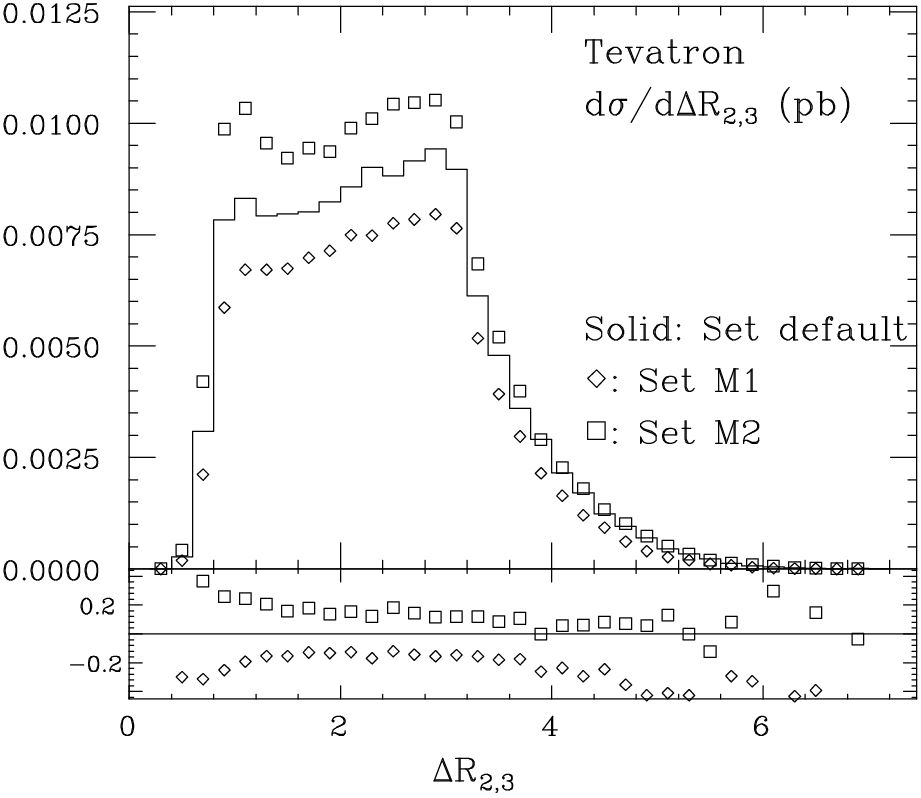}
\\
\ccaption{}{\label{fig:Njet_GM} Comparison between the three
  alternative sets of generation and matching parameters given in
  table~\ref{tab:cuts_syst}, for multijet distributions at the Tevatron.}
\end{center}
\end{figure}

Then we proceed to study some distributions, following the template
of the comparisons between the $S_1$ and $S_3$ samples discussed
earlier. For the Tevatron, the observables dominated by contributions with up
to 1 hard parton are shown in fig.~\ref{fig:pt_GM}, and those relative
to multijet final states in fig.~\ref{fig:Njet_GM}.
Even the rates for large jet multiplicities are extremely
stable. Consider for example the 3-jet bin. At the Tevatron, 
the contribution of the 3\inc\ sample varies from 10fb, for the
default generation/matching cuts, to 24fb for set G1 and 2fb for set
G2. Nevertheless the total 3-jet rates show a stability at the level
of 10\%, as do  the shapes of the distributions (see for example the 
$\Delta R_{2,3}$ case in the figure). 
\begin{table}
\begin{center}
\begin{tabular}{|l|cc|cc|}
\hline
&   \multicolumn{2}{c|}{Generation parameters} &
       \multicolumn{2}{c|}{Matching parameters} \\
Param set & \ptmin & \Rmin & min \etclus & \Rmat\\
\hline
Tevatron, default  & 20 &  0.7 & 25 & 1.5 $\times$ 0.7 \\
Tevatron, Set G1   & 15 &  0.7 & 20 & 1.5 $\times$ 0.7 \\
Tevatron, Set G2   & 30 &  0.7 & 36 & 1.5 $\times$ 0.7 \\
Tevatron, Set M1   & 20 &  0.7 & 20 & 1.5 $\times$ 0.7 \\
Tevatron, Set M2   & 20 &  0.7 & 25 & 1.5 $\times$ 1.0 \\
\hline
LHC, default  & 30 &  0.7 & 36 & 1.5 $\times$ 0.7 \\
LHC, Set G1   & 25 &  0.7 & 30 & 1.5 $\times$ 0.7 \\
LHC, Set G2   & 40 &  0.7 & 48 & 1.5 $\times$ 0.7 \\
LHC, Set M1   & 30 &  0.7 & 30 & 1.5 $\times$ 0.7 \\
LHC, Set M2   & 30 &  0.7 & 36 & 1.5 $\times$ 1.0 \\
\hline
\end{tabular}
\ccaption{}{\label{tab:cuts_syst}
Variations of  the generation and matching parameters used for the
study of the systematics.}
\end{center}
\end{table}

\begin{table}
\begin{center}
\begin{tabular}{|l|l|l|l|l|l|}
\hline
Tevatron & Default & Set G1 & Set G2 & Set M1 & Set M2 \\
\hline
0\exc  & 3.42  &  3.15  &  3.79  & 3.14 & 3.33 \\
1\exc  & 0.66  &  0.82  &  0.42  & 0.78 & 0.74 \\
2\exc  & 0.09  &  0.15  &  0.036 & 0.13 & 0.11  \\
3\inc  & 0.010 &  0.024 &  0.002 & 0.021 & 0.012  \\
\hline
Total  & 4.18  &  4.14  &  4.25  & 4.08 & 4.19  \\
\hline
\end{tabular}
\ccaption{}{\label{tab:xsect-tev} Cross sections (in pb),
for $t\bar t$ production at the Tevatron. The contribution of the
different parton samples, for various generation and matching
options. The columns are labeled according to the parameter
definitions introduced in table~\ref{tab:cuts_syst}.
The relative numerical integration precision is at the permille level.}
\end{center}
\end{table}

\begin{table}
\begin{center}
\begin{tabular}{|l|l|l|l|l|l|}
\hline
LHC & Default & Set G1 & Set G2 & Set M1 & Set M2 \\
\hline
0\exc & 217 & 185  & 267 & 185 & 203 \\
1\exc & 150 & 156  & 134 & 148 & 160 \\
2\exc & \phantom{0}66 &  \phantom{0}81   & \phantom{0}44  &
\phantom{0}74 & \phantom{0}76 \\
3\inc & \phantom{0}30 &  \phantom{0}45   & \phantom{0}15  &
\phantom{0}40 & \phantom{0}35 \\ 
\hline
Total & 462 & 467 & 460  & 447 & 475 \\
\hline
\end{tabular}
\ccaption{}{\label{tab:xsect-lhc}
Same as table~\ref{tab:xsect-tev}, for the LHC.}
\end{center}
\end{table}

The analysis at the LHC leads to qualitatively and quantitatively
similar results, which will not be shown here. 
\clearpage

\section{Comparisons with MC@NLO}
\label{sec:MC@NLO}
We shall now compare in detail the description of $t\bar t$ events as
provided by \ALPGENs and \mcnlo.  For consistency with the \mcnlo\
approach, where only the \oacube\ ME effects are included, we use
\alpgens samples obtained summing the 0\exc\ and 1\inc\ contributions.
As in the case of the results shown before, all plotted quantities
refer to $t$ and $\bar t$ quarks regarded as {\em stable} and after
performing the showering of the event but without including any non
perturbative effect (non perturbative gluon splitting, hadronization,
underlying event,...).
 
\begin{figure}
\begin{center}
\includegraphics[height=0.22\textheight,clip]{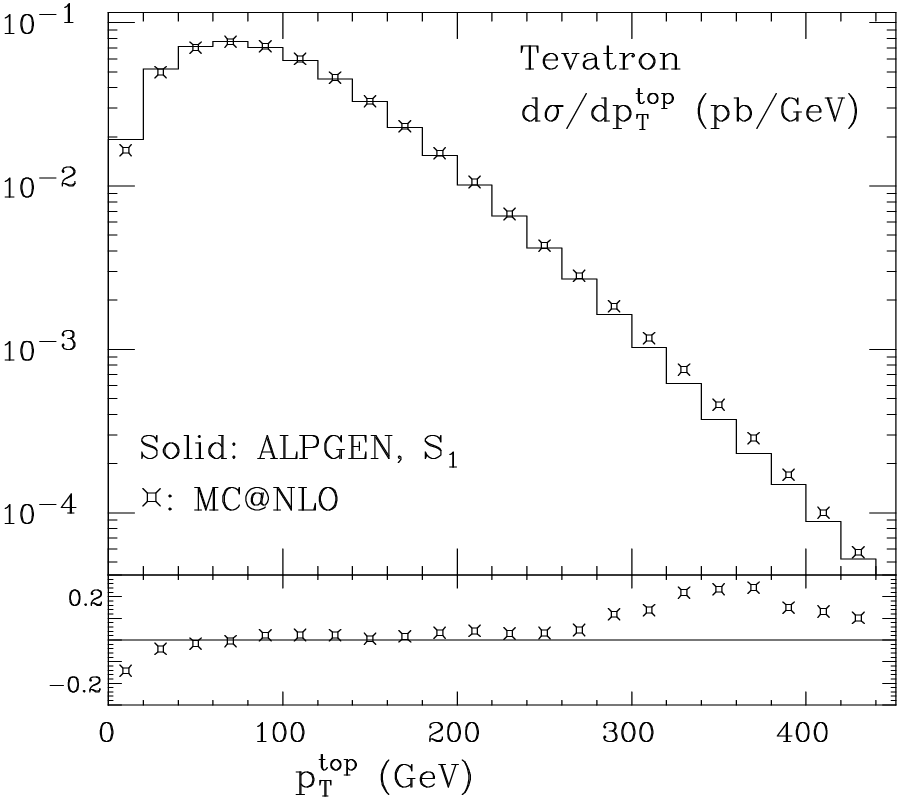}
\hspace*{0.5cm}
\includegraphics[height=0.22\textheight,clip]{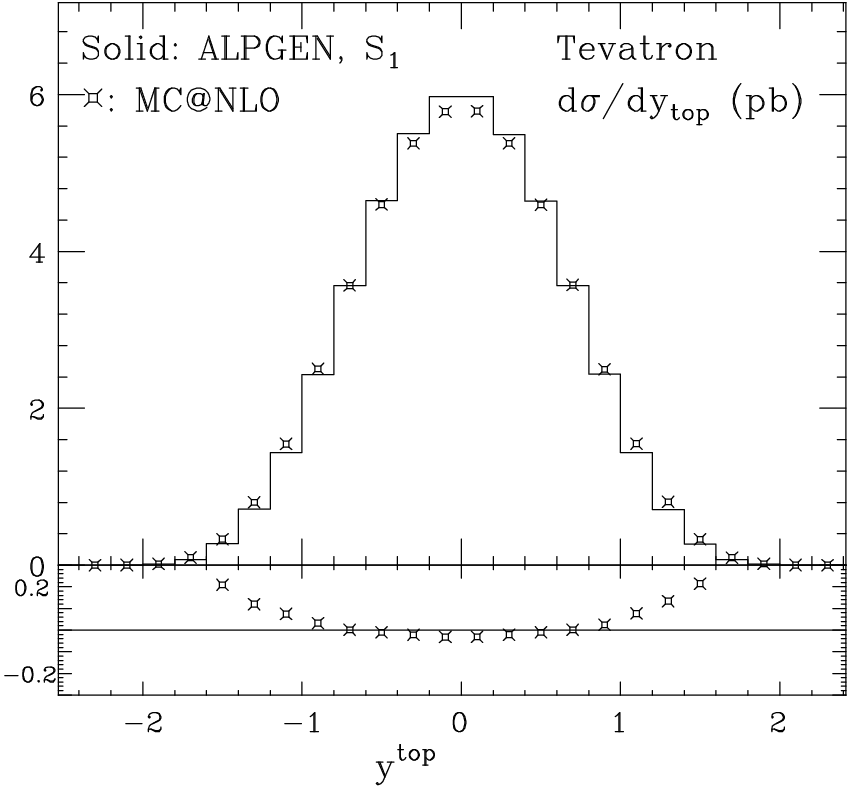}
\\
\includegraphics[height=0.22\textheight,clip]{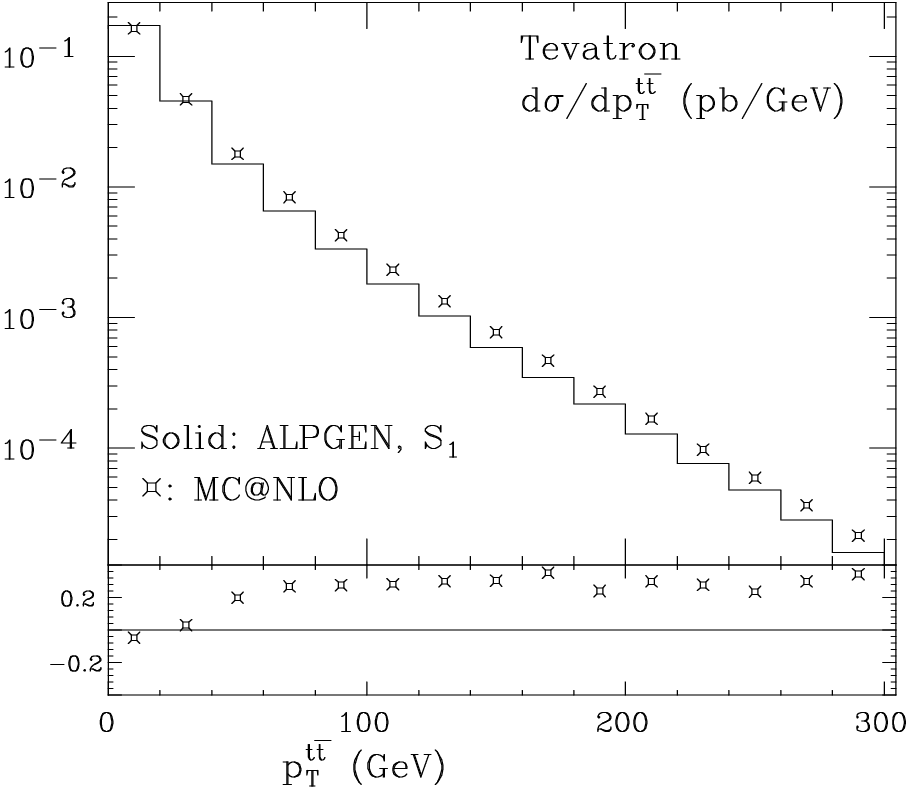}
\hspace*{0.5cm}
\includegraphics[height=0.22\textheight,clip]{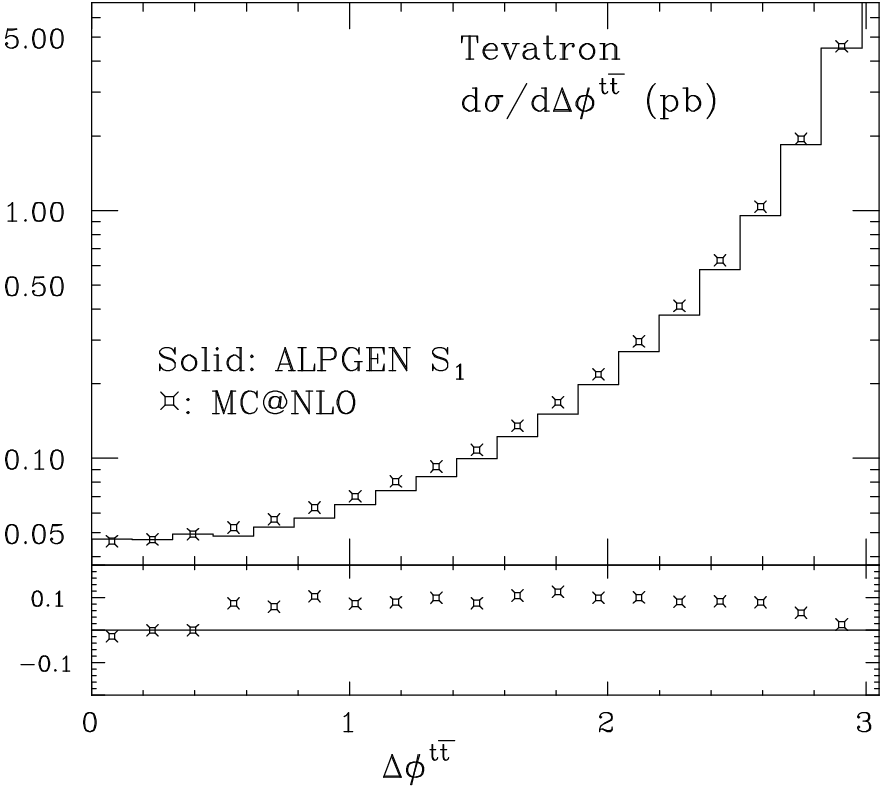}
\\ 
\includegraphics[height=0.22\textheight,clip]{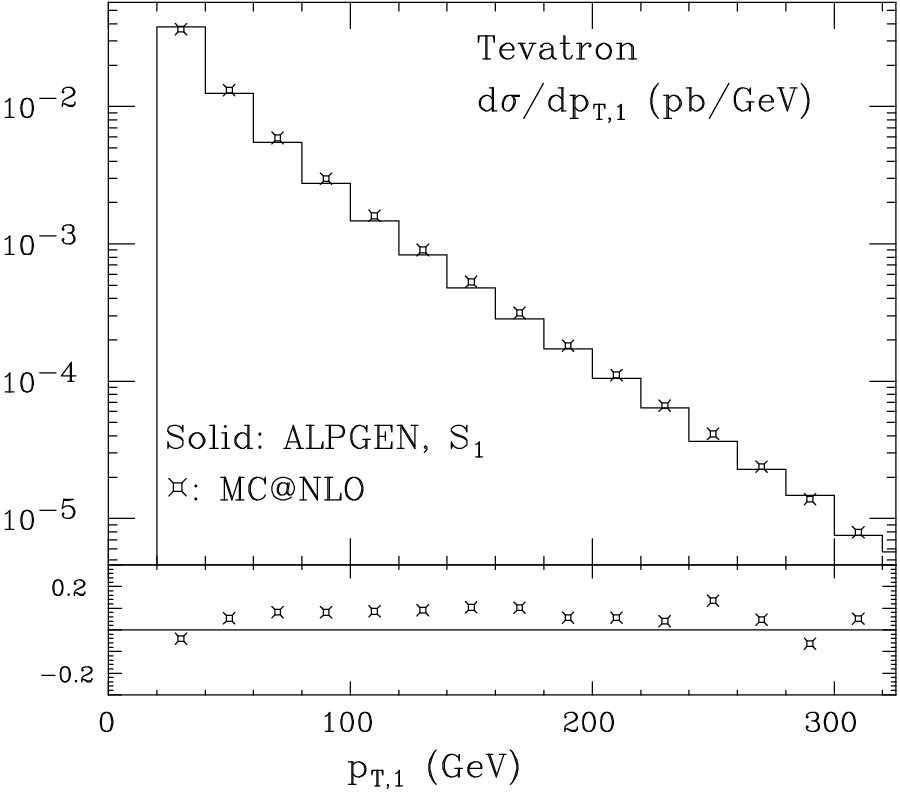}
\hspace*{0.5cm}
\includegraphics[height=0.22\textheight,clip]{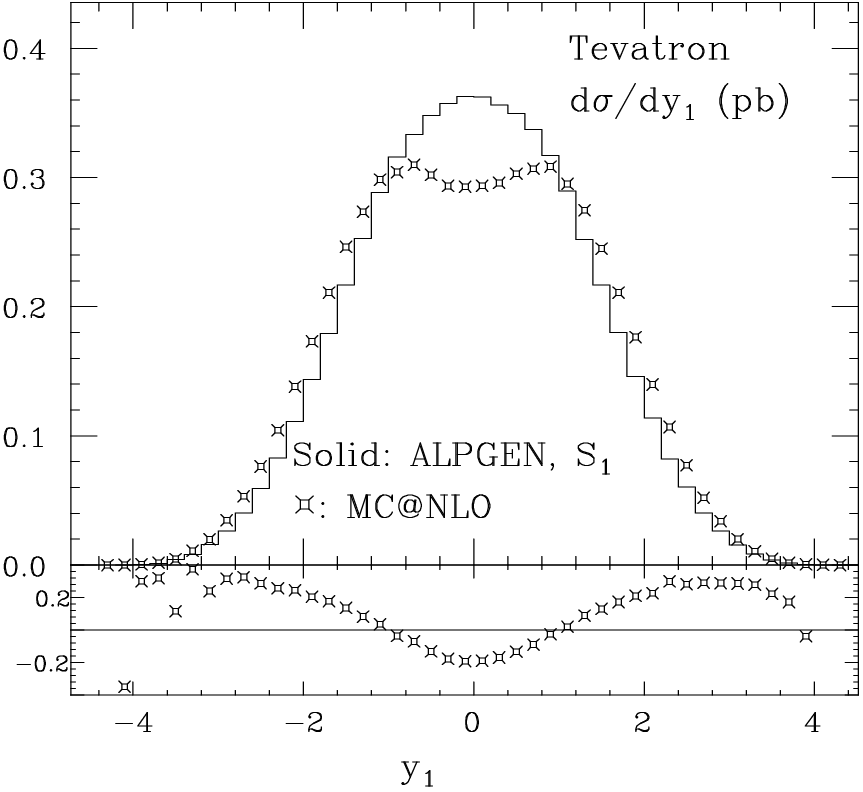} 
\\
\includegraphics[height=0.22\textheight,clip]{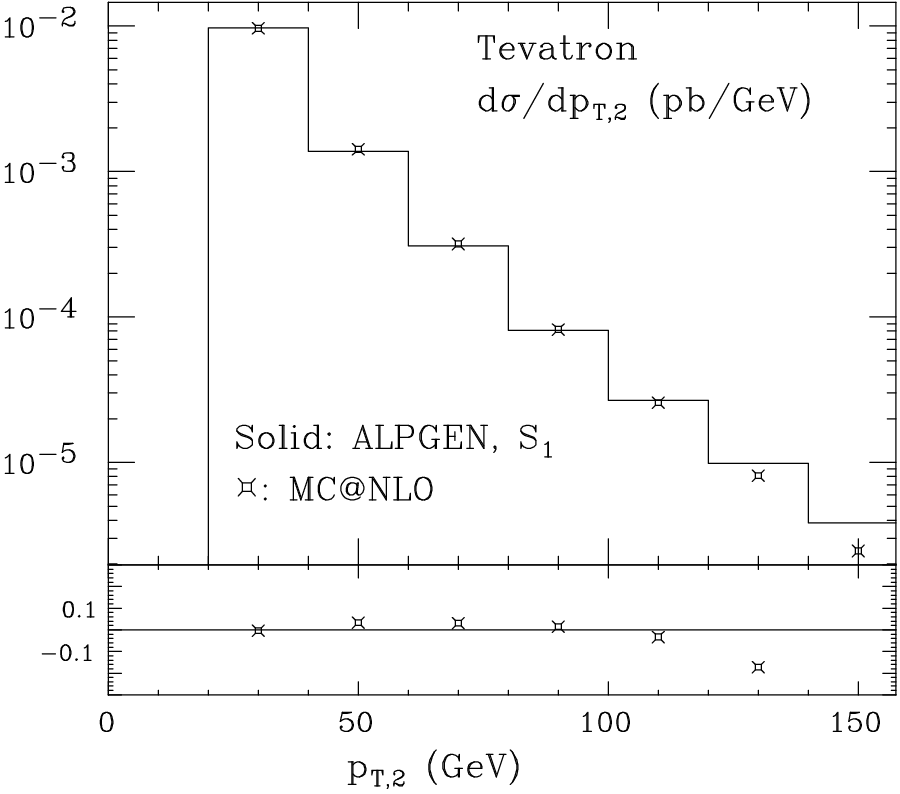}
\hspace*{0.5cm}
\includegraphics[height=0.22\textheight,clip]{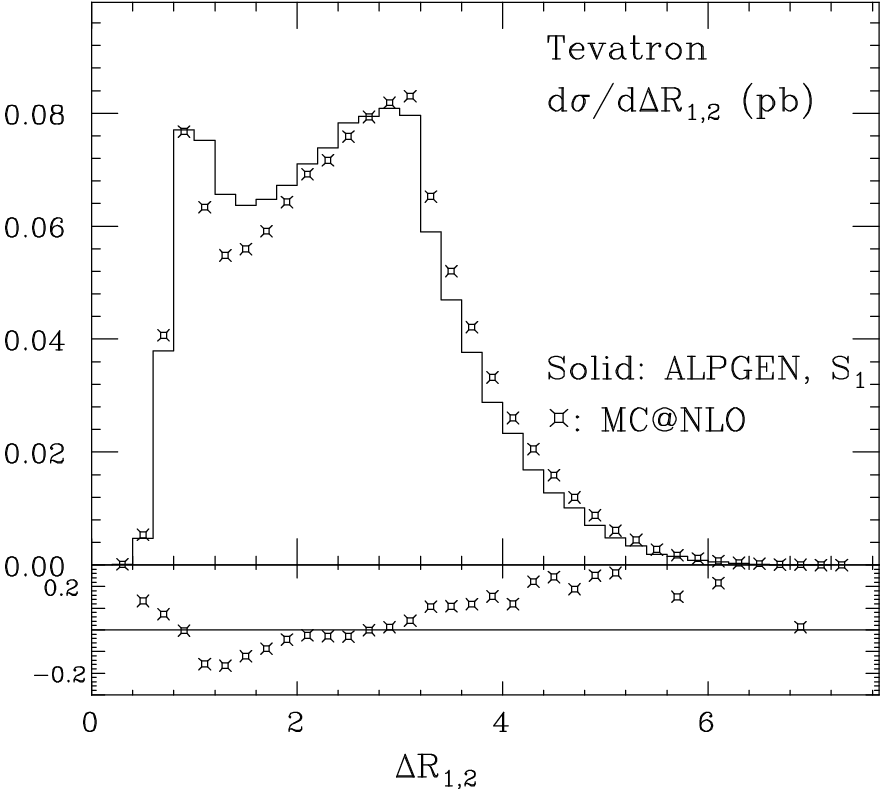}
\ccaption{}{\label{fig:inclusiveTEV} Comparison of \alpgen\
(histogram) and \mcnlo\ (plot) distributions, at the Tevatron. The
\alpgen\ results are rescaled to \mcnlo, using the K factor of 1.36.
The relative difference (\mcnlo-\alpgen)/\alpgen) is shown at the
bottom of each plot.  }
\end{center}
\end{figure}

\begin{figure}
\begin{center}
\includegraphics[height=0.22\textheight,clip]{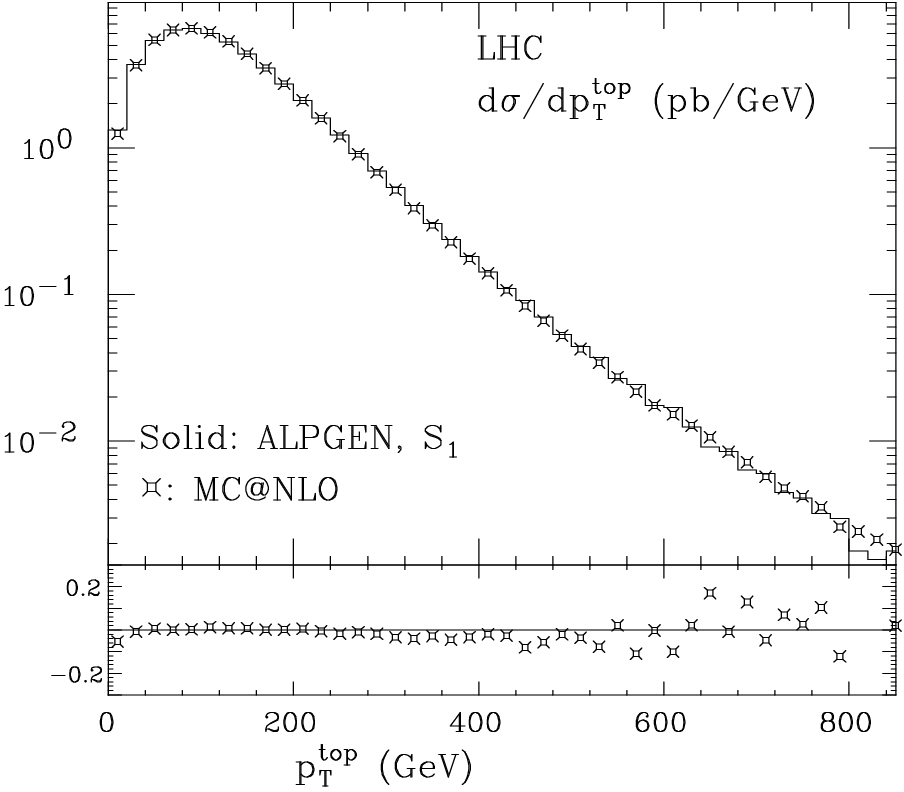}
\hspace*{0.5cm}
\includegraphics[height=0.22\textheight,clip]{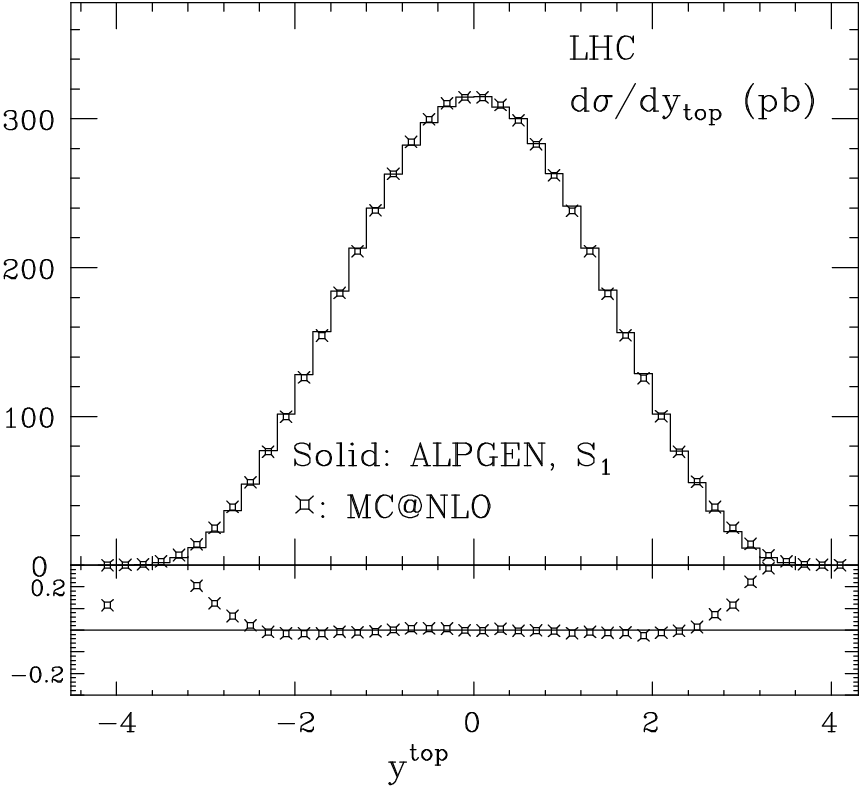}
\\
\includegraphics[height=0.22\textheight,clip]{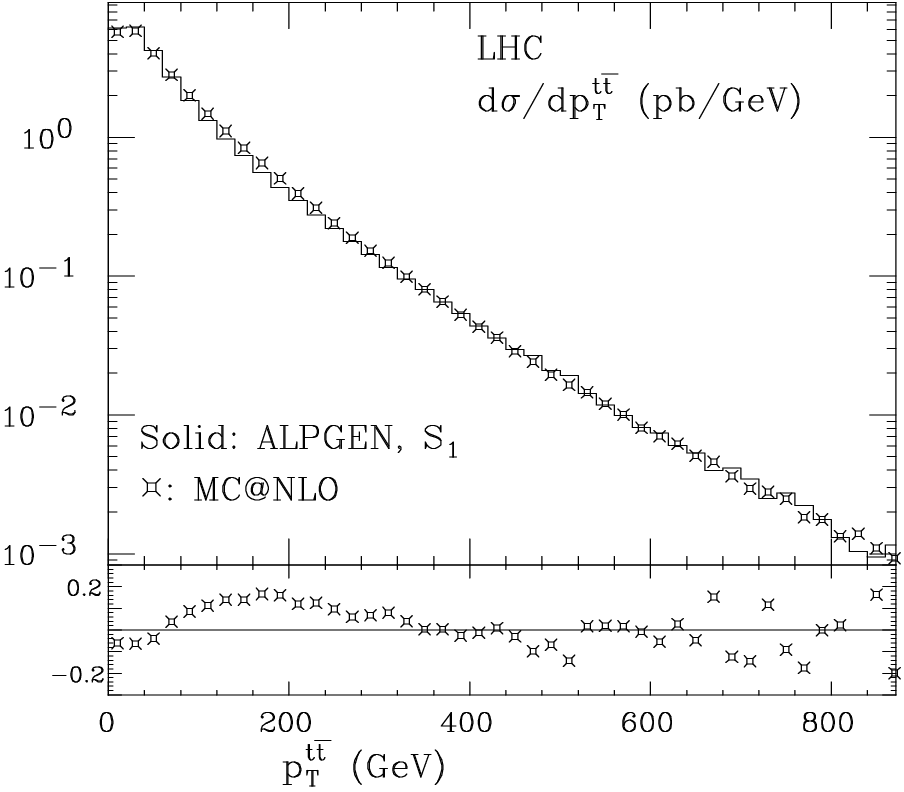}
\hspace*{0.5cm}
\includegraphics[height=0.22\textheight,clip]{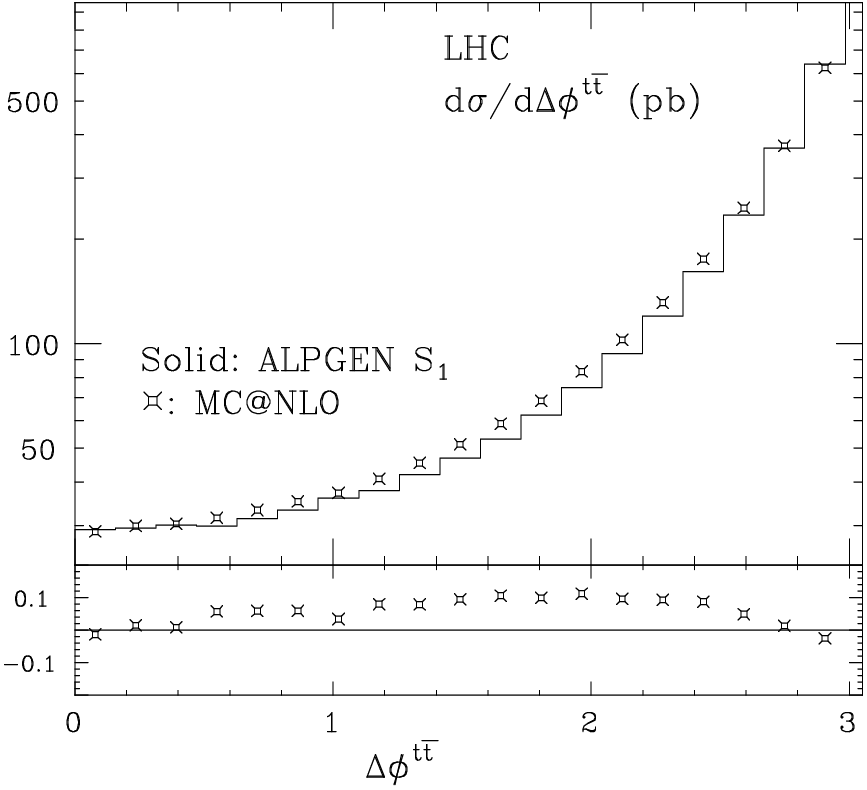}
\\ 
\includegraphics[height=0.22\textheight,clip]{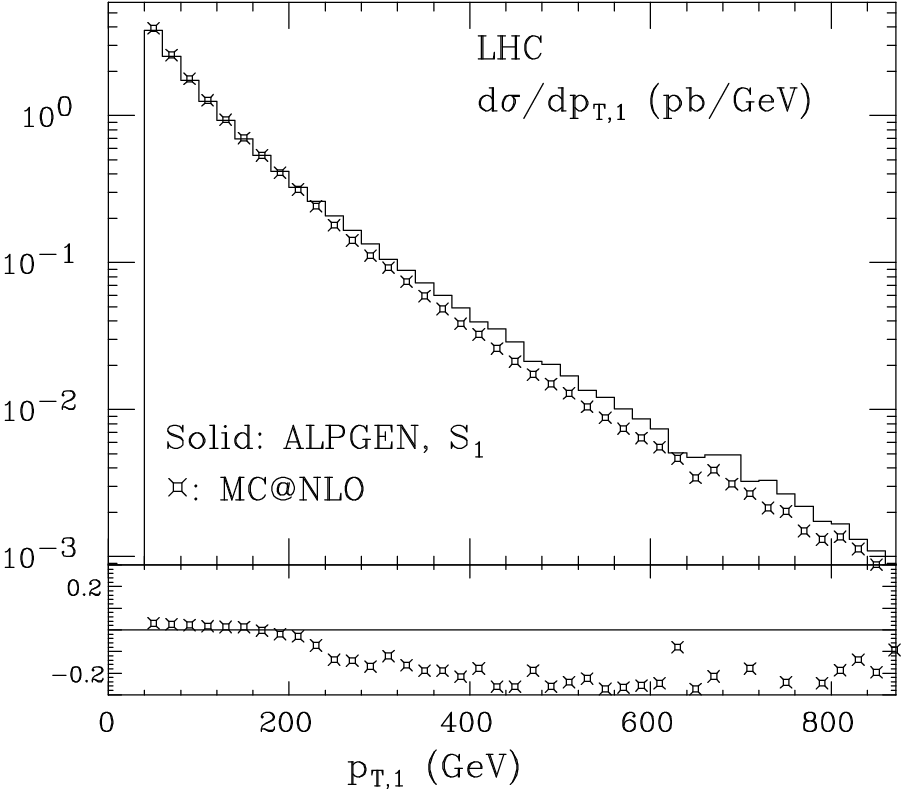}
\hspace*{0.5cm}
\includegraphics[height=0.22\textheight,clip]{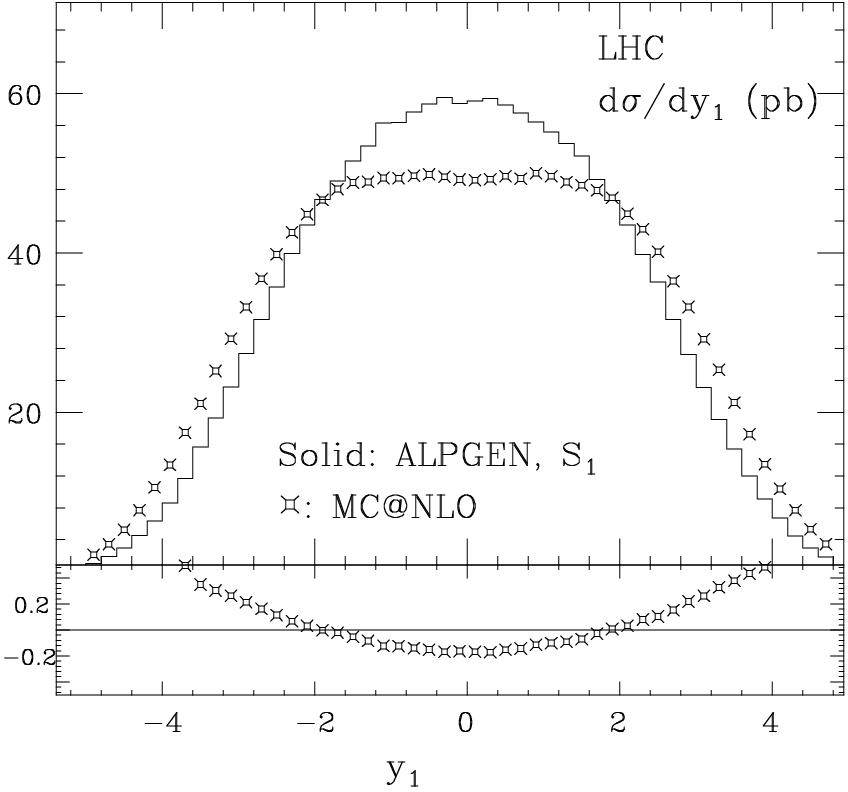} 
\\
\includegraphics[height=0.22\textheight,clip]{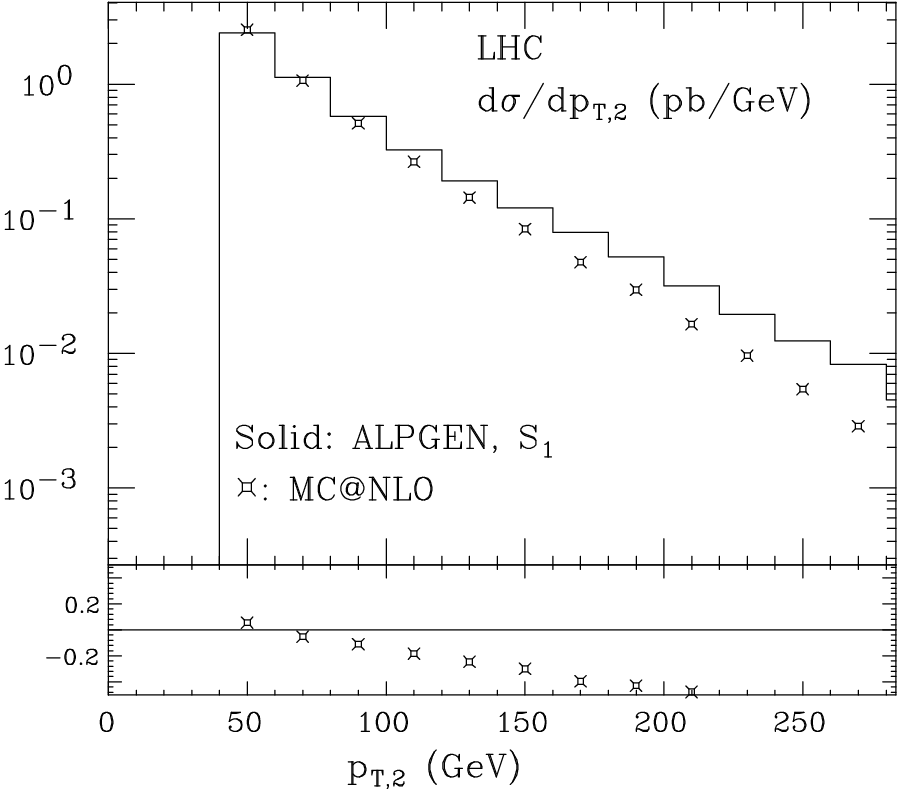}
\hspace*{0.5cm}
\includegraphics[height=0.22\textheight,clip]{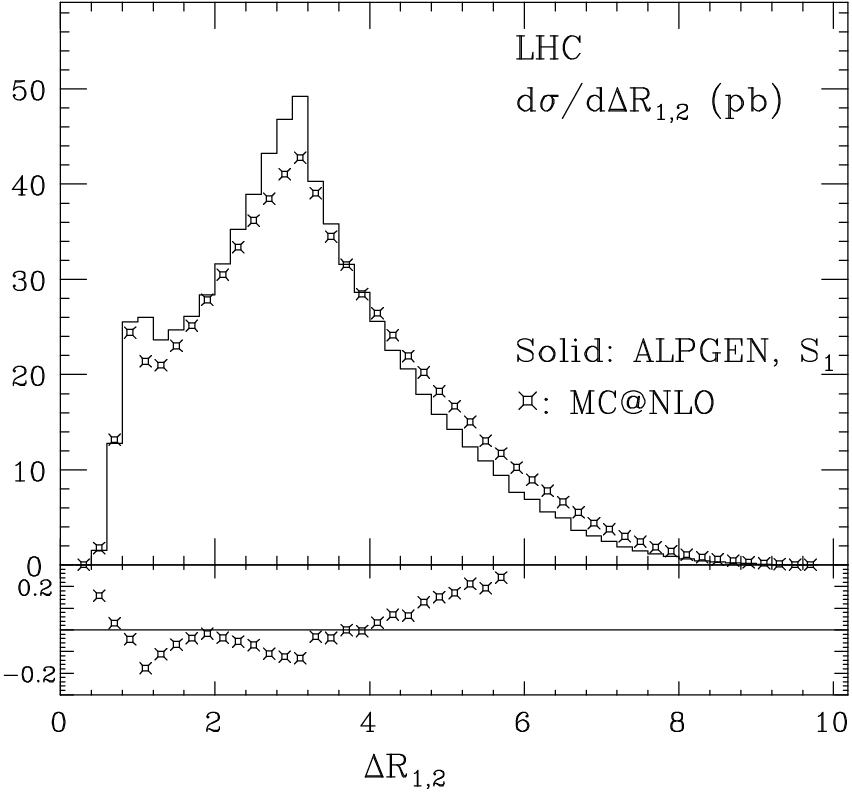}
\ccaption{}{\label{fig:inclusiveLHC} Same as
  fig.~\ref{fig:inclusiveTEV} for the LHC, using the K
  factor of 1.51.}
\end{center}
\end{figure}
To match \mcnlo's default we have used, for both codes, 
the same factorization and renormalization scale
\[
\mu^2 =\frac{1}{2} \left ({\pttopsq+\mtsq} + 
{\pttbarsq+\mtsq} \right )
\]
the same set of PDF  {\small MRST2001J}\cite{Martin:2001es}
and the same value for the top mass (175 GeV).

The upper two rows of 
plots in figs.~\ref{fig:inclusiveTEV} and \ref{fig:inclusiveLHC}
refer to inclusive properties of the $t\bar{t}$ system, 
namely the transverse momentum and rapidity
of the top and anti-top quark, the transverse momentum of the 
$t \bar t$ pair, and the azimuthal angle $\phitt$
between the top and anti-top quark.
The overall agreement is good, once \alpgens is corrected with the proper
K-factor (1.36 for the Tevatron, and 1.51 for the LHC), 
and no large  discrepancy is seen between the two 
descriptions of the chosen distributions.
The most significant differencies (10 to 20\%) are seen in the 
$\pttop$ distribution, \ALPGEN's one being slightly softer.

The study of jet quantities reveals instead 
one important difference: the rapidity of the
leading jet, \yjet, is different in the two descriptions, where \mcnlos
exhibits a dip at $\yjet=0$. This difference is particularly
marked at the Tevatron, but is very visible also at the LHC.
This is shown in the right figure of the third row in
figs.~\ref{fig:inclusiveTEV}
and \ref{fig:inclusiveLHC}. 

Furthermore, while the \pt\ spectrum of the 2nd jet is rather 
similar in the two approaches at the Tevatron, the agreement 
becomes worse at the LHC. Visible differences are also present 
in the distribution of the 1st and 2nd jet separation in 
$(\eta,\phi)$ space, $\Delta R_{1,2}$.

\begin{figure}
\begin{center}
\includegraphics[height=0.25\textheight,clip]{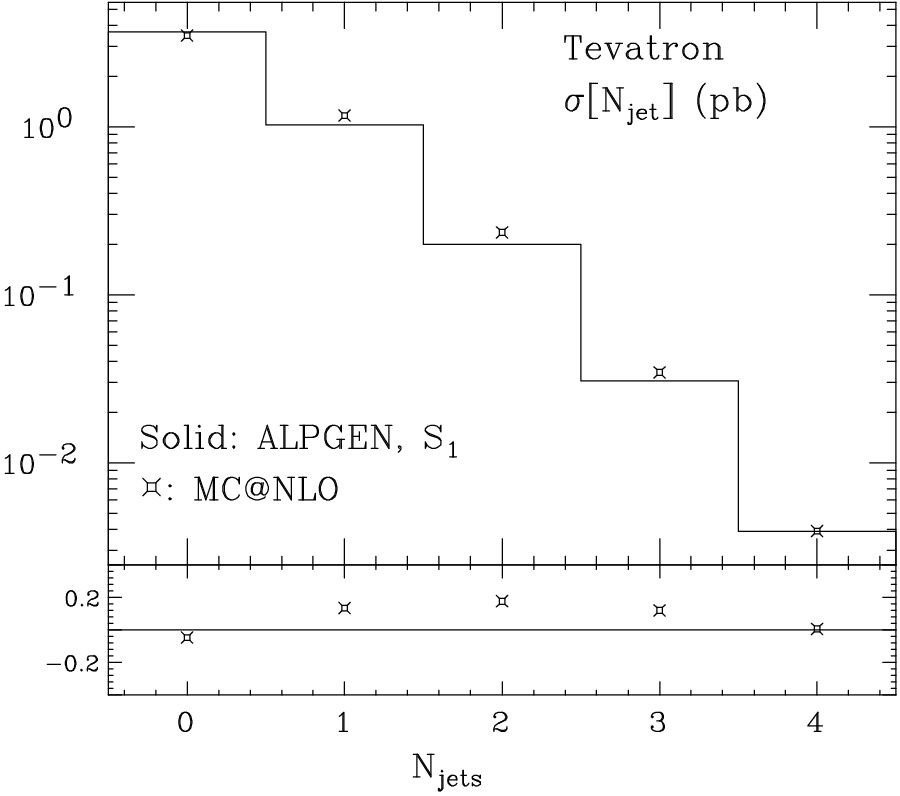}
\hspace*{0.5cm}
\includegraphics[height=0.25\textheight,clip]{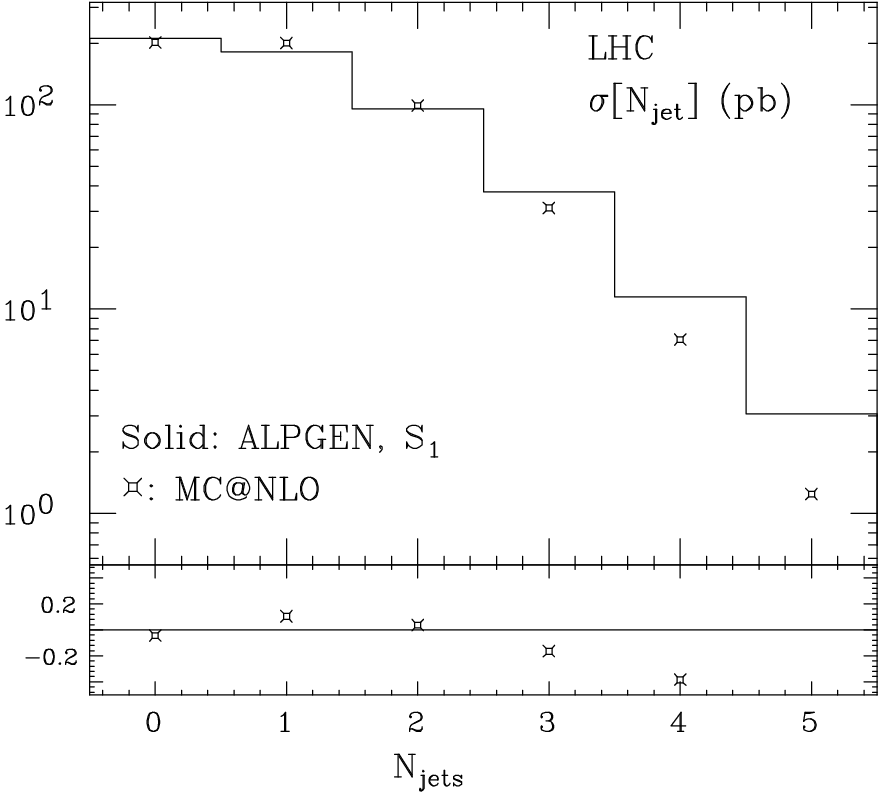}
\ccaption{}{ \label{fig:Njets}
Jet multiplicity from \alpgen\ and \mcnlo, at the Tevatron (left) and
at the LHC (right).
The relative difference (\mcnlo-\alpgen)/\alpgen\
is shown at the bottom of each plot.}
\end{center}
\end{figure}

Figure~\ref{fig:Njets}, finally, shows the comparison of the jet
multiplicity distributions.

\clearpage

\noindent

\begin{figure}
\begin{center}
\includegraphics[height=0.25\textheight,clip]{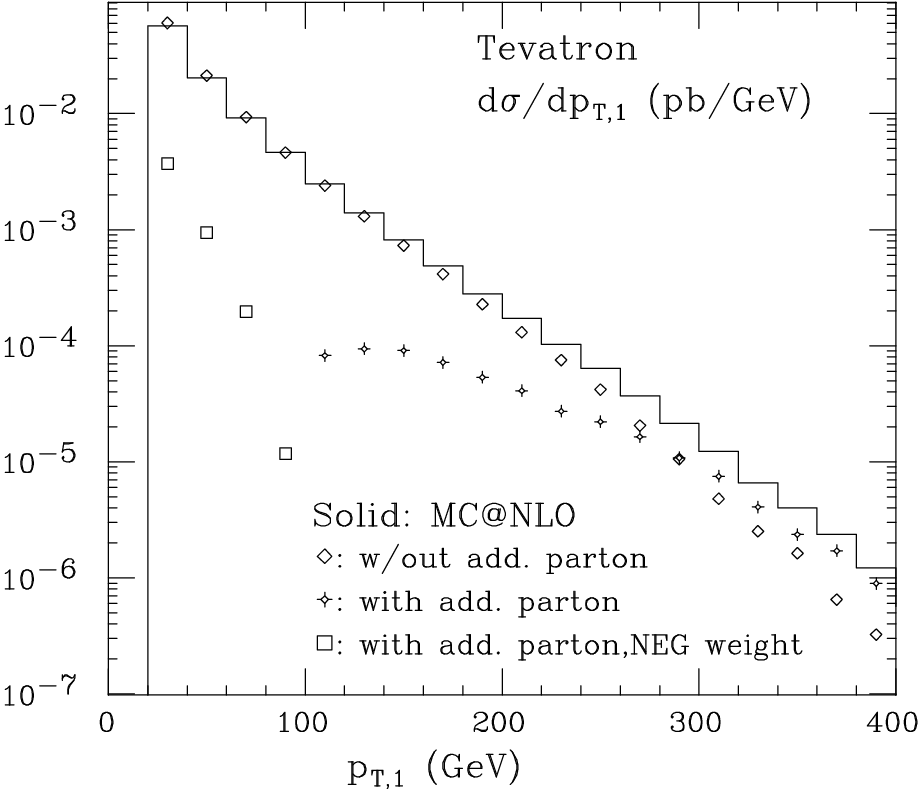}
\hspace*{0.5cm}
\includegraphics[height=0.25\textheight,clip]{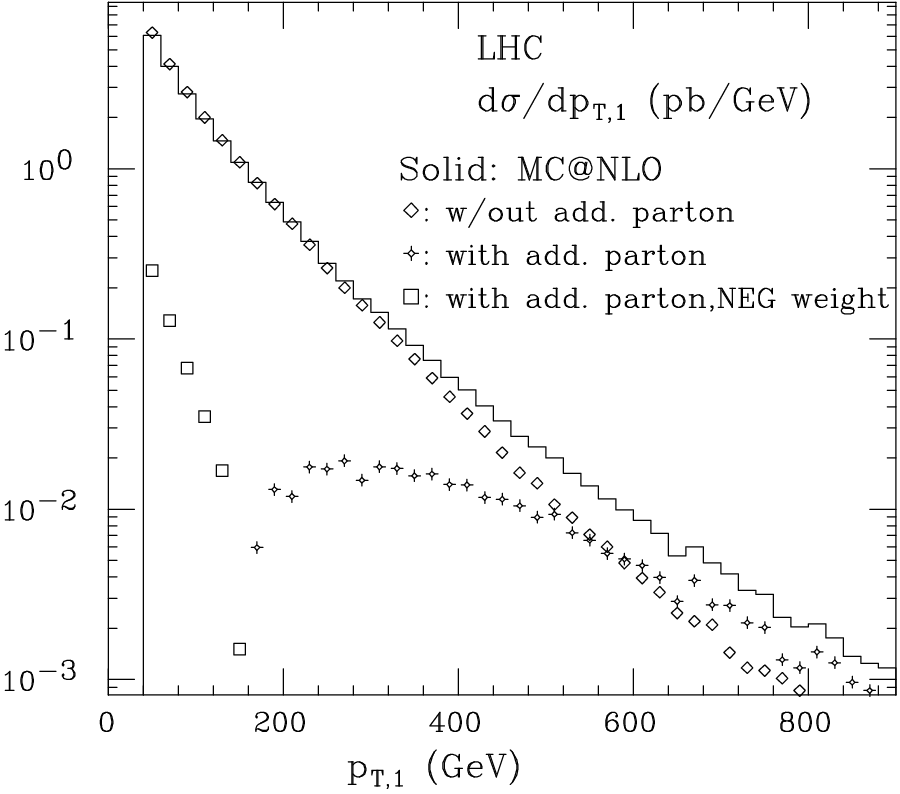}
\ccaption{}{\label{fig:01mcnlo} Contributions to the transverse
  momentum of the leading jet in \mcnlo. 
Tevatron (left) and LHC (right). }
\end{center}
\end{figure}

\begin{figure}
\begin{center}
\includegraphics[width=0.25\textwidth,clip,]{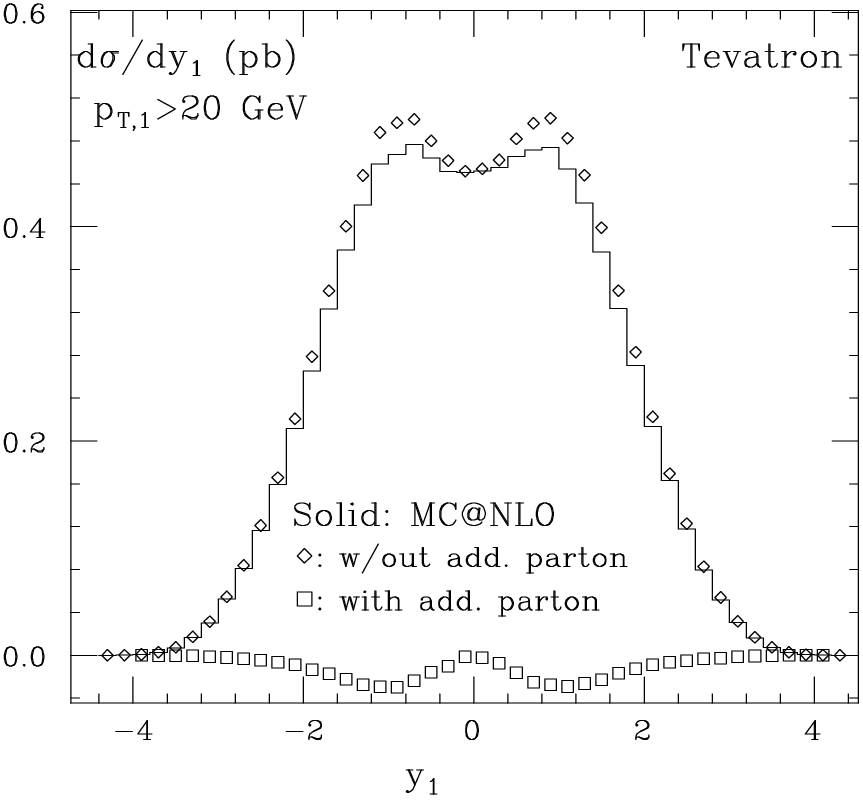}
\hspace*{0.5cm}
\includegraphics[width=0.25\textwidth,clip,]{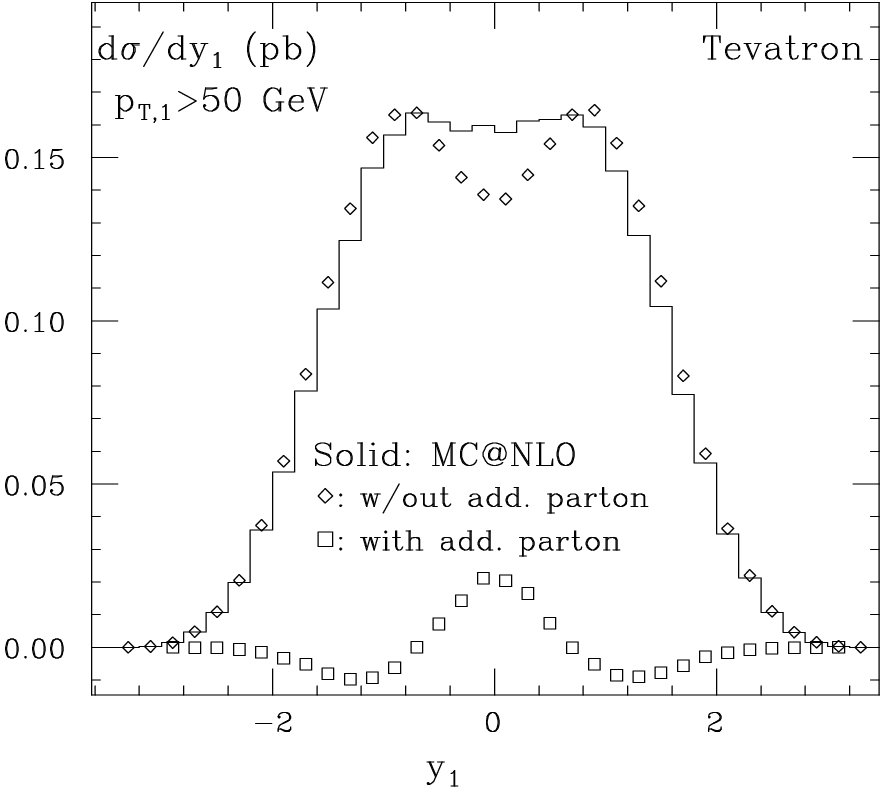}
\hspace*{0.5cm}
\includegraphics[width=0.25\textwidth,clip,]{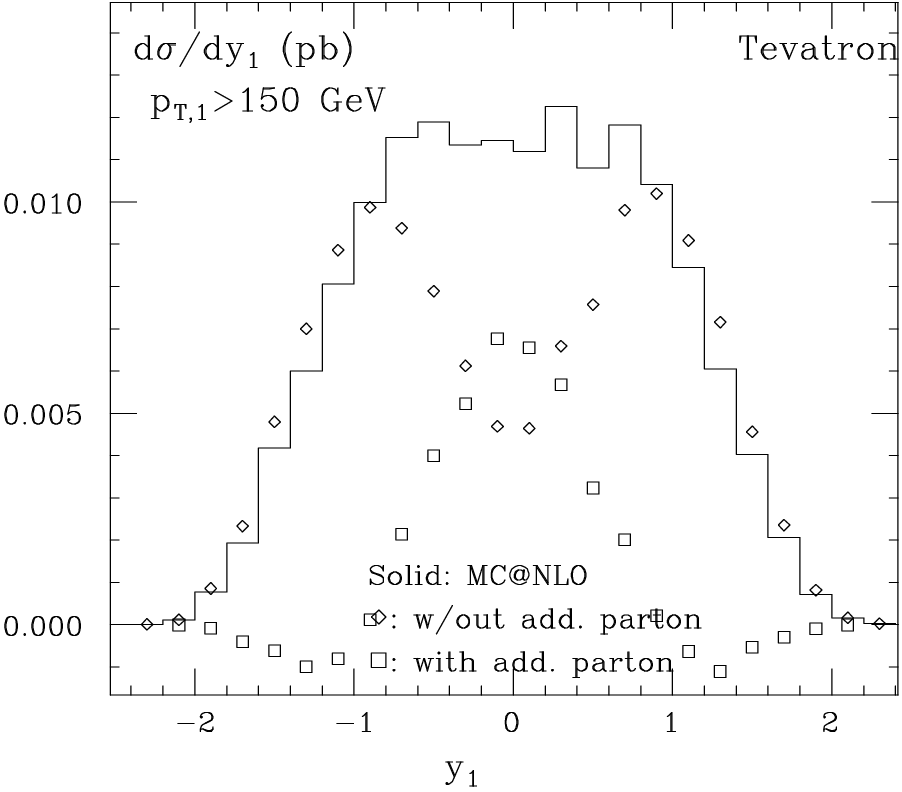}
\\
\includegraphics[width=0.25\textwidth,clip,]{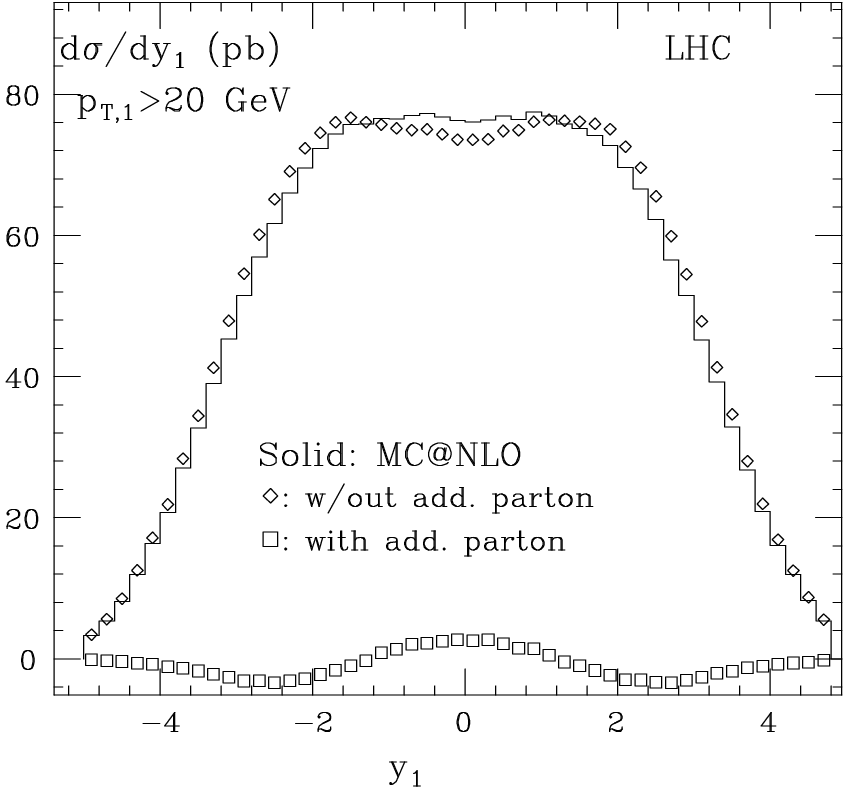}
\hspace*{0.5cm}
\includegraphics[width=0.25\textwidth,clip,]{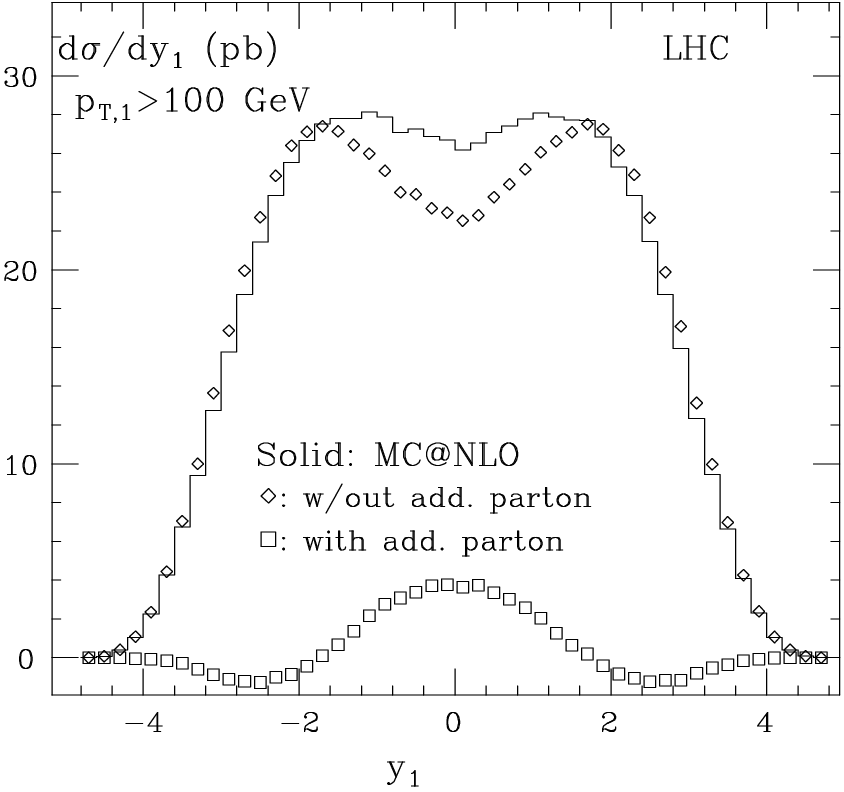}
\hspace*{0.5cm}
\includegraphics[width=0.25\textwidth,clip,]{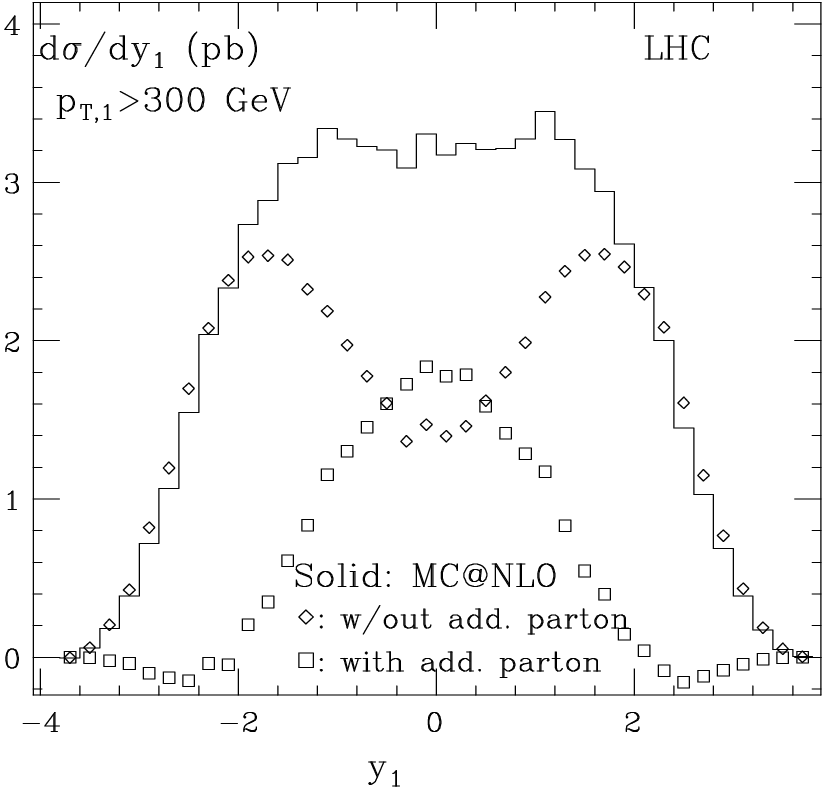}
\\
\ccaption{}{\label{fig:01y} 
Rapidity of the leading jet $\yjet$
as described by \mcnlo. The plots show the results
for various jet \pt\ thresholds. Upper set: Tevatron, lower set: LHC} 
\end{center}
\end{figure}
\begin{figure}
\begin{center}
\includegraphics[width=0.25\textwidth,clip,]{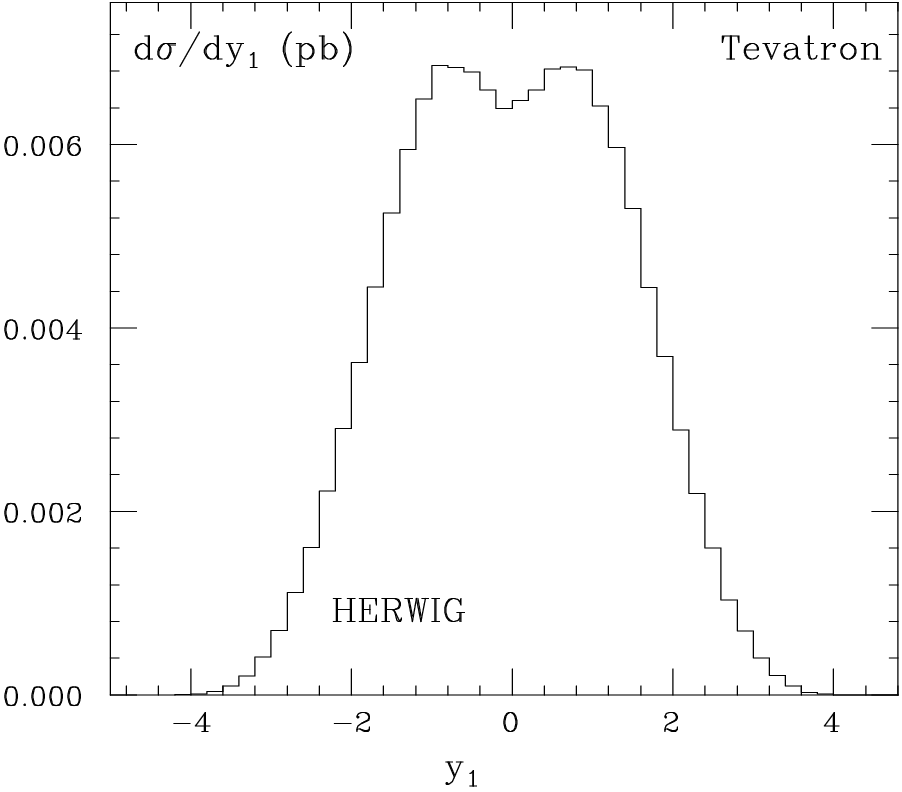}
\hspace*{0.5cm}
\includegraphics[width=0.25\textwidth,clip,]{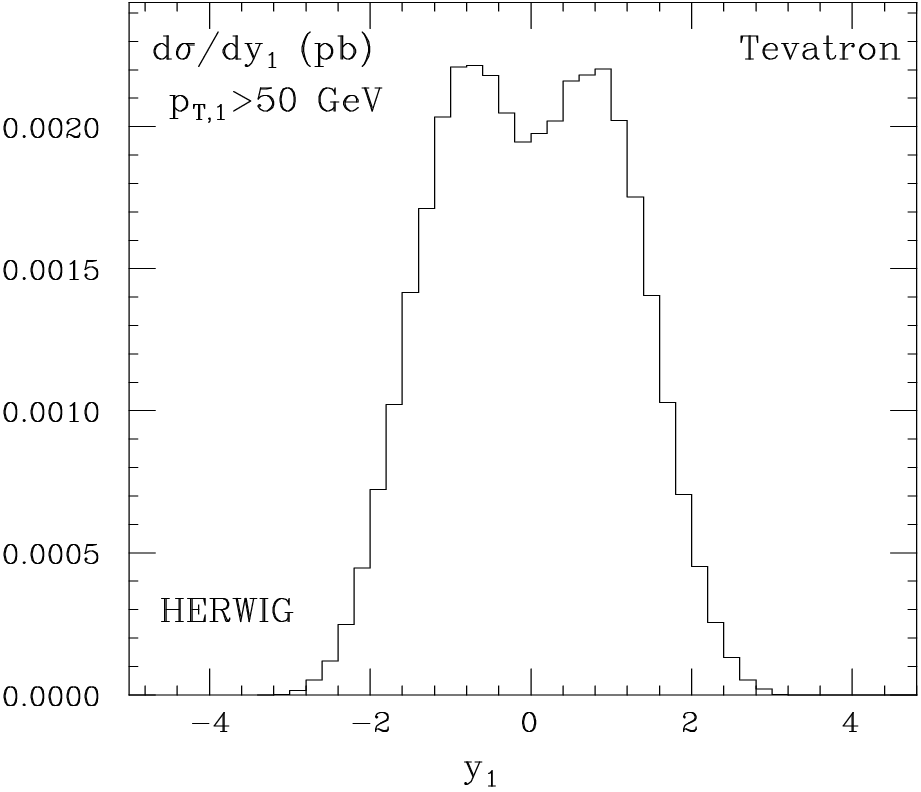}
\hspace*{0.5cm}
\includegraphics[width=0.25\textwidth,clip,]{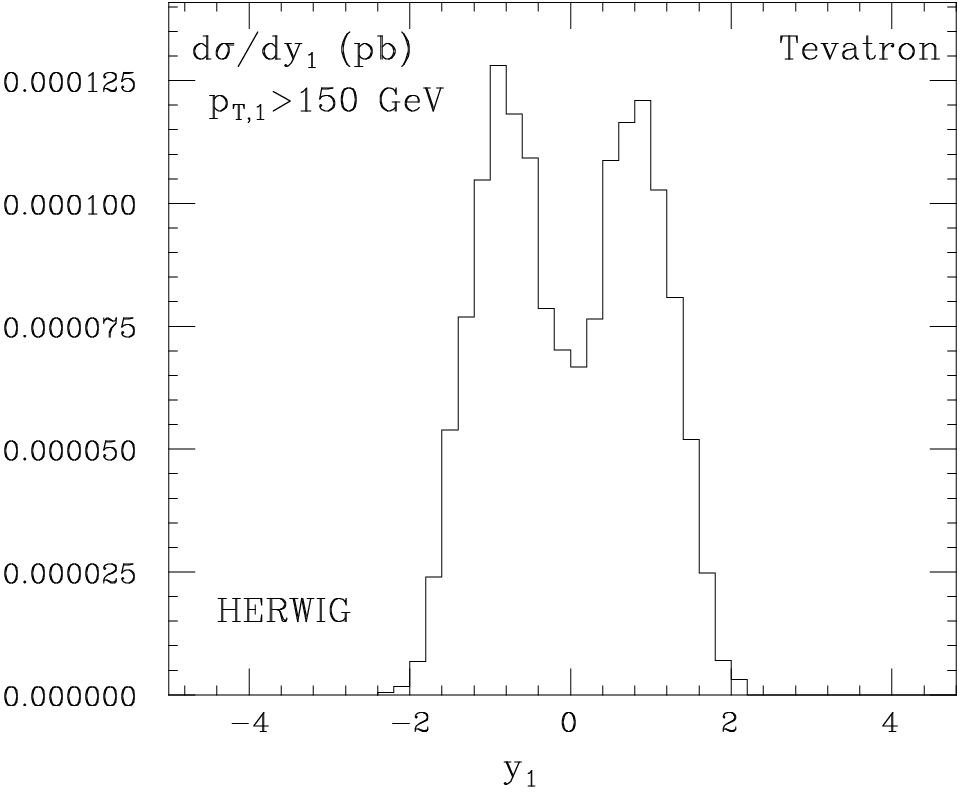}
\\
\includegraphics[width=0.25\textwidth,clip,]{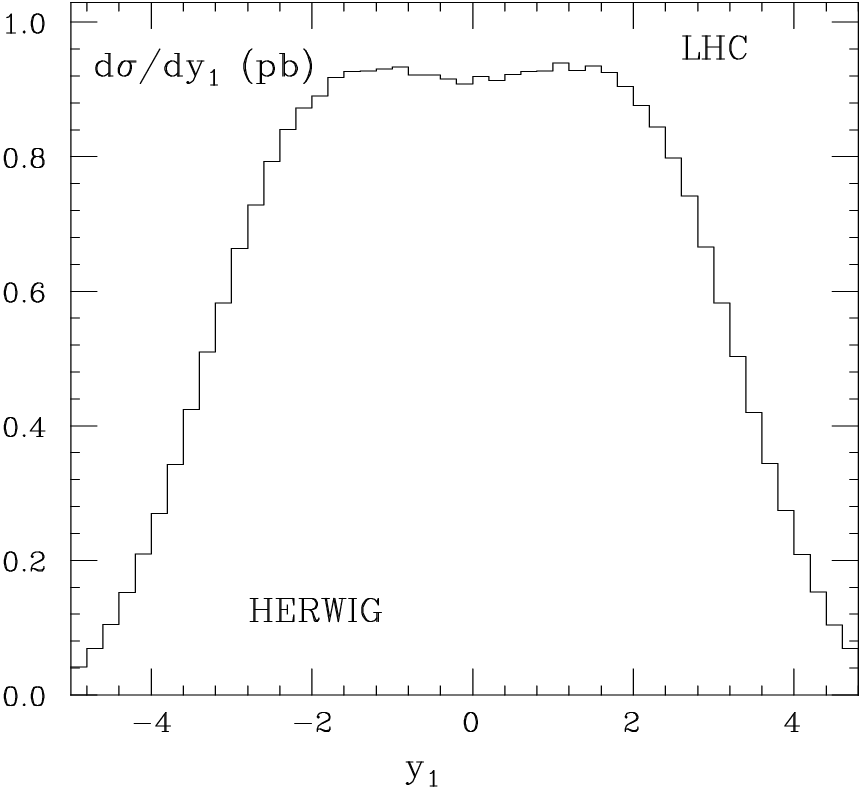}
\hspace*{0.5cm}
\includegraphics[width=0.25\textwidth,clip,]{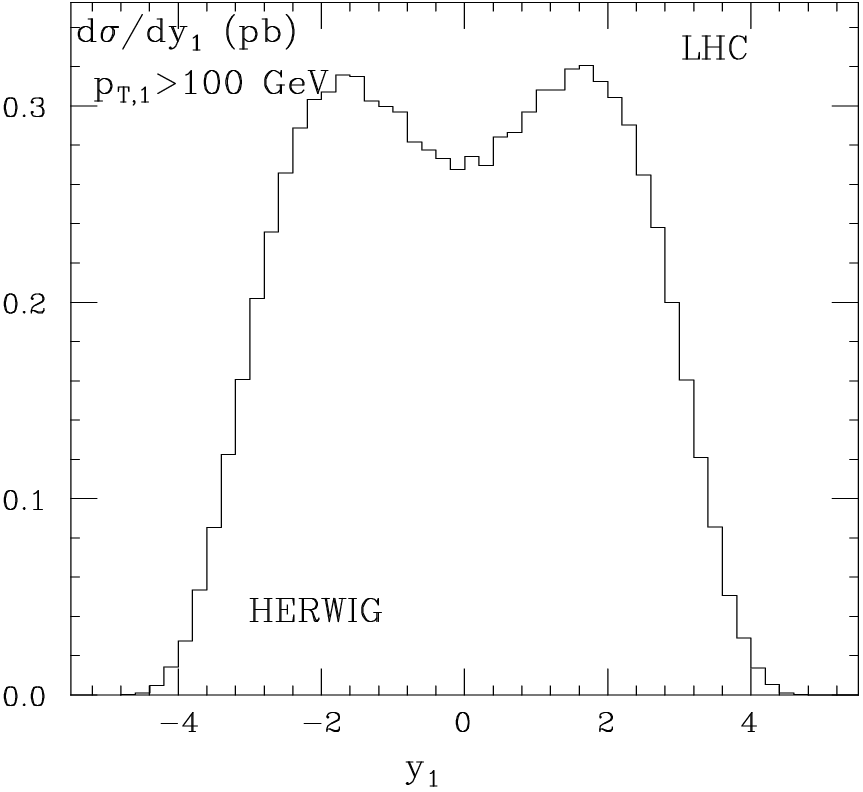}
\hspace*{0.5cm}
\includegraphics[width=0.25\textwidth,clip,]{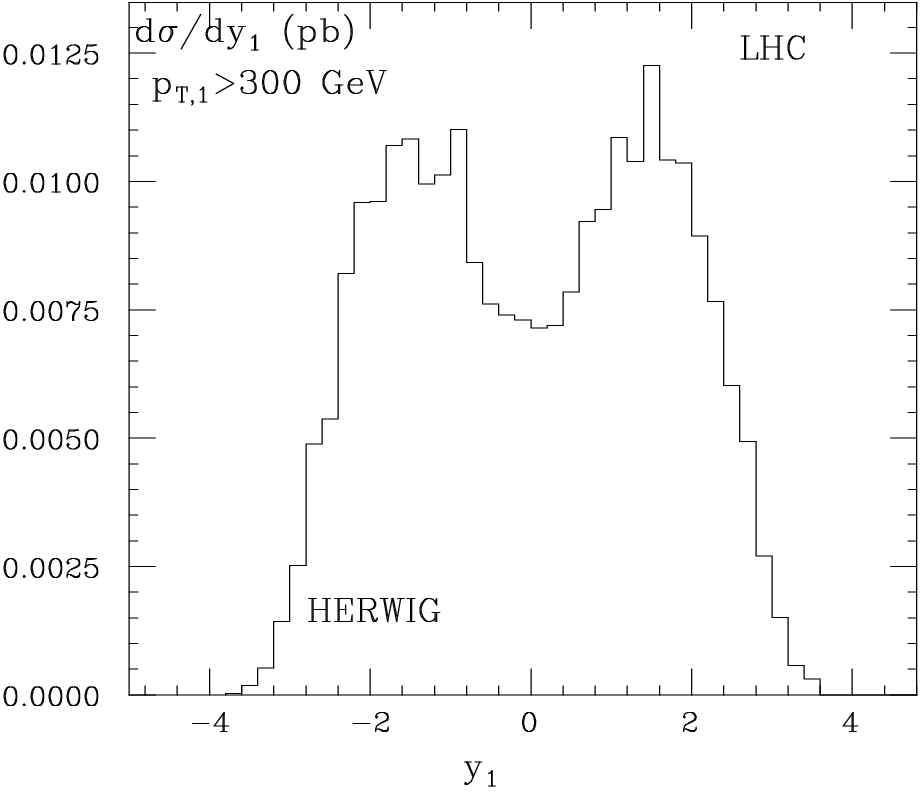}
\\
\ccaption{}{\label{fig:01y-her} 
Rapidity of the leading jet $\yjet$
as described by \herwig. The plots show the results
for various jet \pt\ thresholds. Upper set: Tevatron, lower set: LHC} 
\end{center}
\end{figure}
To understand the difference in the rapidity distribution,
we look in more detail in
fig.~\ref{fig:01mcnlo} at some
features in the \mcnlo\ description of the leading jet.
For the \pt\ of the leading jet, \ptja,  we plot separately the contribution from the various components
of the \mcnlo\ generation: 
events in which the shower is initiated by the LO $t \bar t$ hard process,
and  events in which the shower is initiated by 
a $t \bar t + q (g)$ hard process. In this last case, we separate the
contribution of positive- and negative-weight events, where the
distribution of negative events is shown in absolute value.
The plots show that for \mcnlos
the contribution of the $t \bar t + q (g)$ hard process is 
almost negligible over most of the relevant range and becomes appreciable
only for very large values of $\ptja$.
This hierarchy is stronger at the LHC than at the Tevatron. 

Figure.~\ref{fig:01y} shows the various contributions to the rapidity
distribution \yjet\ for different
jet \pt\ thresholds. It
appears that the \yjet\ distribution resulting from the shower evolution
of the \ttbar\ events in \mcnlo\ has a strong dip at \yjet=0, a dip that
cannot be compensated by the more central distributions of the jet
from the  $t \bar t + q (g)$ hard process, given its marginal role in
the overall jet rate. 

\begin{figure}
\begin{center}
\includegraphics[height=0.15\textheight,clip]{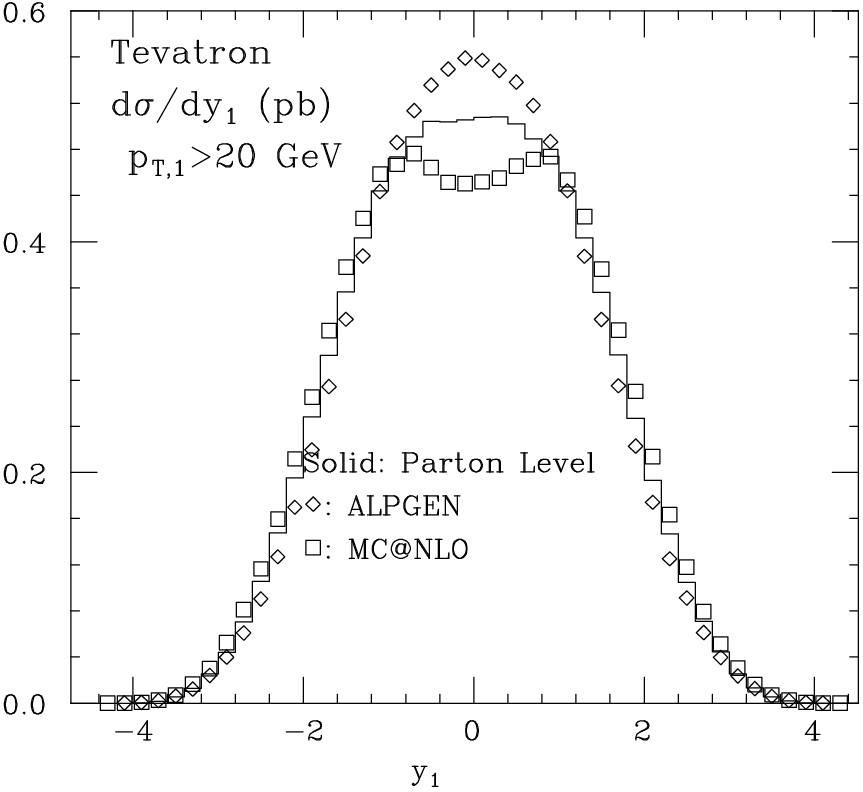}
\includegraphics[height=0.15\textheight,clip]{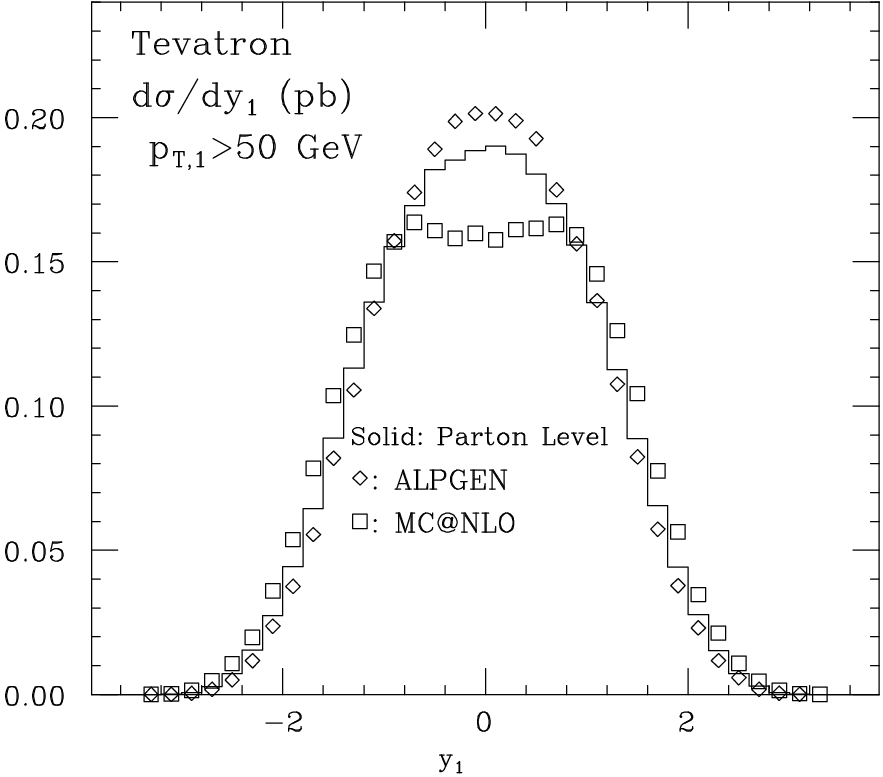}
\includegraphics[height=0.15\textheight,clip]{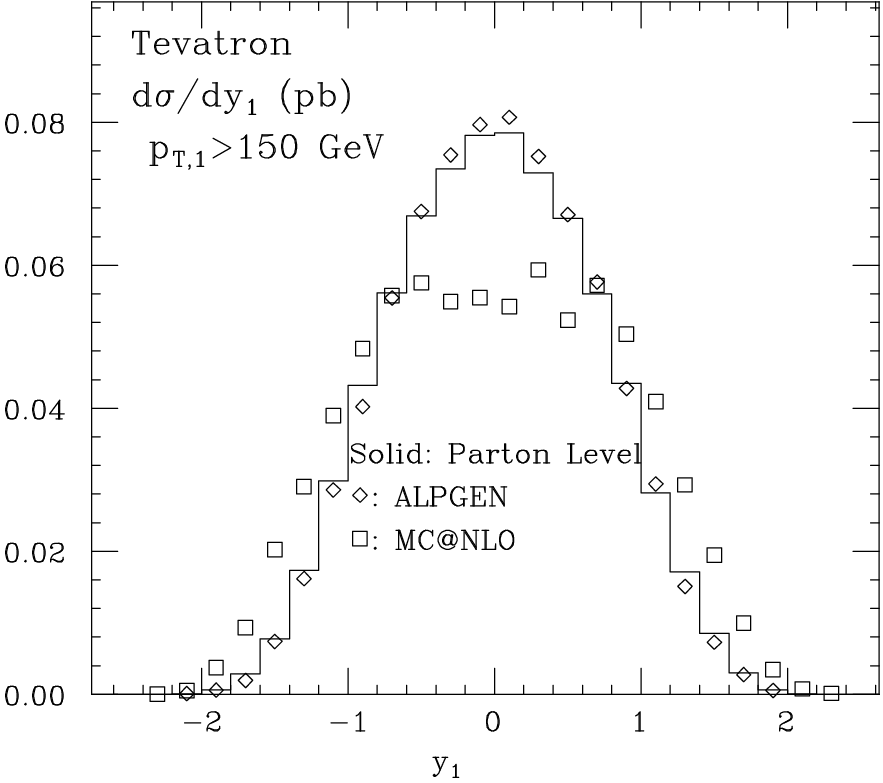}
 \\
\includegraphics[height=0.15\textheight,clip]{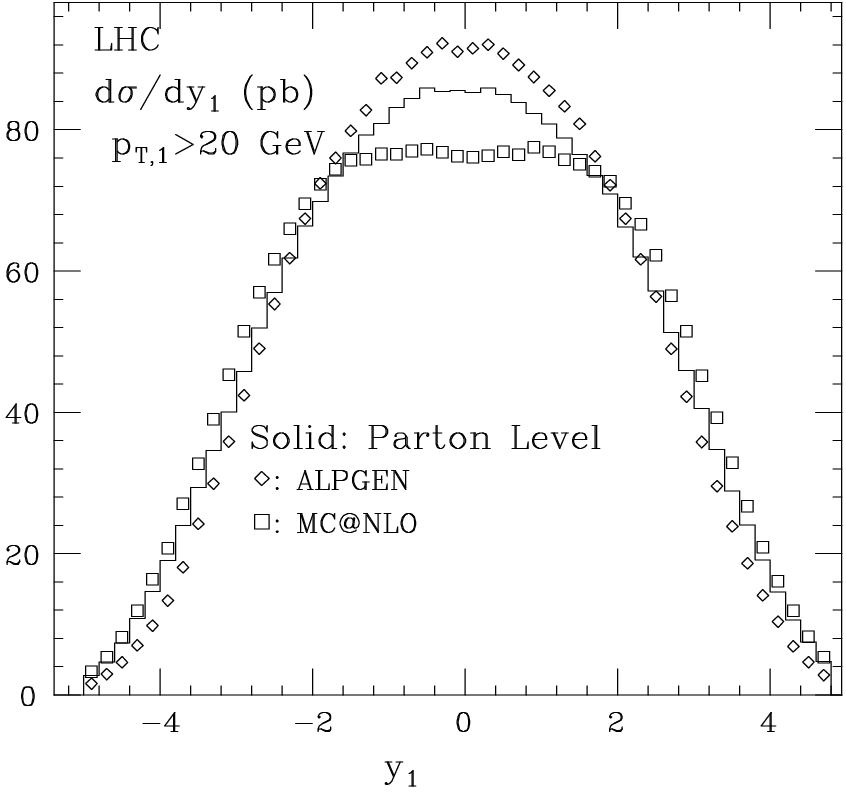}
\includegraphics[height=0.15\textheight,clip]{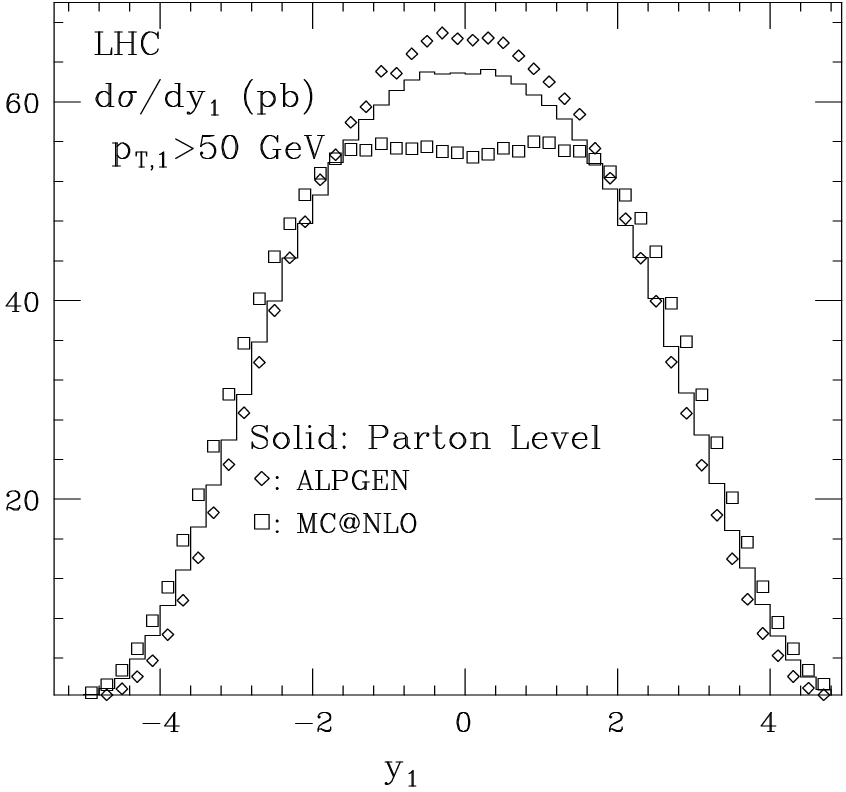}
\includegraphics[height=0.15\textheight,clip]{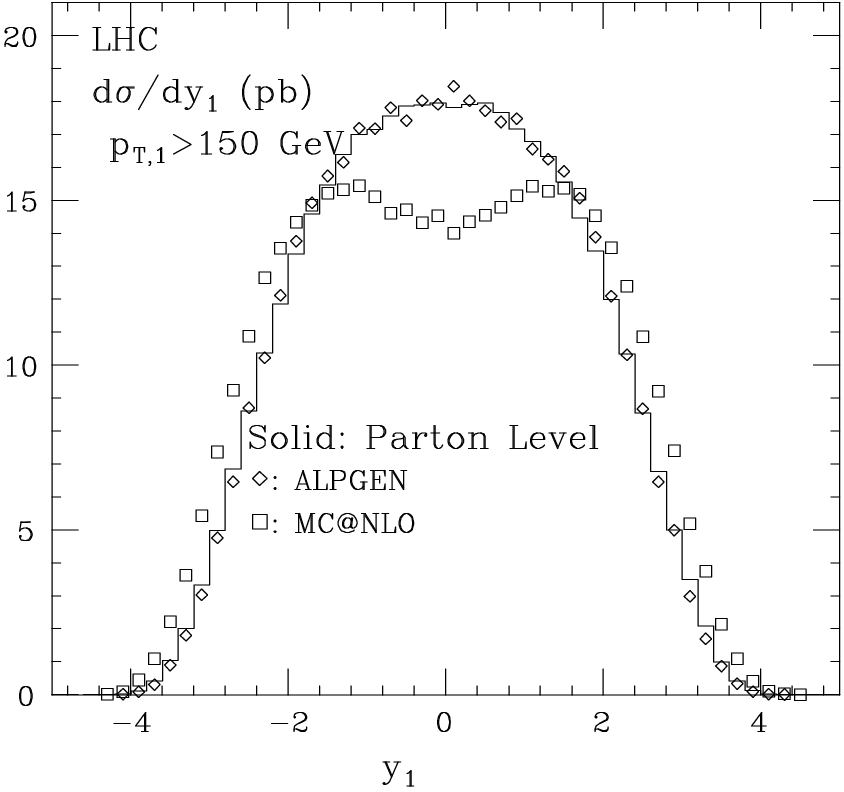}
 \\
\ccaption{}{\label{fig:MCvsALPvsPL} Rapidity spectrum of the leading
 jet, as predicted by \alpgen, \mcnlo, and by the parton
 level, for various \pt\ thresholds of the jet. Upper curves:
 Tevatron; lower curves: LHC.}
\end{center}
\end{figure}

That the
dip at \yjet=0 is a feature typical of jet emission from the \ttbar\
state in \herwig\ is shown in fig.~\ref{fig:01y-her}, obtained from
the standard \herwig\ code rather than from \mcnlo. 
We speculate that this feature is a consequence of the dead-cone
description of hard emission from heavy quarks implemented in the
\herwig\ shower algorithm.
To complete our analysis, we show in fig.~\ref{fig:MCvsALPvsPL} the
comparison between the \alpgen, \mcnlo\ and the parton-level \yjet\
spectra, for different jet \pt\ thresholds. We notice that at large
\pt, where the Sudakov effects that induce potential differences
between the shower and the PL results have vanished, the \alpgen\
result reproduces well the PL result, while still differing
significantly from the \mcnlo\ distributions.

\clearpage


\subsection {Spin correlations in top decays}
\label{sec:topdec}
Top decays are described differently in the two codes. In \mcnlo\
the top quark is assumed stable at the parton level and it is then \herwigs
that models the decay: production and decay are thus uncorrelated, and
spin correlations are missing. In \alpgen, on the other hand, spin
correlations are taken into account in the evaluation of the matrix
elements, and the proper correlations are then preserved by the shower
evolution. This is a minor issue for \mcnlo, which is being
addressed in its forthcoming releases\footnote{S. Frixione, private
  communication.}.  We show here nevertheless a study
of the impact of spin correlations, to conclude that indeed it is
important to keep track of them for a reliable simulation of the final
state kinematics.

\begin{figure}[h]
\begin{center}
\includegraphics[height=0.2\textheight,clip,]{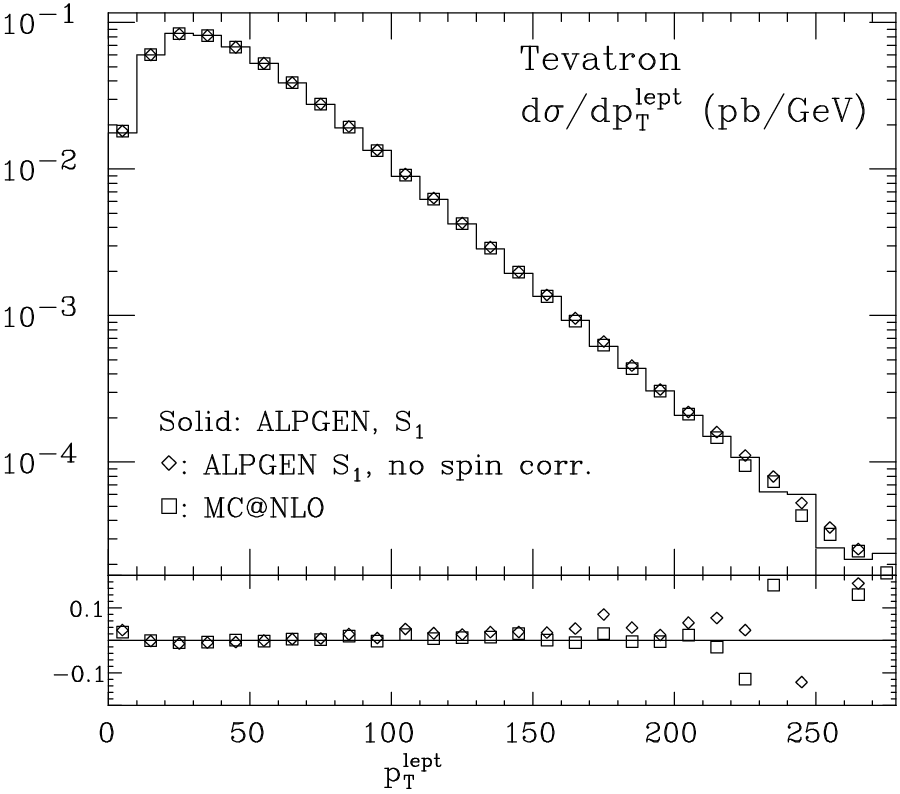}
\hspace*{0.5cm}
\includegraphics[height=0.2\textheight,clip,]{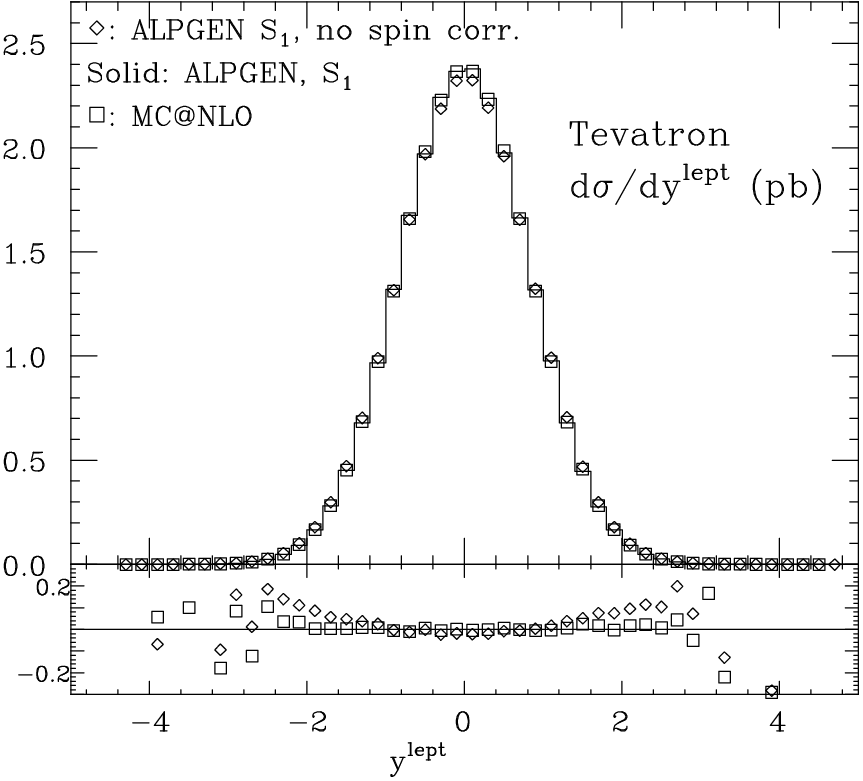}
\\
\includegraphics[height=0.2\textheight,clip,]{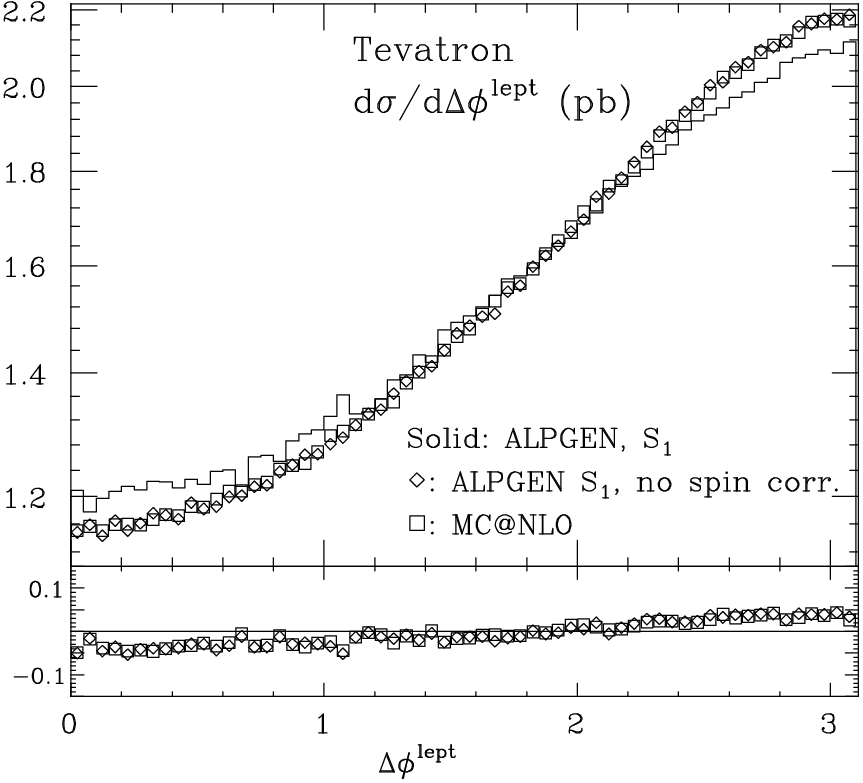}
\hspace*{0.5cm}
\includegraphics[height=0.2\textheight,clip,]{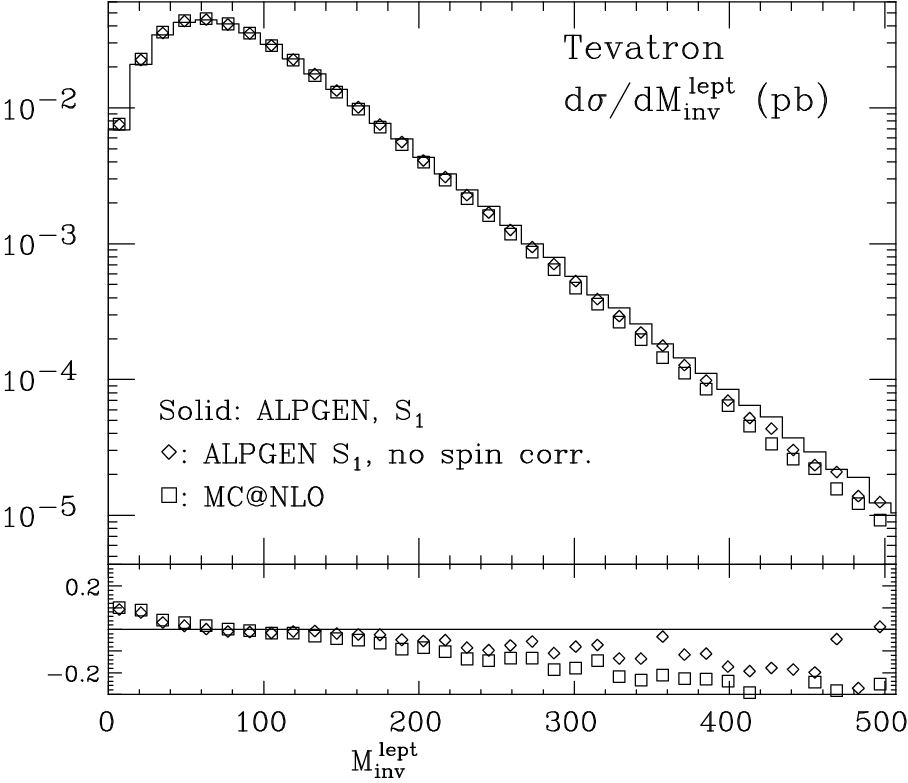}
\ccaption{}{\label{fig:lepTEV}
Leptonic distributions for the default \alpgen, for \alpgen\ without
spin correlations in top decays, and for \mcnlo.}
\end{center}
\end{figure}

To this end we have  selected the leptonic decay channel for both
top and antitop, and studied, after showering, several dilepton
distributions.  For simplicity we just present the results for the
Tevatron, since those for the LHC exhibit the same features.  In
fig.~\ref{fig:lepTEV} we plot the transverse momentum $p_T^{\sss{lept}}$ and
the rapidity $y^{\sss{lept}}$ of the leading lepton, the invariant dilepton
mass and the azimuthal difference $\dphil$ between the two leptons in the
tranverse plane. For \alpgen\ we plot the distribution with and
without spin correlation taken into account.  The angular separation
$\dphil$ and the invariant charged dilepton mass exhibit some
sensitivity to spin correlation which is more evident at higher
energies. The other quantities look fairly insensitive to spin
correlations. Notice that, as expected, \mcnlo\ behaves exactly like
\alpgen\ without spin correlations.

\section {Conclusions}
The study presented in this paper examines the predictions of \alpgen\ and
its matching algorithm for the description of \ttbar+jets
events. Several checks of the algorithm have
shown its internal consistency, and indicate a rather mild dependence
of the results on the parameters that define it. 
The consistency of the approach is confirmed by the comparison with
\mcnlo. In particular, inclusive variables sensitive to the matching
at the transition between the \oatwo\ and \oacube\ matrix elements
(such as the transverse momentum of the \ttbar\ pair) show excellent
agreement, once the NLO/LO $K$ factor is included. 

We found, on the other hand, a rather surprising difference between
the predictions of two codes for the rapidity distribution of the
leading jet accompanying the \ttbar\ pair. At large \pt\ one expects
the jet spectrum to agree with the LO, \oacube, parton level
calculation. This agreement is verified in the \alpgen\ calculation,
but is not present in the case of \mcnlo. In view of the relevance of
this variable for the study at the LHC of new physics signals including jets in
association with top quark pairs (such as \ttbar $H$), it is important to
further pursue the origin of this discrepancy, with independent
calculations, and with a direct comparison with data. Preliminary
results~\cite{Nason:2006XX} obtained with the
new positive-weight NLO shower MC introduced
in~\cite{Nason:2004rx,Nason:2006hf}, appear to support the
distributions predicted by \alpgen. It would also be very interesting to
verify whether the Tevatron statistics is sufficient to directly probe
this observable, and conclusively resolve this issue.

\section*{Acknowledgements}

M. Moretti, F. Piccinini and M. Treccani wish to thank the CERN Theory 
unit for its kind hospitality and support.


\begin{thebibliography}{99}                                            
\bibitem{Mangano:2005dj}
  M.~L.~Mangano and T.~J.~Stelzer,
  Ann.\ Rev.\ Nucl.\ Part.\ Sci.\  {\bf 55} (2005) 555.
\bibitem{Dobbs:2004qw}
  M.~A.~Dobbs {\it et al.},
  arXiv:hep-ph/0403045.
\bibitem{Frixione:2002ik}
  S.~Frixione and B.~R.~Webber,
  JHEP {\bf 0206} (2002) 029
  [arXiv:hep-ph/0204244].
\bibitem{Nason:2004rx}
  P.~Nason,
  JHEP {\bf 0411} (2004) 040
  [arXiv:hep-ph/0409146].
\bibitem{Kuhn:2000cr}
  R.~Kuhn, A.~Schalicke, F.~Krauss and G.~Soff,
  arXiv:hep-ph/0012025.
\bibitem{Catani:2001cc}
  S.~Catani, F.~Krauss, R.~Kuhn and B.~R.~Webber,
  JHEP {\bf 0111} (2001) 063
  [arXiv:hep-ph/0109231].
\bibitem{Lonnblad:2001iq}
  L.~Lonnblad,
  JHEP {\bf 0205} (2002) 046
  [arXiv:hep-ph/0112284].
\bibitem{Krauss:2002up}
  F.~Krauss,
  JHEP {\bf 0208} (2002) 015
  [arXiv:hep-ph/0205283].
\bibitem{mlmfnal}
M.L. Mangano, presentation at the FNAL Matrix Element/Monte Carlo
Tuning Working Group, 15 Nov 2002,
http://www-cpd.fnal.gov/personal/mrenna/tuning/nov2002/mlm.pdf .
\bibitem{Mangano-ta}
M.L. Mangano, to appear.
\bibitem{Mrenna:2003if}
  S.~Mrenna and P.~Richardson,
  JHEP {\bf 0405} (2004) 040
  [arXiv:hep-ph/0312274].
\bibitem{Lavesson:2005xu}
  N.~Lavesson and L.~Lonnblad,
  JHEP {\bf 0507} (2005) 054
  [arXiv:hep-ph/0503293].
\bibitem{Hoche:2006ph} S.~Hoche, F.~Krauss, N.~Lavesson, L.~Lonnblad,
  M.~Mangano, A.~Schalicke and S.~Schumann, 
arXiv:hep-ph/0602031.  
\bibitem{Nason:2006hf}
  P.~Nason and G.~Ridolfi,
  JHEP {\bf 0608} (2006) 077
  [arXiv:hep-ph/0606275].
\bibitem{Frixione:2003ei}
  S.~Frixione, P.~Nason and B.~R.~Webber,
  JHEP {\bf 0308} (2003) 007
  [arXiv:hep-ph/0305252].
\bibitem{Frixione:2003vm}
  S.~Frixione and B.~R.~Webber,
  arXiv:hep-ph/0309186.
\bibitem{Nason:2006XX}
  P.~Nason, presented at the 3rd meeting of the
  ``Workshop sui Monte Carlo, la fisica e le simulazioni a LHC'',
  Frascati, 23-25 October 2006,
 {\tt  http://moby.mib.infn.it/\~nason/mcws3/Nason-10-2006.pdf}.
\bibitem{Frixione:2005vw}
  S.~Frixione, E.~Laenen, P.~Motylinski and B.~R.~Webber,
  JHEP {\bf 0603} (2006) 092
  [arXiv:hep-ph/0512250].
\bibitem{Mangano:2002ea}
  M.~L.~Mangano, M.~Moretti, F.~Piccinini, R.~Pittau and A.~D.~Polosa,
  JHEP {\bf 0307} (2003) 001
  [arXiv:hep-ph/0206293];\\
F.~Caravaglios, M.~L.~Mangano, M.~Moretti and R.~Pittau,
Nucl.\ Phys.\ B {\bf 539} (1999) 215
[hep-ph/9807570].

\bibitem{Marchesini:1988cf}
G.~Marchesini and B.~R.~Webber,
Nucl.\ Phys.\ B {\bf 310} (1988) 461.
G.~Marchesini, B.~R.~Webber, G.~Abbiendi, I.~G.~Knowles, M.~H.~Seymour and L.~Stanco,
Comput.\ Phys.\ Commun.\  {\bf 67} (1992) 465.
G.~Corcella {\it et al.},
JHEP {\bf 0101} (2001) 010
[hep-ph/0011363].
\bibitem{Sjostrand:1994yb}
T.~Sjostrand,
Comput.\ Phys.\ Commun.\  {\bf 82} (1994) 74.
T.~Sjostrand, P.~Eden, C.~Friberg, L.~Lonnblad, G.~Miu, S.~Mrenna and
E.~Norrbin,
Comput.\ Phys.\ Commun.\  {\bf 135} (2001) 238
[hep-ph/0010017].
\bibitem{Corcella:2002jc}
  G.~Corcella {\it et al.},
  arXiv:hep-ph/0210213.

\bibitem{Martin:2001es}
  A.~D.~Martin, R.~G.~Roberts, W.~J.~Stirling and R.~S.~Thorne,
  Eur.\ Phys.\ J.\ C {\bf 23} (2002) 73
  [arXiv:hep-ph/0110215].
\bibitem{getjet}
F.E.~Paige and S.D.~Protopopescu, in {\it Physics of the SSC}, Snowmass, 1986, 
Colorado, edited by R. Donaldson and J.~Marx.

\bibitem{Mangano:1991jk}
  M.~L.~Mangano, P.~Nason and G.~Ridolfi,
  Nucl.\ Phys.\ B {\bf 373}, 295 (1992).



\end{thebibliography}
\end{document}